\documentclass[aps,prd,twocolumn,superscriptaddress,preprintnumbers,nofootinbib, floatfix]{revtex4-2}
\usepackage{booktabs}
\usepackage{bigstrut}
\usepackage{xspace}
\usepackage{multirow}
\usepackage{amsmath}
\usepackage{graphicx}
\usepackage{threeparttable}
\usepackage{changepage}
\usepackage{amsfonts}
\usepackage{float}
\usepackage{lineno}
\usepackage{amsmath}
\usepackage{subfigure}
\usepackage{threeparttable} 
\usepackage{bm}
\usepackage{multirow}
\usepackage{array}
\usepackage{xcolor}

\definecolor{linkcolor}{rgb}{0.6,0,0}
\definecolor{citecolor}{rgb}{0,0,0.75}
\definecolor{urlcolor}{rgb}{0.12,0.46,0.7}
\usepackage[breaklinks, colorlinks, urlcolor=urlcolor, linkcolor=linkcolor,citecolor=citecolor,linktocpage=true,pdfencoding=auto]{hyperref}

\newcommand {\beq} {\begin{equation}}
	\newcommand {\eeq} {\end{equation}}
\newcommand*{\LCDM}{$\Lambda$CDM}

\newcommand{\sptg}{{\mbox{\sc SPT-3G}}\xspace}
\newcommand{\sptpol}{\mbox{{SPTpol}}\xspace}
\newcommand{\actpol}{ACT\xspace}
\newcommand{\sptsz}{\mbox{{\sc SPT-SZ}}\xspace}

\newcommand{\lcdm}{$\Lambda$CDM\xspace}
\newcommand{\planck}{\textit{Planck}\xspace}
\newenvironment{rcases}
{\left.\begin{aligned}}
	{\end{aligned}\right\rbrace}
\newcommand{\bea}{\begin{eqnarray}}
\newcommand{\eea}{\end{eqnarray}}
\newcommand{\orcidlink}[1]{\href{https://orcid.org/#1}{\includegraphics[height=6.7pt]{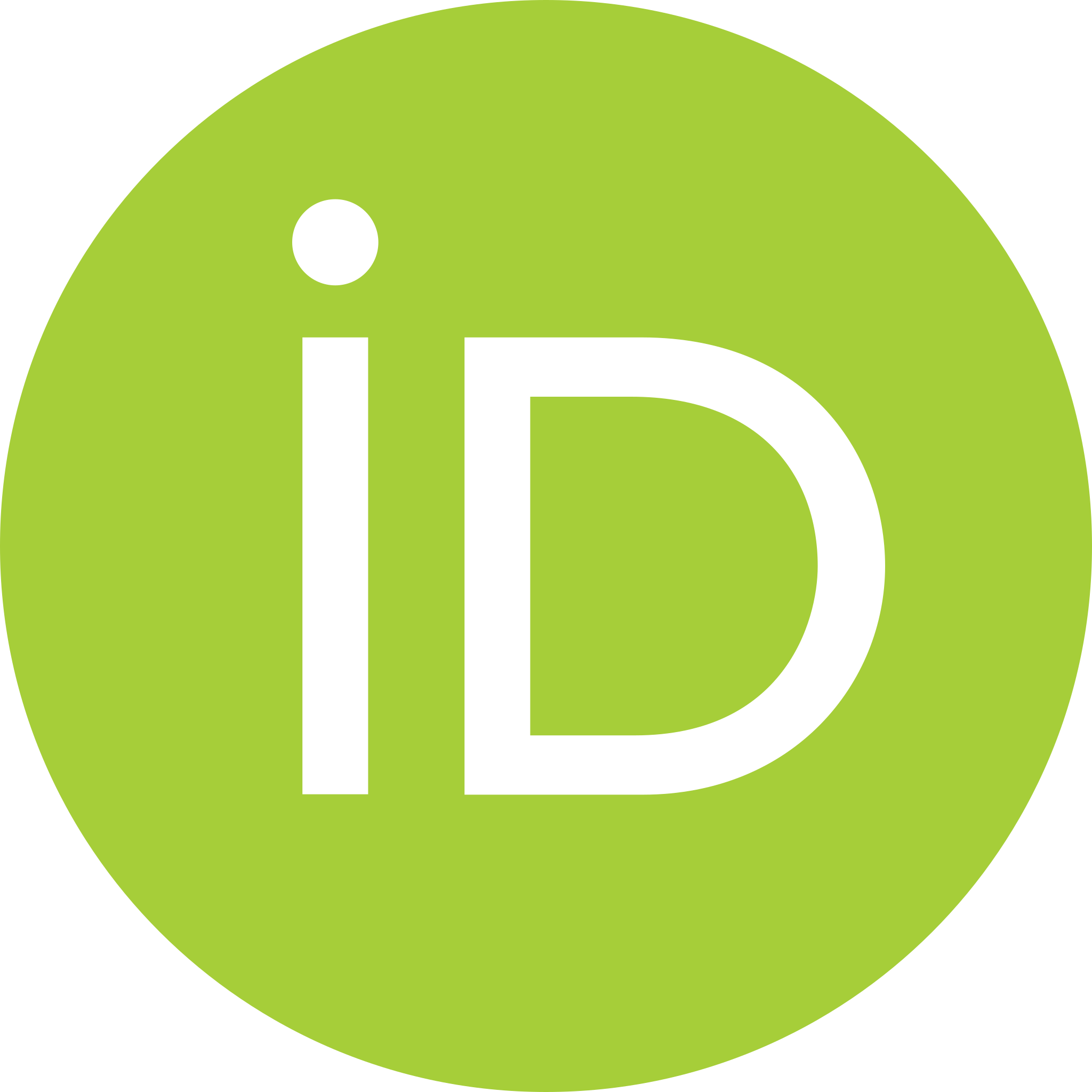}}}

\def\balign{\begin{align}}
\def\ealign{\end{align}}
\def\bl{\pmb{\ell}}
\def\bL{\bm{L}}
\def\mbell{\pmb{\ell}} 

\def\mbL{\bm{L}} 

\def\clpphat{\ensuremath{C^{\hat{\phi} \hat{\phi}}_{L}}}
\def\hatclpp{\ensuremath{\hat{C}^{\phi \phi}_{L}}}

\defcitealias{wu19}{W19}
\defcitealias{omori17}{O17}
\defcitealias{reichardt21}{R21}
\defcitealias{dutcher21}{D21}

\graphicspath{{./}{figures/}}
\begin{document}
\title{A Measurement of Gravitational Lensing of the Cosmic Microwave Background Using SPT-3G 2018 Data}
\author{Z.~Pan\hspace{1pt}\orcidlink{0000-0002-6164-9861}}\email{panz@anl.gov}
\affiliation{High-Energy Physics Division, Argonne National Laboratory, 9700 South Cass Avenue., Lemont, IL, 60439, USA}
\affiliation{Kavli Institute for Cosmological Physics, University of Chicago, 5640 South Ellis Avenue, Chicago, IL, 60637, USA}
\affiliation{Department of Physics, University of Chicago, 5640 South Ellis Avenue, Chicago, IL, 60637, USA}
\author{F.~Bianchini\hspace{1pt}\orcidlink{0000-0003-4847-3483}}\email{fbianc@slac.stanford.edu}
\affiliation{Kavli Institute for Particle Astrophysics and Cosmology, Stanford University, 452 Lomita Mall, Stanford, CA, 94305, USA}
\affiliation{Department of Physics, Stanford University, 382 Via Pueblo Mall, Stanford, CA, 94305, USA}
\affiliation{SLAC National Accelerator Laboratory, 2575 Sand Hill Road, Menlo Park, CA, 94025, USA}
\author{W.~L.~K.~Wu\hspace{1pt}\orcidlink{0000-0001-5411-6920}}
\affiliation{Kavli Institute for Particle Astrophysics and Cosmology, Stanford University, 452 Lomita Mall, Stanford, CA, 94305, USA}
\affiliation{SLAC National Accelerator Laboratory, 2575 Sand Hill Road, Menlo Park, CA, 94025, USA}
\author{P.~A.~R.~Ade}
\affiliation{School of Physics and Astronomy, Cardiff University, Cardiff CF24 3YB, United Kingdom}
\author{Z.~Ahmed}
\affiliation{Kavli Institute for Particle Astrophysics and Cosmology, Stanford University, 452 Lomita Mall, Stanford, CA, 94305, USA}
\affiliation{SLAC National Accelerator Laboratory, 2575 Sand Hill Road, Menlo Park, CA, 94025, USA}
\author{E.~Anderes}
\affiliation{Department of Statistics, University of California, One Shields Avenue, Davis, CA 95616, USA}
\author{A.~J.~Anderson\hspace{1pt}\orcidlink{0000-0002-4435-4623}}
\affiliation{Fermi National Accelerator Laboratory, MS209, P.O. Box 500, Batavia, IL, 60510, USA}
\affiliation{Kavli Institute for Cosmological Physics, University of Chicago, 5640 South Ellis Avenue, Chicago, IL, 60637, USA}
\author{B.~Ansarinejad}
\affiliation{School of Physics, University of Melbourne, Parkville, VIC 3010, Australia}
\author{M.~Archipley\hspace{1pt}\orcidlink{0000-0002-0517-9842}}
\affiliation{Department of Astronomy, University of Illinois Urbana-Champaign, 1002 West Green Street, Urbana, IL, 61801, USA}
\affiliation{Center for AstroPhysical Surveys, National Center for Supercomputing Applications, Urbana, IL, 61801, USA}
\author{K.~Aylor}
\affiliation{Department of Physics \& Astronomy, University of California, One Shields Avenue, Davis, CA 95616, USA}
% ...

\begin{abstract}

We present a measurement of gravitational lensing over 1500 deg$^2$ of the Southern sky using \sptg
 temperature data at 95 and 150~GHz taken in 2018.
The lensing amplitude relative to a fiducial \planck{} 2018 Lambda-Cold Dark Matter (\lcdm{}) cosmology is found to be $1.020\pm0.060$,
excluding instrumental and astrophysical systematic uncertainties.  
We conduct extensive systematic and null tests to check the robustness of the lensing measurements, and
report a minimum-variance combined lensing power spectrum over angular multipoles of $50<L<2000$, which we use to
constrain cosmological models.
When analyzed alone and jointly with primary cosmic microwave background (CMB) spectra within the \lcdm{} model, our lensing amplitude measurements are consistent with measurements from \sptsz{}, \sptpol{}, \actpol{}, and \planck{}.
 Incorporating loose priors on the baryon density and other parameters including uncertainties on a foreground bias template, we obtain a $1\sigma$
 constraint on $\sigma_8 \Omega_{\rm m}^{0.25}=0.595 \pm 0.026$ using
 the \sptg 2018 lensing data alone, where $\sigma_8$ is a common measure of the amplitude of structure today and $\Omega_{\rm m}$ is the matter density parameter.
Combining \sptg 2018 lensing measurements with baryon acoustic oscillation (BAO) data, we derive parameter constraints of  $\sigma_8 = 0.810 \pm 0.033$, $S_8 \equiv \sigma_8(\Omega_{\rm m}/0.3)^{0.5}=  0.836 \pm 0.039$, and Hubble constant $H_0 =68.8^{+1.3}_{-1.6}$\text{ km s$^{-1}$ Mpc$^{-1}$}.
Our preferred $S_8$ value is higher by 1.6 to 1.8$\sigma$ compared to cosmic shear measurements from DES-Y3, HSC-Y3, and KiDS-1000 at lower redshift and smaller scales.
We combine our lensing data with CMB anisotropy measurements from both \sptg{} and
 \planck{} to constrain extensions of \lcdm{}.  Using CMB anisotropy and lensing measurements from \sptg{}  only, we provide independent constraints on the spatial curvature of
 $\Omega_{K} = 0.014^{+0.023}_{-0.026}$ (95\% C.L.) and the dark energy density of
 $\Omega_\Lambda = 0.722^{+0.031}_{-0.026}$ (68\% C.L.). When combining \sptg{} lensing data with \sptg{} CMB anisotropy and BAO data, we find an upper limit on the sum of the neutrino masses of $\sum m_{\nu}< 0.30$~eV (95\% C.L.).
Due to the different combination of angular scales and sky area, this lensing analysis provides an independent check on lensing measurements by  \actpol and \planck.

\end{abstract}

\keywords{Cosmic microwave background --- Gravitational lensing --- South Pole Telescope --- Growth of structure}
\maketitle

\section{Introduction} \label{sec:intro}

Photons from the cosmic microwave background (CMB) are deflected by the intervening
gravitational potentials of the large-scale structure (LSS) as
they travel to us from the surface of last scattering~\citep[e.g.,][]{lewis06}.
The distortion of the primordial CMB by gravitational lensing
provides a unique way to map the projected matter distribution of the universe,
as lensing introduces correlations between CMB fluctuations on different angular scales.
We can leverage these correlations to reconstruct
 the underlying projected matter over- and under-densities and
measure the CMB lensing potential power spectrum, from which we can infer the underlying matter power spectrum.
Thanks to the high redshift ($z\simeq 1100$) of the CMB, lensing measurements contain
 LSS information
from the last scattering surface to the present day, with the maximum of the lensing kernel around redshift of 2.
Lensing measurements can therefore probe the large-scale structure and can inform many key topics in cosmology, including the amplitude of matter density fluctuations  \citep{bianchini20a,qu23}, the mass of the neutrinos \citep{kaplinghat03, lesgourgues06, lesgourgues06b, deputter09, pan15}, the nature of dark energy \citep{sherwin11}, and gravity \citep{calabrese09, namikawa18, singh18, zhang20}.

Lensing measurements have been made with data from several experiments, including ACT \citep{das11, sherwin17, darwish20, qu23},
BICEP/Keck \citep{bicep2keck16b, ade23},
\planck \citep{planck13-17, planck15-15, planck18-8, carron22},
POLARBEAR \citep{polarbear14c, polarbear19b}, and
SPT \citep{vanengelen12, story15, omori17, wu19, millea21, omori23}.
The tightest lensing amplitude measurements currently come from the DR6 dataset
by Advanced ACT~\citep[\actpol hereafter; 2.3\%,][]{qu23} and from \planck{} \texttt{NPIPE} maps~\citep[2.4\%, ][]{carron22}.

This work presents the first lensing measurement from \sptg{}, the current camera  
on the South Pole Telescope, using data taken during the abbreviated 2018 season when only 
a subset of the detectors in the focal plane were fully operational. CMB primary anisotropy cosmology 
results from the \sptg{} 2018 data are published in \citet[hereafter 
\citetalias{dutcher21}]{dutcher21} and Balkenhol et al. \cite{balkenhol21, balkenhol23}.

The focuses of this paper are lensing power spectrum measurements from the \sptg{} 2018 
data and their cosmological implications, though we also show the reconstructed lensing 
maps. Compared to previous SPT lensing measurements, our input maps have higher noise 
than those used in the \sptpol{} measurements presented in \citet[hereafter 
\citetalias{wu19}]{wu19} and cover a smaller patch than the temperature-only \sptsz{} 
measurements in \citet[hereafter \citetalias{omori17}]{omori17}.\footnote{In \citetalias{omori17}, \sptsz{} and \planck{} maps are inverse-variance combined over the 2500 square degree \sptsz{}  observing field before lensing reconstruction. Most of the 
lensing $S/N$ comes from the \sptsz{} maps.} 
We use temperature data for this 
lensing reconstruction, and the resulting \sptg{} lensing map's $S/N$ per mode is lower than 
that from \sptpol{} and higher than that from \sptsz{}. However, because of the larger area of 
\sptg{} compared to \sptpol{} and lower noise compared to \sptsz{}, we are able to constrain 
the lensing amplitude with uncertainties similar to both previous measurements at 
$\simeq$6\%.

With this stringent lensing measurement, the \sptg{} 2018 data already enables competitive 
constraints on cosmological parameters, alone and in combination with external datasets. 
The constraints are particularly interesting in light of the current tensions in cosmology, in 
which cosmological parameters inferred using different probes, each with high precision, do 
not agree with each other. Specifically, measurements of $H_0$ from the Cepheid-calibrated 
local distance ladder and the CMB from \planck are in tension at the 5.7$\sigma$ level \citep{planck18-6, 
	freedman21, riess22, murakami23}. 
Additionally, the structure growth parameter, $S_8$, inferred from weak lensing 
measurements from optical galaxy surveys, shows a discrepancy with the value suggested by 
CMB data \citep{asgari21, joudaki19, abbott22a}. Experiments that are 
relatively independent, such as \sptg{}, \actpol{}, and \planck{}, provide cross-checks, allowing for 
a more detailed investigation of these tensions.

This paper is organized as follows. 
We describe the data used in this analysis in Sec.~\ref{sect:data} and the simulations in Sec.~\ref{sect:simulations}.
We then summarize the lensing analysis steps in Sec.~\ref{sect:method}.
We present the lensing maps, power spectra, and amplitude parameters in Sec.~\ref{sect:results}. 
We also discuss the robustness of the results in the same section.
In Sec.~\ref{sect:params}, we explore the cosmological implications of our lensing measurements for the \lcdm{} model and extensions, first in isolation and then in combination with BAO and primary CMB data. We conclude in Sec.~\ref{sect:conclusions}.

\section{Data}
\label{sect:data}
In this section, we introduce the telescope and receiver used to collect the raw data,
the data reduction, and the map-level processing for this lensing analysis.

\subsection{Instrument and CMB Observations}\label{section:instrument}

The South Pole Telescope (SPT) is a 10-meter diameter submillimeter-quality
telescope located at the Geographical South Pole \citep{carlstrom11}.
The currently operating third-generation receiver, SPT-3G \citep{ sobrin22},
has polarization sensitivity and three frequency bands centered at 95, 150, and 220~GHz.
The  combination of high sensitivity from about 16,000 detectors and arcminute angular resolution given the 10-meter primary mirror makes the resulting maps an ideal dataset for CMB lensing analysis.

The main \sptg survey field covers a 1500~deg$^2$ patch of sky extending in right ascension
from  $20^\textrm{h}40^\textrm{m}0^\textrm{s}$ to $3^\textrm{h}20^\textrm{m}0^\textrm{s}$
and in declination from $-42^\circ$ to $-70^\circ$.
We divide this survey field into four subfields centered
at $-44.75^\circ$, $-52.25^\circ$, $-59.75^\circ$, and $-67.25^\circ$
declination to minimize the change in optical loading and detector responsivity during the observation of any one subfield.
The telescope observes using a raster scanning strategy, where it completes a left and right scan over the full azimuth range at constant elevation and then moves in elevation by approximately 12~arcmin until the full elevation range of the subfield is complete.

We use data from the 2018 observing season for this paper.
During 2018, problems with the telescope drive system and receiver resulted in a half-season of observation with only ~50\% of the detectors operational.
The remaining operable detectors had excess low frequency noise due to detector wafer temperature drifts, which can be filtered out during data processing.
Subsequent repairs were undertaken during the 2018 Austral summer, successfully restoring instrument performance and observation efficiency to the anticipated level.
We describe processing of all three bands and use only data from 95 and 150~GHz for
cosmological inference since the 220~GHz channel is three times noisier and the inclusion of data at 220~GHz does not significantly improve the
lensing reconstruction $S/N$ ratio. We do not include polarization
data for the same reason.
As we will show, the 2018 dataset has sufficient depth for competitive measurements of lensing and cosmological parameters.

\subsection{Time-ordered Data to Maps}
\label{subsect:map_making}
The data processing methods are similar to those in \citetalias{dutcher21} with a few major differences. We summarize the steps and differences in this section.

We start with the time-ordered data (TOD) from the detectors and calibrate them
using observations of two Galactic star-forming $\textsc{HII}$ regions: RCW38
and MAT5a (NGC3576).
To reduce the low-$\ell$  noise from atmospheric fluctuations and 
detector wafer temperature drifts, we subtract a 19th-order Legendre polynomial
and remove modes corresponding to multipoles $\lesssim 300$ along the scan direction
from the TOD for each constant-elevation scan.
Additionally, we apply a common-mode filter
where we calculate the averaged signal across all detectors
within the same wafer and frequency band and subtract the averaged signal
from each detector's TOD for the corresponding detector group.
The common-mode filter is more efficient at removing atmosphere noise correlated among the detectors, 
as compared to individual detector filtering, which primarily addresses uncorrelated low-frequency noise. 
We apply a filter that passes frequencies corresponding to $\ell < 6000$ to the TOD
 for individual scans to avoid
aliasing of high-frequency signal and noise beyond the spatial Nyquist frequency of the
two-arcminute map pixel.
To avoid undesired oscillatory features when we fit a polynomial or other filtering templates 
to the TOD around bright sources, we mask
point sources matching one or more criteria of above 6~mJy, 6~mJy, and
12~mJy at 95~GHz, 150~GHz, and 220~GHz in the TOD when constructing the filter templates for the above filters.
The masks for all frequency bands are the same and contain all the sources mentioned above. 
This masking is applied to the TOD for each detector, zero-weighting samples within a certain
radius from the location of the point source while leaving other weights at unity.
The TOD masking radius is $3'$ for sources with maximum flux across the frequency bands between 5 and 20 mJy, $5'$ for sources greater than 20 mJy, and $5'$ for galaxy clusters.
The point source regions, just as the rest of the TOD, have the filtering templates subtracted,
and are then binned to maps. 
This is different from the
map-level inpainting and masking discussed in Sec.~\ref{subsect:mask_inpaint}. 

After filtering, detector weights are calculated based on their noise in the frequency range
 corresponding to the angular multipole range of $300< \ell < 2000$ with our telescope scanning speed.
We calculate the weights in multipole space instead of frequency space and with
the low side of the multipole range set lower compared to \citetalias{dutcher21}.
This effectively down-weights observations with high low-$\ell$ noise,
allowing the noise properties for different observations
 to be statistically similar among themselves.

We perform data quality checks and cut data on several levels: individual detectors in a single scan or all observations, all data in a scan, and all data in a subfield observation.

Many cuts are done at the detector level. Data from a detector is cut from a scan if there are sharp spikes 
 in the TOD (one glitch over 20$\sigma$ or more than 7 glitches over 5$\sigma$),
oscillations from unstable bolometer operation, anomalously low TOD variance, or
response less than $S/N$ of 20 to a chopped thermal source during calibration.
A detector is also cut if the bias point is not in the superconducting transition or if the
readout is beyond its dynamic range.
While the above reasons constitute most detector cuts, there are cuts due to technical
reasons in data processing that cause a detector to have unphysical values or miss
identifying information.
We remove detectors with anomalously high or low weights beyond 3$\sigma$ of the mean and exclude some bolometers because of their noisy behavior, fabrication defects, or readout issues. 
We also cut one of ten detector wafers because of excess noise power at 1.0~Hz, 1.4~Hz (pulse tube refrigerator frequency), 10~Hz, and their harmonics.
Out of the remaining operable detectors, the other cuts discussed above removed $\approx$20\% of detectors,
which results in around 8340 detectors contributing to the final map.

Cuts are also done at the level of complete scans. All data for a scan is cut if fewer than
50\% of the operable detectors survive the detector cuts or the telescope pointing range does not match the intended survey field's range.

Additionally, cuts are done to entire observations at the level of
subfields. We cut subfield observations without complete calibration information or detector
mapping information.
Out of 602 subfield observations in 2018, we retain 569 for a total of $\sim$1420 observing hours.
We coadd all observations corresponding to one subfield with  inverse-variance weights to reduce the map noise.

We convert TOD into maps with detector pointing information, detector weights, and
detector polarization properties following the same procedure as discussed
in \citetalias{dutcher21}.
We make maps in the oblique Lambert azimuthal equal-area projection
first with square one-arcminute pixels to avoid aliasing.
We then apply an anti-aliasing filter in Fourier space, which removes information beyond the Nyquist frequency corresponding to the map resolution, and average every four-pixel unit into one two-arcminute pixel.

\subsection{Beam}
\label{sect:beam}
We measure the telescope beam using a combination of Mars and point source observations
(\citetalias{dutcher21}, also similar to \citet{keisler11}).  
The Mars observations have high $S/N$ out to
tens of arcminutes away from the peak response but show signs of detector nonlinearity near the peak.
We therefore mask the data obtained during a scan
around Mars within $\approx$ 1 beam FWHM.
To fill the hole around the peak planet response ($\approx$ 1~arcmin radius),
we stitch the Mars beam with observations of fainter point sources convolved with the Mars disk.
The planet disk and pixel window function are later corrected after the stitching to obtain the beam profile.
The beam uncertainty and correlations across multipoles are estimated by varying the
combinations of point sources and Mars observations used for estimating different angular scales of
the beam while also changing the parameters used to stitch the two types of maps together.
The beam profiles are used to obtain the calibration factors.

\subsection{Temperature Calibration}
\label{sect:tcal}
We obtain the absolute
temperature calibration of the coadded maps by comparing the \sptg 95 and 150~GHz
maps against the 100 and 143~GHz maps from the \planck satellite (PR3
dataset\footnote{Planck Legacy Archive, \url{https://pla.esac.esa.int}}) over the
angular multipole range of $400<\ell < 1500$.
We compute the per-subfield calibration factor 
by dividing the \sptg
cross-spectra between two half-depth SPT maps by the cross-spectra between full-depth
\sptg and \planck maps, with correction factors including the beam, pixel window
function, and transfer function from map making applied (see Sec.~\ref{sect:sim_tf}).
We mask bright point sources and galaxy clusters before computing the cross-spectra to avoid biases.
The uncertainties of the per-subfield calibration factors are generated by 
repeating the same analysis on 20 \sptg and \planck  simulation realizations, with power spectra and noise spectra matching the    
original datasets (see also Sec. \ref{sect:sim_tf}), and taking the standard deviation of the distribution for each subfield.
The uncertainties are at the level of $\approx$0.3\% and 0.2\% for 95 and 150~GHz, respectively.
We divide each subfield map by the corresponding calibration factor
before stitching them to get the full 1500~deg$^2$ field map.

\begin{figure}
	\centering
	\includegraphics[width=1.0\linewidth]{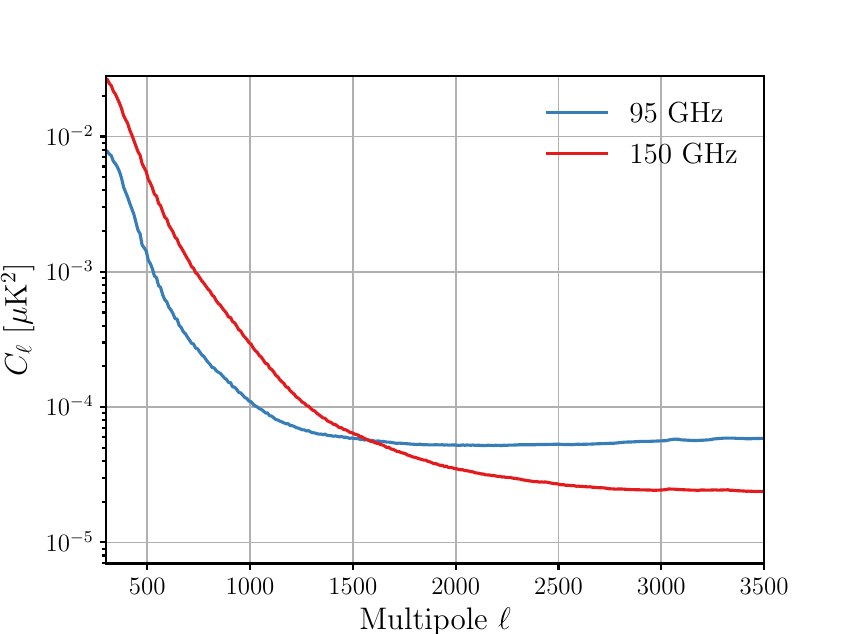}
	\caption{Noise spectra of coadded temperature maps
		for the 95 and 150~GHz frequency bands.  }
	\label{fig:noise_curve}
\end{figure}

The noise spectra as a function of multipole, $\ell$, are plotted in
Fig.~\ref{fig:noise_curve}. These curves are calibrated and corrected for the transfer function and beam.
Compared to 150~GHz, the 95~GHz data has less low-frequency noise from atmospheric fluctuations, but higher noise at $\ell>1600$.
The resulting statistical uncertainty for the lensing spectra is similar for 95 and 150~GHz.
The white noise levels are 26~$\mu$K-arcmin and 17~$\mu$K-arcmin for 95 and 150~GHz at $\ell \approx 3000$.

\subsection{Source Inpainting and Masking}
\label{subsect:mask_inpaint}

To mitigate the lensing biases from point sources and galaxy clusters,  we can remove them in map space using a combination of source inpainting and masking. The inpainting process replaces the pixels around the source with samples drawn from a Gaussian random field with power spectrum consistent with that of the CMB. The samples are constrained to have correlations with the surrounding pixels that follow the predicted CMB correlation function~\citep{hoffman91}. For source masking, we multiply the map with a mask that effectively zeroes the source pixels using cosine tapers, which smoothly decrease from 1 to 0, following the shape of a cosine curve.
Masking holes introduce a mean field that can be estimated using simulations and subtracted from the lensing map (Sec.~\ref{sect:lensing_estimation}).
To reduce the mean field and its associated uncertainties, we inpaint most of the sources and mask bright ones that are above $50$~mJy at 150~GHz. 
Inpainting or masking sources and clusters with our thresholds discussed below corresponds to cutting 4\% of the map area.

The inpainting method used here is similar to that used in \citet{benoitlevy13} and \citet{raghunathan19c}.
We define two regions around the source or cluster center, $R\le R_1$
and $R_1<R\le R_2$, where $R_1$ is the inpainting radius and $R_2$ is fixed to be 25$^{\prime}$.
We fill values within $R_1$ based on values in the $R_1<R\le R_2$
annulus using constrained Gaussian realizations
\beq
\hat T_1= \tilde T_1 + \hat{\boldsymbol{C}}_{12} \hat{\boldsymbol{C}}_{22}^{-1}(T_2-  \tilde T_2)\, ,
\eeq
where 1 indicates the $R\le R_1$ region, 2 indicates the $R_1<R\le R_2$ region,
$T$ is the original map, $\hat T$ is the inpainted map, $\tilde T$ is
the simulated Gaussian map realization, and $\hat{\boldsymbol{C}}_{XY}$
is the covariance matrix of the CMB fields between two regions X and Y.
We generate Gaussian realizations in a fixed $200 ^\prime \times 200 ^\prime$
box with the same CMB, foreground, noise spectra, and  transfer function as the data to be inpainted.
We estimate the covariance matrices $\hat{\boldsymbol{C}}_{12}$ and
$\hat{\boldsymbol{C}}_{22}$ with 5000 Gaussian realizations following
the method in \citet{raghunathan19c}.

We inpaint or mask all sources detected above $S/N$ of 5 in any of the three frequency bands.
The $S/N$ threshold roughly corresponds to minimum fluxes of
2.7, 3.3, and 12.0~mJy at 95, 150, and 220~GHz.
We set the inpainting radii based on the maximum flux across the three frequency bands for each source. The radii are summarized in Tab.~\ref{tab:inpaint_mask_radii}.
We inpaint all detected sources, including the brighter sources ($>50$~mJy) that will be later zeroed by masking to reduce the impact of their variance in the covariance matrix on neighboring regions to be inpainted.
Similarly, we inpaint clusters detected with $S/N$ $>$ 5.
The cluster-finding process was performed using two years of \sptg data, resulting in a
$S/N$  for the detected clusters that is 1.5 times higher compared to the $S/N$ reported for the
same clusters in the SPT-SZ survey \citep{bleem15b}. Note that this is a preliminary cluster list,
and the $S/N$ for a full analysis will be much higher.

\begin{table}
	\caption{Inpainting and masking radii. }
	\begin{adjustwidth}{.4cm}{}
    \begin{center}
			\begin{tabular}{l | m{2.4cm} | m{1.5cm}| m{1.5cm}}
				\hline\hline
				&Flux ($S$ in mJy) or cluster $S/N$   & Inpainting radius  & Masking radius\\ [0.5ex]
				\hline
				\addlinespace[.2ex]
				\multirow{5}{4em}{Point sources} & \hspace{.2cm} $S\leq 6$ & \hspace{.5cm} $2'$& \hspace{.5cm} $-$\\
				& \hspace{.2cm} $6<S\leq20$&\hspace{.5cm} $3'$&\hspace{.5cm} $-$\\
				& \hspace{.2cm} $20<S\leq50$&\hspace{.5cm} $5'$&\hspace{.5cm} $-$\\
				& \hspace{.2cm} $50<S\leq500$&\hspace{.5cm} $7'$&\hspace{.5cm} $7'$\\				
				&\hspace{.2cm}  $S>500$&\hspace{.43cm} $10'$&\hspace{.43cm} $10'$\\				
				\hline
				\addlinespace[.2ex]
				\multirow{2}{4em}{Galaxy clusters} & \hspace{.2cm} $5<S/N\leq9$ &\hspace{.5cm} $5'$&\hspace{.5cm} $-$ \\
				& \hspace{.2cm} $S/N > 9$&\hspace{.5cm} $6'$&\hspace{.5cm} $6'$\\
				\hline
			\end{tabular}
	\end{center}
	\end{adjustwidth}
	\label{tab:inpaint_mask_radii}
\end{table}

In addition to the source inpainting, we apply a mask that zeros the region around the brightest point
sources in the map.
The masking radii are set by flux at 150~GHz for point sources and detection $S/N$ for galaxy clusters (see Tab.~\ref{tab:inpaint_mask_radii}). 
The source mask has cosine tapers with a radius of $10'$.
With $\sim$4\% of the area lost due to masking and inpainting, 
we expect the bias from masking locations correlated with
the underlying $\phi$ field to be negligible~\citep[see, e.g.,][]{lembo22}.
The masking thresholds correspond to 361 sources being masked,
and the resulting mean field is well-behaved.
We show that the analysis is robust to
these inpainting and masking choices in Sec.~\ref{sect:consistency_tests}.

Besides source masking, we apply a boundary mask with $30'$-cosine taper to downweight the noisy field edges.

\subsection{Inverse-variance Filtering}
\label{subsect:ivf}

To minimize the variance of the reconstructed lensing map, the weights applied
in the quadratic estimator include inverse-variance filtering the input CMB maps (Sec.~\ref{sect:lensing_estimation}).
To construct the filter, we model the data maps as consisting of three components: the CMB sky
signal, referred to as $T_{\boldsymbol{\ell}}$; ``sky noise," $N_{\boldsymbol{\ell}}$,  which includes
astrophysical foregrounds and atmospheric noise; and pixel-domain noise, $n_j$, modeled as
white, uncorrelated, and spatially non-varying within the mask. We express the
relationship between the data maps and the three components as follows:

\beq
T_j = \sum_{\boldsymbol{\ell}} P_{j\boldsymbol{\ell}}T_{\boldsymbol{\ell}}+\sum_{\boldsymbol{\ell}} P_{j\boldsymbol{\ell}}N_{\boldsymbol{\ell}}+ n_j \, .
\eeq
Here, the matrix operator $P_{j\boldsymbol{\ell}}$ incorporates the transfer function
and Fourier transform, enabling the conversion from  Fourier space to map space. It is
defined as $P_{j\boldsymbol{\ell}} = e^{i\boldsymbol{\ell}\mathbf{x}_j}F_{\boldsymbol{\ell}}$, where
$F_{\boldsymbol{\ell}}$ contains the beam, pixelization effects, and timestream
filtering. The position vector $\mathbf{x}_j$ represents the coordinates of pixel $j$.

The inverse-variance filtered field  $\bar{T}$ is given by:
\beq
\bar T = S^{-1}  [S^{-1}+P^{\dagger}n^{-1}P]^{-1} P^{\dagger}n^{-1}T\, ,
\label{eq:ivf}
\eeq
where the total signal covariance matrix $S=C_{\boldsymbol{\ell}}^{TT} +C_{\boldsymbol{\ell}}^{NN}$
has contributions from the CMB, foregrounds, and anisotropic noise.
$C_{\boldsymbol{\ell}}^{TT}$ represents the sum of CMB and foreground spectra
interpolated to 2D, while $C_{\boldsymbol{\ell}}^{NN}$ corresponds to the 2D
anisotropic noise spectrum. The term $n$ denotes the pixel-space
noise variance multiplied by the mask.

For $C_{\boldsymbol{\ell}}^{TT}$, we use the same CMB and foreground spectra 
 (namely a CMB TT spectrum and extragalactic foreground spectra representing
tSZ, kSZ, CIB, and diffuse radio sources)
that were used in
generating the CMB simulations discussed in Sec.~\ref{sect:simulations}.
To estimate $C_{\boldsymbol{\ell}}^{NN}$, we generate 500 noise
realizations by subtracting the left-going and right-going CMB maps and adding the
difference maps with random signs.
We average the noise spectra over these 500 noise realizations
and subtract a white noise level $n$, modeled in pixel space, from the averaged
2D noise spectrum for each band. We solve for $\bar{T}$ with a conjugate-gradient solver.

\subsection{Fourier Space Masking}

In order to mitigate the contamination from instrumental and atmospheric noise at low $\ell$, as well as
astrophysical foregrounds dominating over the CMB at high $\ell$, we apply a Fourier mask that includes modes in the multipole range of $\ell$ between 300 and 3000 for the 95~GHz map and between 500 and 3000 for the 150~GHz map. 
In addition, we apply
a cut excluding data with $\ell_x <300$ and $\ell_x < 500$\footnote{Here $\ell_{x}$ refers to the axis along the $\mathrm{x}$-direction in Fourier space after a 2D Fourier transform of the map given the map projection we have chosen in Fig.~\ref{fig:lensing_maps_combined}, with x being the horizontal direction.}
(referred as $\ell_{\mathrm{xmin}}$ cut hereafter)  for the 95 and 150~GHz frequency
bands, respectively, to reduce some
noisy modes along the scan direction below $\ell_{\mathrm{xmin}}$.
We show in Sec.~\ref{sect:consistency_tests} that our analysis is
robust to different cut choices.

\section{Simulations}
\label{sect:simulations}
We use simulations to estimate the transfer function, the response (normalization) and
mean field correction to the lensing map,
the lensing spectrum noise biases ($N^0_L$, $N^1_L$),
and biases to the lensing spectrum from extragalactic foregrounds.

\subsection{CMB and Noise Components}
\label{sect:sim_overview}

We base the simulated CMB skies on a fiducial cosmology from the best fit of \planck~2018
\texttt{TT,TE,EE+lowE+lensing}~\citep{planck18_archive}.
We use \texttt{CAMB}~\citep{lewis11b} to generate CMB and lensing potential
angular power spectra from the fiducial cosmology, and \texttt{HEALPix} \citep{gorski05} to synthesize spherical harmonic realizations
 of the unlensed CMB and lensing potential.
We use \texttt{Lenspix} \citep{lewis05} to create lensed CMB realizations.
The instrument and sky noise realization generation is described in Sec.~\ref{subsect:ivf}.

\subsection{Foreground Components}
\label{sect:sim_fgnds}
The millimeter-wave sky, while dominated by the CMB at high galactic latitude,
also contains signals from the cosmic
infrared background (CIB), thermal and kinetic Sunyaev-Zel'dovich effects (tSZ, kSZ),
and radio sources.
We model the sub-inpainting-threshold/diffuse foreground emissions below 6.4~mJy at 150~GHz
as Gaussian described by
 their measured angular power spectra.
The measured foreground spectra are based on \citet[hereafter \citetalias{reichardt21}]{reichardt21}.
The CIB in the simulation consists of a Poisson-distributed component
with $D_\ell \propto \ell^2$ from faint dusty star-forming galaxies
and a clustered part with $D_\ell \propto \ell^{0.8}$.
Here $D_\ell$ is related to angular power spectrum
$C_\ell$ by $D_{\ell}  = \frac{1}{2\pi} \ell(\ell+1)C_\ell$.
The spectral shape of the CIB is set to be $\nu^{\beta}B_{\nu}(T_{\mathrm{dust}})$,
where $B_{\nu}$ is the blackbody spectrum, $T_{\mathrm{dust}}$ is 25~K,
and $\beta$ is $1.48$ for the Poisson term and  $2.23$ for the clustered term.
At 150~GHz and $\ell=3000$, the amplitude of $D_\ell^{\mathrm{CIB}}$
is $D_{3000}^{\mathrm{CIB, P}} = 7.24\,{\rm \mu K}^2$ for the Poisson term
and $D_{3000}^{\mathrm{CIB, cl}} = 4.03\, {\rm \mu K}^2$ for the clustered term.
The clustered term includes the contributions from one- and two-halo terms \cite{viero13a}.
The shapes of the tSZ and kSZ angular power spectra follow the tSZ template
in \citet{shaw10} and the kSZ template in \citet{shaw12} and \citet{ zahn12}.
The amplitude at 143~GHz and $\ell=3000$ is
$D_{3000}^{\mathrm{tSZ}} =3.42\,{\rm \mu K}^2$ for tSZ
and $D_{3000}^{\mathrm{kSZ}} =3.0\,{\rm \mu K}^2$ for kSZ.
The radio source component has a spectrum shape of $D_\ell \propto \ell^2$
and amplitude of $D_{3000}^{\mathrm{radio}} =1.01\,{\rm \mu K}^2$ at 150~GHz.
The population spectral index of the radio sources is set to be $\nu^{-0.76}$.
The tSZ and radio spectra are adjusted from the measured values in \citetalias{reichardt21} given the
different masking and inpainting thresholds in this analysis described in Sec.~\ref{subsect:mask_inpaint}. 

We use these Gaussian foreground simulations to account for the contribution of foregrounds to the
disconnected bias term ($N^0_L$) in the lensing spectrum.
However, this set of simulations does not account for the non-zero trispectrum and primary
and secondary bispectrum biases~\citep{vanengelen14, omori22} one expects from these
extragalactic foregrounds.
We discuss our approach to estimating the foreground biases in Sec.~\ref{sect:astro_fgnds}. 

Besides the diffuse foregrounds, the observed maps also contain point sources and galaxy clusters.
We identify point sources and galaxy clusters in the data maps using simplified versions of methods
in \citet{everett20} and \citet{bleem15b}.
We check that the point source fluxes and the galaxy clusters' peak amplitudes are
unbiased and accurate to within 10\% based on previous measurements for overlapping detections.
We include point sources and clusters at their detected positions,
amplitudes, and profiles in the simulated maps for all three frequencies.
We use the beam profile for point sources and a beta profile \citep{cavaliere78} convolved with the beam for galaxy clusters. 
The point sources and clusters are the same between data and simulations, which
allows us to use the same masks and inpainting for both.

\subsection{Simulation Processing and Transfer Function}
\label{sect:sim_tf}
We convolve the simulated CMB and foreground maps with the corresponding beams
 of the three frequency bands.
We then mock-observe the simulations using the same methods for data processing so that
the mock-observed simulations have the same filter transfer function and
mode-coupling as the data maps.
We also add the noise maps discussed in Sec.~\ref{subsect:ivf} to the mock-observed maps.

The transfer function is obtained as the square root of the ratio of the 2D power spectra of the mock-observed map and the noise-free simulated map.
To reduce noise scatter in the estimate, we average and smooth the transfer functions obtained from 160 simulations.
We note that the transfer function shows a small dependence on the power spectra of the
input maps. Small changes in the transfer function lead to variations in the
weighting of the CMB modes at the inverse-variance filtering step (Sec.~\ref{subsect:ivf}),
affecting the optimality of the filtered map. We
show the effect of using different input spectra to estimate the transfer function
 on the reconstructed lensing spectra to be negligible
in Sec.~\ref{sect:consistency_tests}.

\subsection{Simulations for Estimating  Foreground Biases}
\label{sect:ng_sims}

While the simulations described so far
are needed for pipeline checks and estimating the transfer function (Sec. \ref{sect:sim_tf}) and lensing biases (Sec. \ref{sect:bias_est_with_sims}), they also assume no other astrophysical sources
of statistical anisotropy besides lensing.
However, extragalactic foregrounds are non-Gaussian themselves and correlated with the lensing field.
Therefore, we expect an extragalactic foreground bias to our measurement.
The galactic foregrounds have negligible effects on our lensing reconstruction
since our field is chosen to have low galactic foregrounds.

To estimate the foreground bias, we use the \texttt{AGORA} simulation~\citep{omori22},
an N-body-based simulation with tSZ, kSZ, CIB, radio sources, and weak lensing components.
A CMB map lensed with the $\phi$ field obtained by ray-tracing through the lightcone  is combined with appropriately scaled foreground components (that are correlated with $\phi$) to produce mock 95 and 150~GHz maps.

We create a parallel set of simulations
with the same CMB field but Gaussian realizations of foregrounds that
have identical power spectra to the sum of all non-Gaussian foregrounds. 
Both the Gaussian and non-Gaussian simulations will later be used to estimate the foreground bias template in Sec. \ref{sect:astro_fgnds}. 
We have one full-sky realization of \texttt{AGORA} simulations at 95 and 150~GHz and
we cut them into 16 patches the size of our observing field
(Fig.~\ref{fig:lensing_maps_combined}). 
The \texttt{AGORA} simulations, along with their Gaussian counterparts,
undergo the same inpainting and masking procedure as the
data described in Sec.~\ref{subsect:mask_inpaint}.
This ensures that point sources and galaxy clusters have consistent
masking thresholds and corresponding radii as the data. The power spectra of the
inpainted and masked \texttt{AGORA} non-Gaussian simulations are within 10-20\% of
 the simulations used for the baseline analysis discussed in Sec.~\ref{sect:sim_fgnds}.

\section{Lensing Analysis}
\label{sect:method}
\subsection{Quadratic Lensing Estimator}
\label{sect:lensing_estimation}

The unlensed CMB is well described by a statistically isotropic Gaussian random field
with zero off-diagonal covariance.
Lensing breaks the statistical isotropy and introduces off-diagonal
correlations across CMB temperature and polarization modes in harmonic space.
In the general case where $X$ and $Y$ $\in[T, E, B]$, the covariance in the flat-sky approximation is
\beq
\label{eq:lens_expansion}
\langle X_{\pmb{\ell\phantom{'}}} Y^{*}_{\pmb{\ell'}} \rangle_{\rm CMB} = \delta(\pmb{\ell}-\pmb{\ell}') C_{\ell}^{XY} + W^{XY}_{\pmb{\ell}, \pmb{\ell'}} \phi_{\pmb{\ell}-\pmb{\ell}'} + \mathcal{O}(\phi^2) \,,
\eeq
where $\boldsymbol{\ell}$ ($\boldsymbol{\ell}^{\prime}$) is a vector in Fourier space,
$\phi$ is the lensing potential, and $C_{\ell}^{XY}$ is the power spectrum of $XY$.
For the temperature-based estimator we use in this paper, $W$ is
  derived as the leading-order coefficient of the CMB correlation in terms of the lensing potential induced  by lensing.
In this work, we only include temperature,
therefore, in the remainder of the paper, we will replace $X$ and $Y$ with $T_{\nu}$ and $T_{\mu}$,
the temperature fields at frequencies $\nu,\mu \in [95,150]$ GHz.

Using these off-diagonal correlations, we can estimate the unnormalized lensing potential
at $\bm{L}$ by calculating the weighted sum of the inverse-variance filtered lensing modes
separated by $\boldsymbol{L}=\boldsymbol{\ell}-\boldsymbol{\ell}^{\prime}$ \citep{hu02a}
\beq
\label{eq:phi_bar}
\bar{\phi}^{T_\nu T_\mu}_{\bm{L}} = \int{d^2\pmb{\ell} W^{TT}_{\pmb{\ell},\pmb{\ell}-\bm{L}}}
\bar{T}_{\nu,\pmb{\ell}}\, \bar{T}^{*}_{\mu,\pmb{\ell}-\bm{L}}\,.
\eeq
Here we use an over-bar to denote an inverse-variance-weighted quantity.
$W^{TT}$ in Eq.~\ref{eq:phi_bar} is designed to
maximize the sensitivity to the lensing-induced signal while minimizing
noise. For temperature, $W^{TT}$ for lensing reconstruction takes the same form as the correlation coefficient derived from Eq. \ref{eq:lens_expansion}.

In realistic cases, $\bar{\phi}^{T_\nu T_\mu}_{\bm{L}}$ contains
biases from other statistically anisotropic sources unrelated to lensing,
such as the map mask and inhomogeneous sky noise.
We estimate this map-level bias, which we call the mean field (MF)
$\bar{\phi}^{T_{\nu}T_{\mu},{\rm MF}}_{\bm{L}}$, by averaging
the lensing estimations of 160 simulations with different realizations of CMB,
lensing potential, and noise:
\beq
\label{eq:phi_MF}
\bar{\phi}^{T_{\nu}T_{\mu},{\rm MF}}_{\bm{L}} = \langle\int{d^2\pmb{\ell} W^{TT}_{\pmb{\ell},\pmb{\ell}-\bm{L}}}
 \bar{T}_{\nu,\pmb{\ell}}\, \bar{T}^{*}_{\mu,\pmb{\ell}-\bm{L}} \rangle \,.
\eeq
The lensing potentials from different simulations are independent and average to zero,
so the averaged lensing estimation only contains the MF from
common non-lensing features shared among the simulations.
We subtract the MF from $\bar{\phi}^{T_{\nu}T_{\mu}}_{\bm{L}}$.

We normalize the mean-field-subtracted lensing potential by the inverse of the response.
We obtain the total response by combining an analytic and a Monte-Carlo (MC) response estimate.
The analytic response is given by
\beq
\label{eq:analytical_resp}
\mathcal{R}^{T_{\nu}T_{\mu},{\rm Analytic}}_{\bL} = \int{ d^2\bl\, W^{TT}_{\bl,\bl-\bL}\times W^{TT}_{\bl,\bl-\bL} \mathcal{F}^{T_{\nu}}_{\bl} \mathcal{F}^{T_{\mu}}_{\bl-\bL} } \,.
\eeq
Here $ \mathcal{F}^{T_{\nu}}_{\bl} T_{\nu,\mbell}  = [C^{T_{\nu}T_{\nu}}_{\mbell} + N^{T_{\nu}T_{\nu}}_{\mbell}]^{-1} T_{\nu, \mbell} $ is
an approximation of the inverse-variance filter  in Sec.~\ref{subsect:ivf},
where the approximation is exact if there is no masking, and
$N_{\boldsymbol{\ell}}^{T_{\nu}T_{\mu}}$ captures all anisotropic noise.
For the general case,
we apply an MC response correction
$\mathcal{R}_{\mbL}^{T_{\nu}T_{\mu},{\rm MC}}$ to account for the deviation from this approximation.
We divide the cross-spectrum between the estimated lensing potential and
the input lensing potential by the input auto-spectrum and average this ratio
over many simulation realizations to get the MC response
\beq
\label{eq:mc_resp}
\mathcal{R}_{\mbL}^{T_{\nu}T_{\mu},{\rm MC}} =
\frac{\langle \hat{\phi}^{T_{\nu}T_{\mu}}_{\mbL}\,\, \phi^{I*}_{\mbL}\rangle}
{\langle \phi^I_{\mbL} \phi^{I*}_{\mbL}\rangle} \,.
\eeq
Here $\phi^{I}$ is the input lensing potential and $\hat{\phi}^{T_{\nu}T_{\mu}}_{\mbL}$
is the mean-field-subtracted lensing potential with analytic response normalization. 
We use a hat ($\hat{\phi}$) to denote debiased quantities.
We also note that the response is
the Fisher matrix for the lensing potential \citep{story15, planck13-17},
making Eq.~\ref{eq:phi_bar} an inverse-variance-weighted quantity.
We average the
MC response into 1D to
reduce noise and get
$\mathcal{R}_{L}^{T_{\nu}T_{\mu},{\rm MC}} = \langle \mathcal{R}_{\mbL}^{T_{\nu}T_{\mu},{\rm MC}} \rangle$.
Here $\mbL$ is a vector in Fourier space, and $\langle \rangle$ means averaging over an annuli in 2D Fourier space corresponding to the same $L$.
The MC response 
correction $\mathcal{R}_{L}^{T_{\nu}T_{\mu},{\rm MC}}$ is $\lesssim$10\% across the range of scales used in this work.

The total response combining the analytic and MC response is
\mbox{$\mathcal{R}_{\mbL}^{T_{\nu}T_{\mu}} =\mathcal{R}_{L}^{T_{\nu}T_{\mu},{\rm MC}} \mathcal{R}_{\mbL}^{T_{\nu}T_{\mu},{\rm Analytic}}$},
and the lensing potential estimate with the full correction is
\beq
\label{eq:phi_hat_full_correction}
\hat{\phi}^{T_{\nu}T_{\mu}}_{\mbL} = \frac{1}{\mathcal{R}^{T_{\nu}T_{\mu}}_{\mbL}}
(\bar{\phi}^{T_{\nu}T_{\mu}}_{\mbL} - \bar{\phi}^{T_{\nu}T_{\mu},{\rm MF}}_{\mbL}) \,.
\eeq

\subsection{Lensing Power Spectrum, Biases, and Amplitude}

\label{subsect:lensing_bias}

We calculate the lensing power spectrum with the debiased lensing potentials
$\hat{\phi}^{T_{\nu}T_{\mu}}$ and $\hat{\phi}^{T_{\alpha}T_{\beta}}$ from different frequency pairs.
To account for the border apodization and point source mask applied to the four temperature maps
entering the quadratic lensing spectrum estimate, we divide out a masking factor $f_{\rm mask}$,
which is the average of the mask applied to a single map to the fourth power,
from the spectrum of the debiased lensing potential
\beq
\label{eq:clpp_def}
C^{\hat{\phi}^{T_{\nu}T_{\mu}} \hat{\phi}^{T_{\alpha}T_{\beta}}}_{L} =
f_{\rm mask}^{-1}\langle \hat{\phi}^{T_{\nu}T_{\mu}}_{\mbL} \,\, \hat{\phi}^{*}\,^{T_{\alpha}T_{\beta}}_{\mbL} \rangle.
\eeq

The raw lensing power spectrum is biased by a few sources, including spurious
correlations of the input fields to the zeroth and first order of the lensing spectrum,
$N^{0}_{L}$ and $N^{1}_{L}$, and the foreground bias to be discussed in Sec.~\ref{sect:astro_fgnds}.
There are higher order biases in terms of the lensing spectrum such as the $N^{3/2}$ biases from post-Born corrections and large-scale structure cross-bispectra that are negligible given our noise levels~\citep{fabbian19b}.
The debiased lensing spectrum after  $N^{0}_{L}$ and $N^{1}_{L}$ correction is\footnote{For clarity, here we suppress the frequency dependence of the lensing reconstruction.}
\beq
\label{eq:bias_correction}
\hatclpp = \clpphat - N^{\rm RD, 0}_{L}  - N^{1}_{L} \,,
\eeq
where $N^{\rm RD, 0}_{L}$ is a variant of $N^{0}_{L}$ to be introduced below.

We estimate the $N^0_L$ and $N^1_L$ bias terms with simulations.
$N^0_L$ is estimated using
\beq
\label{eq:n0_bias}
\begin{aligned}
N^0_{L}= &
\Big \langle C^{\hat{\phi} \hat{\phi}}_{L}[\bar{T}^{\rm MC}_\nu,\bar{T}^{\rm MC'}_\mu,\bar{T}^{\rm MC}_\alpha,\bar{T}^{\rm MC'}_\beta] \\
+&  C^{\hat{\phi} \hat{\phi}}_{L}[\bar{T}^{\rm MC}_\nu,\bar{T}^{\rm MC'}_\mu,\bar{T}^{\rm MC'}_\alpha,\bar{T}^{\rm MC}_\beta]
\Big \rangle_{\rm MC,MC'},
\end{aligned}
\eeq
where $ C^{\hat{\phi} \hat{\phi}}_{L}[\bar{T}_\nu,\bar{T}_\mu,\bar{T}_\alpha,\bar{T}_\beta] $
denotes the lensing cross-spectrum between two debiased lensing potentials $\hat{\phi}^{T_\nu T_\mu}$ and $\hat{\phi}^{T_\alpha T_\beta}$. 
We use this format instead of Eq.~\ref{eq:clpp_def} to highlight the superscripts and subscripts when both are present.
MC and MC$'$ denote simulations with different realizations of the CMB, Gaussian foreground, lensing potential and noise.
{Applying Wick's theorem for the contraction in the above equation,
only the Gaussian correlations between fields with
the same superscript  (MC or MC$'$) are non-zero.
The $N^0_L$ estimated this way could be inaccurate because the data may have
slightly different Gaussian power from the simulations depending on the simulation modeling and realization.
To reduce the bias caused by the difference, we adopt a realization-dependent
$N^{\rm RD, 0}_{L}$~\citep{namikawa13} defined by
\beq
\begin{aligned}
	\label{eq:rdn0}
	&N^{\rm RD, 0}_{L} =\\
	\Big \langle
	& \clpphat[\bar{T}^{\rm d}_{\nu},  \bar{T}^{\rm MC}_{\mu},  \bar{T}^{\rm d}_{\alpha},   \bar{T}^{\rm MC}_{\beta}]
	+\clpphat[\bar{T}^{\rm MC}_\nu, \bar{T}^{\rm d}_\mu,   \bar{T}_\alpha^{\rm d},   \bar{T}^{\rm MC}_\beta ]  \\
	& +\clpphat[\bar{T}^{\rm d}_\nu,  \bar{T}^{\rm MC}_\mu,  \bar{T}^{\rm MC}_\alpha,  \bar{T}^{\rm d}_\beta  ]
	+\clpphat[\bar{T}^{\rm MC}_{\nu}, \bar{T}^{\rm d}_{\mu},   \bar{T}^{\rm MC}_\alpha,  \bar{T}^{\rm d}_\beta ]  \\
	& -N^0_{L} \Big \rangle_{\rm MC,MC'} \,\,\,\,\,,
\end{aligned}
\eeq
where $d$ denotes the data. In $N^{\rm RD, 0}_{L} $,
we calculate the lensing spectra from lensing potentials estimated using both the data and simulation.
We then subtract  $N^0_L$  bias defined in Eq.~\ref{eq:n0_bias}.
This method also suppresses off-diagonal contributions to the covariance of the lensing power spectrum~\citep{namikawa13}.

The  $N^1_L$  bias term arises from connected contributions to the trispectrum
and is estimated using simulations with the same lensing field but different CMB realizations. $N^1_{L} $  bias is given by
\beq
\label{eq:n1}
\begin{aligned}
	N^{1}_{L} =& \\
	\Big \langle
	& \clpphat[\bar{T}^{\phi^1,{\rm MC}}_\nu,\bar{T}^{\phi^1,{\rm MC'}}_\mu,\bar{T}^{\phi^1,{\rm MC}}_\alpha,\bar{T}^{\phi^1,{\rm MC'}}_\beta] \\
	& +\clpphat[\bar{T}^{\phi^1,{\rm MC}}_\nu,\bar{T}^{\phi^1,{\rm MC'}}_\mu,\bar{T}^{\phi^1,{\rm MC'}}_\alpha,\bar{T}^{\phi^1,{\rm MC}}_\beta] \\
	& - N^0_{L}
	\Big \rangle_{{\rm MC,MC'}}\,,
\end{aligned}
\eeq
where $\phi^1$ indicates that the simulations share the same lensing field.
Here MC and MC$^{\prime}$ indicate that the simulation components
other than $\phi$ come from independent realizations.
The first two terms contain $N^0_{L} $ bias from Gaussian power and
the $N^1_{L} $ bias from the shared lensing potential among the four subfields.
We subtract the  $N^0_L$ bias from the first two terms to get the $N^1_{L} $ bias.

We present our results in binned bandpowers. Our bin edges are shown in Table \ref{tab:bandpowers}.
We calculate the weighted average of  $\hatclpp$ within each bin:
\beq
\label{eq:binning}
\hat{C}_b^{\phi^{T_{\nu}T_{\mu}} \phi^{T_{\alpha}T_{\beta}}} \equiv \frac{\sum_{{L} \in b} w_{{L}}^{T_{\nu}T_{\mu}T_{\alpha}T_{\beta}} \hat{C}_{{L}}^{\phi^{T_{\nu}T_{\mu}} \phi^{T_{\alpha}T_{\beta}}}}{\sum_{{L} \in b} w_{{L}}^{T_{\nu}T_{\mu}T_{\alpha}T_{\beta}}}. 
\eeq
Here, the subscript $b$ denotes a binned quantity. The weighted average of the $C_L$ inputs, denoted as $C_b$, is calculated using Wiener-filter weights $w$ designed to maximize our sensitivity to departures from the fiducial \LCDM{} expectation:
$w^{T_{\nu}T_{\mu}, T_{\alpha}T_{\beta}}_L = C^{\phi \phi, {\rm theory}}_L / {\rm Var}( \hat{C}^{{\phi}^{T_{\nu}T_{\mu}} {\phi}^{T_{\alpha}T_{\beta}}}_{L} )$.
We obtain the variance ${\rm Var}( \hat{C}_L^{{\phi}{\phi}})$ from analytic estimates of the signal and noise spectrum.

We can then compute the per-bin amplitude, denoted as $A_b^{T_{\nu}T_{\mu},T_{\alpha}T_{\beta}}$, which is defined as the ratio of the unbiased lensing spectrum to the input theory spectrum:
\beq
\label{eq:amp_def}
A_b^{T_{\nu}T_{\mu}T_{\alpha}T_{\beta}} \equiv \frac{\hat{C}^{\phi^{T_{\nu}T_{\mu}} \phi^{T_{\alpha}T_{\beta}}}_b }
{C^{\phi \phi, {\rm theory}}_b } \,.
\eeq
Here ${C^{\phi \phi, {\rm theory}}_b }$ is the theory power spectrum in bin $b$ weighted the same way as in Eq.~\ref{eq:binning}.

The overall lensing amplitude for each estimator is calculated in the same manner as the per-bin amplitude in Eq.~\ref{eq:amp_def}, using the entire range of reported $L$ values.

\subsection{Bias Estimation Using Simulations}
\label{sect:bias_est_with_sims}
To estimate the MF, $N^0_{L} $, and $N^1_{L} $ bias terms, we generate the following set of Gaussian simulations with inputs as detailed in Sec.~\ref{sect:sim_overview} and Sec.~\ref{sect:sim_fgnds}:
\begin{itemize}
	\setlength{\itemsep}{1pt}
	\setlength{\parskip}{0pt}
	\item A: 500 lensed simulations with foregrounds and noise realizations discussed in Sec.~\ref{subsect:ivf},
	\item B: 160 lensed simulations with no foregrounds or noise,
	\item C: 160 lensed simulations with no foregrounds or noise and with the same realizations of lensing potential as set B, but different CMB.
\end{itemize}

We use all simulations in A to estimate $N^{\rm RD, 0}_{L}$
 and 340 simulations in A to evaluate the statistical uncertainty of the lensing power spectrum.
We use 160 simulations in A to estimate the MF (Sec. \ref{sect:lensing_estimation}),
with 80 sims for each of the two lensing potential estimations that form the lensing spectrum.
All of the simulations in set B and C are used to calculate $N^{1}_{L}$,
which is proportional to the first order of lensing power and has no contribution from Gaussian foregrounds.
The number of simulations used for each estimated term is chosen such that the term estimated converges well below the statistical uncertainty level of the lensing spectrum.
In addition to these bias terms, simulations in A are used to estimate the transfer function (Sec.~\ref{sect:sim_tf})
and the covariance matrix for forming minimum-variance bandpowers (Sec.~\ref{sect:mf_combo}). We have performed a pipeline test using Set A to confirm that the reconstructed lensing spectrum is unbiased compared to the input lensing spectrum used to generate the simulations. 

We generate a set of 500 unlensed simulations using lensed CMB power spectrum and the same methods for foregrounds as in Set A.
We use this set for various diagnostic tests, such as validating the mean lensing spectrum using this set of
simulations being consistent with zero for each $L$ bin within a fraction of the statistical uncertainty.

\subsection{Multi-frequency Lensing Spectra Combination}
\label{sect:mf_combo}

We can reconstruct three individual lensing potential maps from the
quadratic combination of the observed temperature fields at the (95, 95), (95, 150), and (150, 150)~GHz frequency combinations.
By correlating these three $\phi$ maps,
we can extract a total of six separate lensing power spectra $C_L^{\phi\phi}$.

In this work, we combine the six individual debiased lensing spectra to produce
a set of minimum-variance lensing band powers from temperature data.
Following previous analyses of primary CMB anisotropies,
\citep[e.g.,][\citetalias{dutcher21}]{mocanu19, planck15-11}, we form a minimum-variance (MV) combination,
in the frequency sense, of the lensing band powers $C_b$ as
\begin{equation}
C^{\mathrm{MV}}=\left(\mathbb{X}^{\top} \mathbf{\mathbb{C}}^{-1} \mathbb{X}\right)^{-1} \mathbb{X}^{\top} \mathbf{\mathbb{C}}^{-1} C.
\label{eq:mv_definition}
\end{equation}
Here, $C$ is a vector of length $6N_{\rm bins}$ formed by concatenating
the lensing power spectra extracted from the various frequency combinations,
while $\mathbb{C}$ denotes their covariance matrix.
$\mathbb{X}$ is a design matrix of shape $6 N_{\rm bins} \times N_{\rm bins}$
in which each column is equal to 1  in the six elements corresponding to a lensing power spectrum
measurement in that $L$-space bin and zero elsewhere.
The band power covariance matrix $\mathbb{C}$ used to weight and combine
the different lensing reconstructions is estimated with a simulation-based approach,
using $N_{\rm sims}=340$ realizations introduced in Sec.~\ref{sect:simulations}.
Given the finite number of simulations, the estimate of the covariance is noisy.
To ameliorate the noise of off-diagonal elements, we condition the underlying covariance matrices
$\mathbb{C}_{bb'}^{\nu\mu\alpha\beta}$ by explicitly setting to zero those entries that
we do not expect to be correlated.
Specifically, we discard elements that are more than one bin away from the diagonal,
i.e. $\mathbb{C}_{bb'}^{\nu\mu\alpha\beta} = 0$ if $|b-b'|>1$.
In each $N_{\rm bins}\times N_{\rm bins}$ block of the full covariance matrix
$\mathbb{C}_{bb'}$, the bins neighboring the diagonal are correlated on average
at the 15\% level and no more than 30\%.
Finally, we note that the average correlation coefficient between different lensing reconstructions
can be as large as 95\% over the $L$ range used in this analysis for two sets of lensing band powers
that share three common frequencies, e.g., $\hat{C}^{\phi^{T_{95}T_{95}} \phi^{T_{95}T_{95}}}_{b} $
and $\hat{C}^{\phi^{T_{95}T_{95}} \phi^{T_{95}T_{150}}}_{b} $
and as low as 56\% for $\hat {C}^{\phi^{T_{95}T_{95}} \phi^{T_{95}T_{95}}}_{b} $
and $\hat{C}^{\phi^{T_{150}T_{150}} \phi^{T_{150}T_{150}}}_{b}$.
We verified that the conditioning applied to the cross-frequency covariance matrices results in a stable estimate of the MV lensing spectrum and its associated covariance matrix. Specifically, we have tested two additional conditioning schemes, one where we only retain the diagonal elements and one based on previous primary CMB SPT analyses \citep{mocanu19,dutcher21} where the on- and off-diagonal covariance blocks are treated differently, and found the differences in the recovered MV lensing spectra to be largely subdominant with respect to statistical uncertainties.

\subsection{Astrophysical Foreground Bias Template}
\label{sect:astro_fgnds}

We expect the lensing spectrum in Eq.~\ref{eq:bias_correction}
to contain residual biases from extragalactic foregrounds.
To address this, we estimate an extragalactic foreground bias template using the simulations
described in Sec.~\ref{sect:ng_sims}.
We use this foreground bias template to model
the shape of the bias spectrum and marginalize over its amplitude in our lensing amplitude measurement
and when estimating cosmological parameters (Sec.~\ref{sec:lens_like}).

To compute this template, we difference the lensing spectra extracted using the \texttt{AGORA} simulations
with the lensing spectra from Gaussian realizations of foregrounds that have the same
power spectra as their non-Gaussian foreground counterparts.
The foreground bias template ($\mathcal{T}$) can be expressed schematically as
\beq
\label{eq:bias_calculation}
\operatorname{\mathcal{T}}=\hat{C}_L^{\phi\phi}\left(\mathrm{CMB}^{\mathrm{NG}}+\mathrm{FG}^{\mathrm{NG}}\right)-\hat{C}_L^{\phi\phi}\left(\mathrm{CMB}^{\mathrm{NG}}+\mathrm{FG}^{\mathrm{G}}\right).
\eeq
Here the superscript NG denotes the non-Gaussian simulations, and G denotes random
Gaussian realizations with  the same power spectrum. $\mathrm{CMB}^{\mathrm{NG}}$ indicates that the CMB is lensed by large-scale structures correlated with the foregrounds  ($\mathrm{FG}^{\mathrm{NG}}$) generated using the same set of N-body simulations.
The first term, $\hat{C}_L^{\phi\phi}\left(\mathrm{CMB}^{\mathrm{NG}}+\mathrm{FG}^{\mathrm{NG}}\right)$, contains both the trispectrum
of the foreground components and the bispectrum with two powers of foreground and one power of the $\phi$ field.
The second term contains neither, so their difference is an estimate of the sum of the foreground
trispectrum- and bispectrum-type biases. 
We have applied similar MF, response, and $N^0_L$ corrections
(Eq.~\ref{eq:phi_hat_full_correction}, \ref{eq:bias_correction}) described in Sec.~\ref{subsect:lensing_bias} for both lensing spectra.
We do not apply the $N^1_L$ correction as it differences out in Eq.~\ref{eq:bias_calculation}.
In addition to Eq.~\ref{eq:bias_calculation}, we also estimate the foreground trispectrum
and the bispectrum biases separately using
$\hat{C}_L^{\phi\phi}\left(\mathrm{CMB}^{\mathrm{NG, 2}} + \mathrm{FG}^{\mathrm{NG}}\right) - \hat{C}_L^{\phi\phi}\left(\mathrm{CMB}^{\mathrm{NG, 2}} + \mathrm{FG}^{\mathrm{G}}\right)$ and $\hat{C}_L^{\phi\phi}\left(\mathrm{CMB}^{\mathrm{NG}} + \mathrm{FG}^{\mathrm{NG}}\right) - \hat{C}_L^{\phi\phi}\left(\mathrm{CMB}^{\mathrm{NG, 2}} + \mathrm{FG}^{\mathrm{NG}}\right)$,
checking and confirming that their sum is consistent with our template constructed using Eq.~\ref{eq:bias_calculation}.
Here the superscript 2 indicates that the CMB comes from a different patch such that its lensing field
does not correlate with the non-Gaussian foreground or contribute to the bispectrum-type bias.

\begin{figure}[htb]
	\centering
	\includegraphics[width=1\columnwidth]{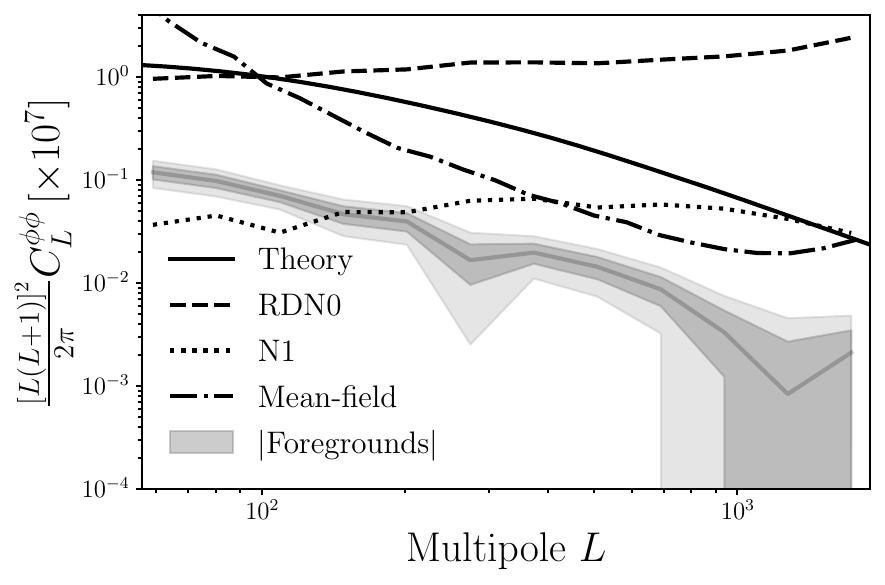}
	\caption{Biases in the CMB lensing auto-spectrum. The solid black line
		denotes the fiducial CMB lensing power spectrum.
		The realization dependent $N^0_L$ and $N^1_L$ biases for our deepest lensing map, $\hat{\phi}^{T_{95}T_{150}}$ as representative noise levels, are shown in black dashed and dotted lines, respectively.
		The power spectrum of the mean field (MF) for $\hat{\phi}^{T_{95}T_{150}}$  is plotted in dash-dotted line.
		The absolute value of the expected mean foreground contamination to the minimum-variance
		combination (see Sec.~\ref{sect:mf_combo}, negative bias) is shown by the solid grey line while the shaded grey areas denote the 1 and 2$\sigma$ scatter with respect to the mean calculated over 16 cutouts from the \texttt{AGORA} simulations.}
	\label{fig:clkk_fg_bias}
\end{figure}

Fig.~\ref{fig:clkk_fg_bias} shows the foreground bias template to the MV combination (see Sec. \ref{sect:mf_combo}), as well as $N^{\rm RD,0}_L$, $N^1_L$, and MF biases for $\hat{\phi}^{T_{95}T_{150}}$.
The 1$\sigma$ and 2$\sigma$ errorbars to the foreground template are estimated from the distribution of
16 bias power spectra estimated using Eq. \ref{eq:bias_calculation} for the 16 patches discussed in Sec.~\ref{sect:ng_sims}. 
We bin the MV foreground bias template using Eq. \ref{eq:binning} to get the binned template $C_b^{\phi\phi,\rm fg}$. 
The mean foreground bias amplitude is negative since the bispectrum-type bias is negative at lower
$L$ and larger in amplitude than the positive trispectrum-type bias term in the $L$
range shown \citep{omori22}. The bias amplitude reduces by $\approx$30\% as we exclude
modes with $\ell>2500$, which is consistent with the foreground
power becoming smaller relative to the CMB at lower $\ell$.

As discussed in Sec.~\ref{sect:mf_combo}, we combine the multifrequency information in debiased lensing power spectrum space, as opposed to lensing map space.
Therefore, in Fig.~\ref{fig:clkk_fg_bias}, we show the mean field spectrum, $N^0_{L}, $ and $N^1_{L}$ noise curves for the auto-spectrum of $\hat{\phi}^{T_{95}T_{150}}$, our deepest lensing map, as a representative noise level.
The mean field spectrum at low $L$ comes from the boundary mask  and
converges sufficiently with the number of simulations we used for averaging.

\section{Lensing Maps, Power Spectra, and Amplitudes}
\label{sect:results}

In this section, we present our main results: the lensing convergence map, lensing power spectrum,
and lensing amplitude measurements. We discuss the relative weights from independent frequency
combinations and compare our results to previous measurements by SPT-SZ (\citetalias{omori17}), \sptpol (\citetalias{wu19}),  POLARBEAR
\citep{polarbear19b},
BICEP/Keck \citep{ade23}, \planck \citep{carron22},
and \actpol \citep{qu23}.
We then discuss the lensing amplitude and its statistical and systematic uncertainties.

\begin{figure*}[htb]
	\centering
	\includegraphics[width=0.98\textwidth]{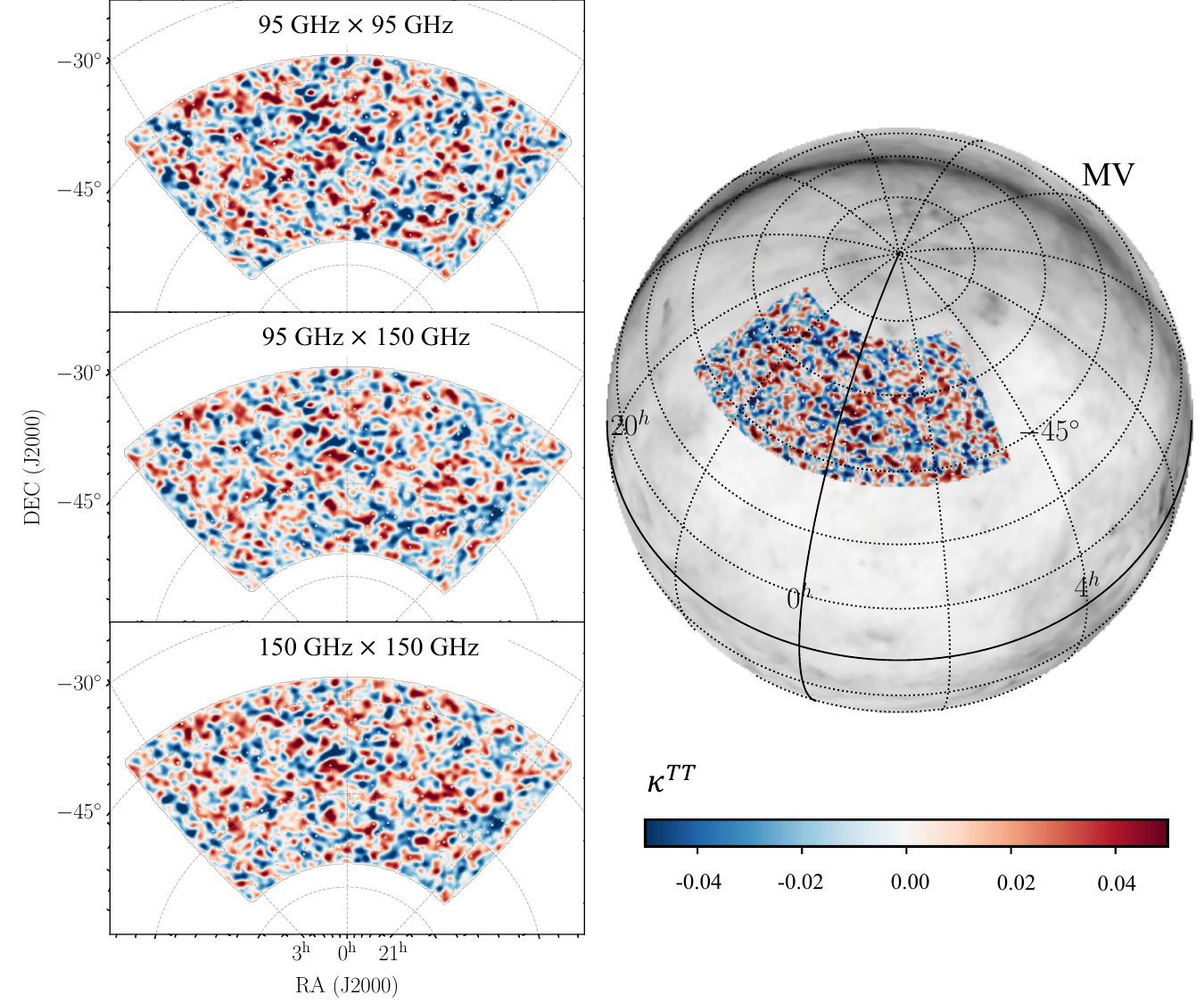}
	\caption {Lensing $\kappa$ maps reconstructed from the \sptg 1500 deg$^2$ field data, smoothed by a
		1-degree FWHM Gaussian to highlight the large-scale modes with higher $S/N$. We have also multiplied the maps by the point source and cluster mask discussed in Sec.~\ref{subsect:mask_inpaint}. The three left panels
		show the lensing convergence map inferred from three independent frequency combinations: 95
		$\times$ 95~GHz, 95 $\times$ 150~GHz, and 150 $\times$ 150~GHz. The right-hand side shows a
		minimum-variance-combined CMB lensing convergence map reconstructed in this work using $95$ and
		$150$~GHz data and projected to Equatorial coordinates. The \sptg footprint covers
		approximately 1500 deg$^2$ of the southern sky. The background shows the Galactic dust map
		from \planck \texttt{Commander} in intensity plotted in a logarithmic color scale.}
	\label{fig:lensing_maps_combined}
\end{figure*}

\subsection{Lensing Maps}
In Fig.~\ref{fig:lensing_maps_combined}, we show the lensing convergence maps $\kappa$
reconstructed from the individual frequency combinations 95 $\times$ 95~GHz,
95 $\times$ 150~GHz, and 150 $\times$ 150~GHz. The convergence is related to the
lensing potential $\phi$ as $\kappa=-\frac{1}{2}\nabla^2\phi$,
which in harmonic space translates to
$\kappa_{\bm{L}}=\frac{L(L+1)}{2}\phi_{\bm{L}}$. The $\kappa$ maps in Fig.~\ref{fig:lensing_maps_combined}
are smoothed using a 1-degree FWHM Gaussian filter to emphasize the large-scale modes
with higher $S/N$.
Our lensing maps are signal dominated at lensing multipoles below $L\approx 70,90,100$ for $\hat{\phi}^{T_{95}T_{95}}$, $\hat{\phi}^{T_{150}T_{150}}$, and $\hat{\phi}^{T_{95}T_{150}}$,  respectively.
The three panels on the left of Fig.~\ref{fig:lensing_maps_combined} have nearly identical
structures and are strongly correlated.  $\hat{\phi}^{T_{95}T_{95}}$, $\hat{\phi}^{T_{95}T_{150}}$, and $\hat{\phi}^{T_{150}T_{150}}$  share
similar $N^0_L$ noise contributions from Gaussian power of the CMB, 
thus their MV combination only slightly reduces the uncertainty
compared to each individual frequency combination. We include the MV-combined lensing map
on the right of Fig.~\ref{fig:lensing_maps_combined}, which has almost the same structure
as the panels on the left. We show the MV lensing map in Equatorial coordinates to highlight
the observing field for this analysis. 
Our lensing map covers three times the area and has higher reconstruction noise per mode compared to \sptpol~(\citetalias{wu19}).
On the other hand, our lensing map is 60\% in size and has lower noise per mode compared to \sptsz{}~(\citetalias{omori17}). 
Our lensing convergence maps show consistent degree-scale structure compared to both the  \sptpol and \sptsz{} lensing maps visually.

\subsection{Lensing Power Spectra and Amplitude}
\label{sec:results_spectra}
The lensing band powers from the six individual debiased lensing spectra and their MV combination, defined    
in Eqs.~\ref{eq:amp_def} and~\ref{eq:mv_definition}, are presented in Fig.~\ref{fig:clkk_multifreq}.
In addition, we have corrected for the foreground bias contamination
estimated in Sec.~\ref{sect:astro_fgnds} using \texttt{AGORA} simulations that closely match
our observing frequencies and map processing steps. The foreground bias template is subject to
uncertainties from sample variance and depends on the accuracy of the associated simulations.
As a result, we do not directly use these foreground-corrected spectra for cosmological
interpretation, but marginalize over the amplitude of the foreground bias template $A_{\rm fg}$ with a conservative uniform prior, as discussed in greater detail in Sec.~\ref{sec:lens_like}.
We report the lensing power spectra in
logarithmically spaced $L$ bins between 50 and 2000.
The corresponding MV bandpowers and uncertainties are provided in Tab.~\ref{tab:bandpowers}.
The uncertainties of the individual bandpowers and the MV are similar because they are dominated by $N^0_L$, which is largely shared between lensing map reconstructed from the 95 and 150~GHz input maps.
In each multipole bin, the observed lensing power differs no more than $\lesssim 1\sigma$ between different reconstructions.
The visual agreement across the spectra recovered from different frequency combinations provides a consistency check and suggests that the foreground biases are not substantially larger in any one combination, though more substantial tests will be done in the next section.

\begin{table}[t]
	\caption{MV lensing bandpowers}
   	\begin{adjustwidth}{1cm}{}
		\begin{tabular}{c c c | c }
				\hline\hline
			$[\,L_{\rm min}$ & $L_{\rm max}\,]$ & $L_b$ & $10^7 [L(L+1)]^2 \hat{C}_L^{\phi\phi} / 2\pi$ \\ [0.5ex]
			\hline
			$[\,50$&$67\,]$&$59$&$ 0.958\pm0.336$\\
			$[\,68$&$91\,]$&$80$&$ 1.261\pm0.233$\\
			$[\,92$&$125\,]$&$109$&$ 1.061\pm0.164$\\
			$[\,126$&$170\,]$&$148$&$ 0.781\pm0.112$\\
			$[\,171$&$232\,]$&$202$&$0.636\pm0.081$\\
			$[\,233$&$315\,]$&$274$&$ 0.442\pm0.064$\\
			$[\,316$&$429\,]$&$373$&$ 0.356\pm0.044$\\
			$[\,430$&$584\,]$&$507$&$ 0.172\pm0.029$\\
			$[\,585$&$794\,]$&$690$&$ 0.084\pm0.022$\\
			$[\,795$&$1080\,]$&$938$&$ 0.061\pm0.018$\\
			$[\,1081$&$1470\,]$&$1276$&$0.059\pm0.015$\\
			$[\,1471$&$1999\,]$&$1735$&$ 0.021\pm0.013$\\
			\hline
			\end{tabular}
			\end{adjustwidth}
			\begin{centering}
			\begin{tablenotes}[flushleft]
			\item MV lensing bandpowers as defined in Eq. \ref{eq:mv_definition}.
	Values are for  $10^7 [L(L+1)]^2 \hat{C}_L^{\phi\phi} / 2\pi$
	given for logarithmically spaced bins between 50 and 2000.
			\end{tablenotes}
			\end{centering}
			\label{tab:bandpowers}
\end{table}

\begin{figure}
	\centering
	\includegraphics[width=1\columnwidth]{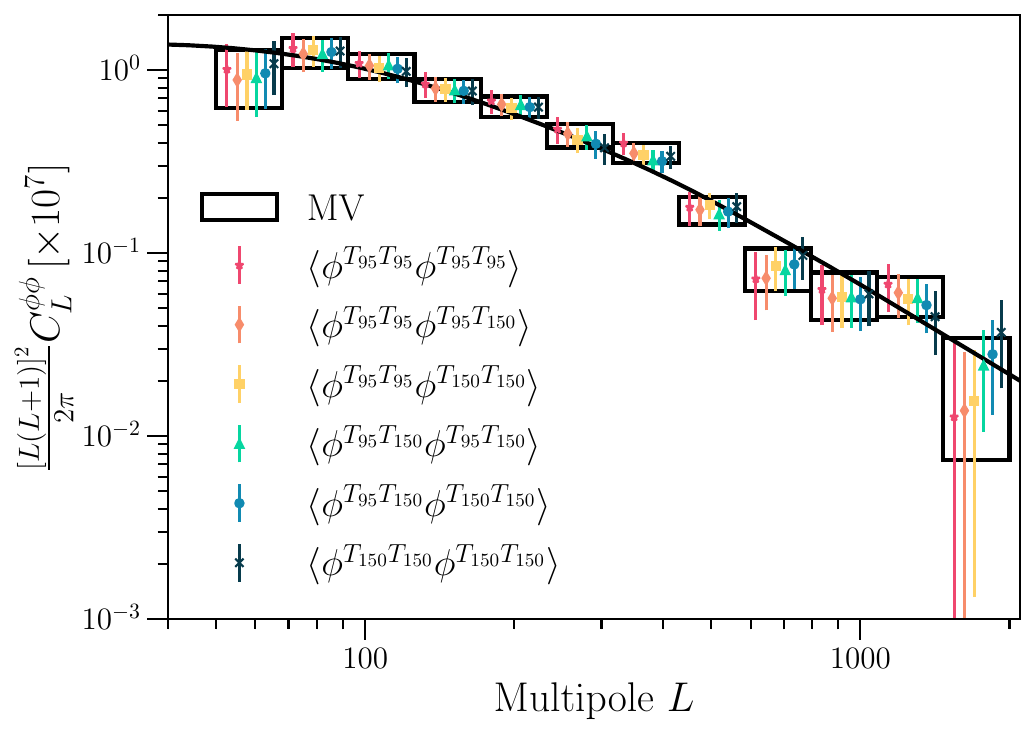}
	\caption{Comparison of the minimum-variance (MV) lensing band powers reconstructed using temperature data at 95 and 150 GHz (black boxes) against band powers from individual frequency combinations (colored points). The solid line is the lensing spectrum from the \planck{} 2018 best-fit \lcdm{} model. }
	\label{fig:clkk_multifreq}
\end{figure}

\begin{table}[ht]
	\caption{Lensing amplitudes and uncertainties}
	\begin{adjustwidth}{0.3cm}{}
		\begin{centering}
			\begin{tabular}{l | c c  c  }
				\hline\hline
				Freq.  comb. & Amp. & Stat. uncert. & Sys. uncert.\\ [0.5ex]
				\hline
				\addlinespace[.2ex]
				$\hat{C}^{\phi^{T_{95}T_{95}} \phi^{T_{95}T_{95}}}_{L}         $&1.085&0.073  & 0.023\\
				$\hat{C}^{\phi^{T_{95}T_{95}} \phi^{T_{95}T_{150}}}_{L}       $&1.018 &0.063 & 0.020\\
				$\hat{C}^{\phi^{T_{95}T_{95}} \phi^{T_{150}T_{150}}}_{L}     $&1.013& 0.060 &0.018\\
				$\hat{C}^{\phi^{T_{95}T_{150}} \phi^{T_{95}T_{150}}}_{L}     $&1.001& 0.061 &0.018\\
				$\hat{C}^{\phi^{T_{95}T_{150}} \phi^{T_{150}T_{150}}}_{L}   $&0.983& 0.062 &0.017\\
				$\hat{C}^{\phi^{T_{150}T_{150}} \phi^{T_{150}T_{150}}}_{L} $&1.003& 0.066 &0.017\\
				MV combined &1.020&0.060& 0.016\\
				\hline
			\end{tabular}
		\end{centering}
	\end{adjustwidth}
	\begin{tablenotes}[flushleft]
	\item 	Lensing amplitudes for individual frequency combinations and their MV combination.
		The statistical and systematic uncertainties are also included. The total systematic uncertainties are evaluated by taking the quadrature sum of the individual contributions.
	\end{tablenotes}
	\label{tab:amplitudes}
\end{table}

We summarize the lensing amplitudes for each estimator and their MV combination, defined also
in Eqs.~\ref{eq:amp_def} and~\ref{eq:mv_definition}, in Tab.~\ref{tab:amplitudes}.
Similar to the bandpowers, we apply a foreground bias correction assuming the foreground template to be exact.
 To determine the statistical uncertainties,
we utilize the standard deviation of the lensing amplitude distribution based on the 340
simulations within set A discussed in Sec.~\ref{sect:bias_est_with_sims}.  
Our lensing amplitude for the MV combination is
\beq
\label{eq:lensamp}
A^{\rm MV} =  1.020 \pm 0.060.
\eeq
Considering solely statistical uncertainty, the lensing amplitude for the MV combination
is measured with an uncertainty of 5.9\%, which is comparable to the uncertainties of  6.7\%
using 95~GHz alone and 6.6\% for 150~GHz alone.
The uncertainty on the amplitude of $\hat{C}^{\phi^{T_{95}T_{95}} \phi^{T_{150}T_{150}}}_{L}$ is the smallest 
among the six combinations because non-common fluctuations (e.g. instrumental noise) 
between the 95GHz and 150GHz maps do not contribute to its $N^0_{L}$ in this case.  
We quantify the systematic uncertainties
associated with the map calibration factor and beam uncertainty,
 in Sec.~\ref{sect:sys_uncertainties}. 
The systematic uncertainty for the MV combination lensing amplitude is $\pm 0.016$,
a fraction of the statistical uncertainties.
Considering both statistical and systematic uncertainties, the measurement of the MV
lensing amplitude has an uncertainty of  7.5\%.
In Sec.~\ref{sec:lensing_amp}, we also explore the lensing amplitude using a loose prior on
the foreground bias template amplitude $A_{\rm fg}$ and derive a constraint on the lensing amplitude of
$A_L^{\phi\phi} = 1.063 \pm 0.090$.
This result is consistent with the result reported in Eq.~\ref{eq:lensamp}.

In Fig.~\ref{fig:global_bp_log}, we compare our MV lensing spectra with other experiments. Our
measurements agree with previous measurements from \sptsz~(\citetalias{omori17}), \sptpol~(\citetalias{wu19}), \planck~\citep{carron22}, and
\actpol~\citep{qu23}.
We constrain the lensing amplitude at 5.9\%, comparable to \sptpol's 6.1\%
(\citetalias{wu19}) and \sptsz{}'s 6.3\% (\citetalias{omori17}).

\begin{figure*}[htb]
	\centering
	\includegraphics[width=\textwidth]{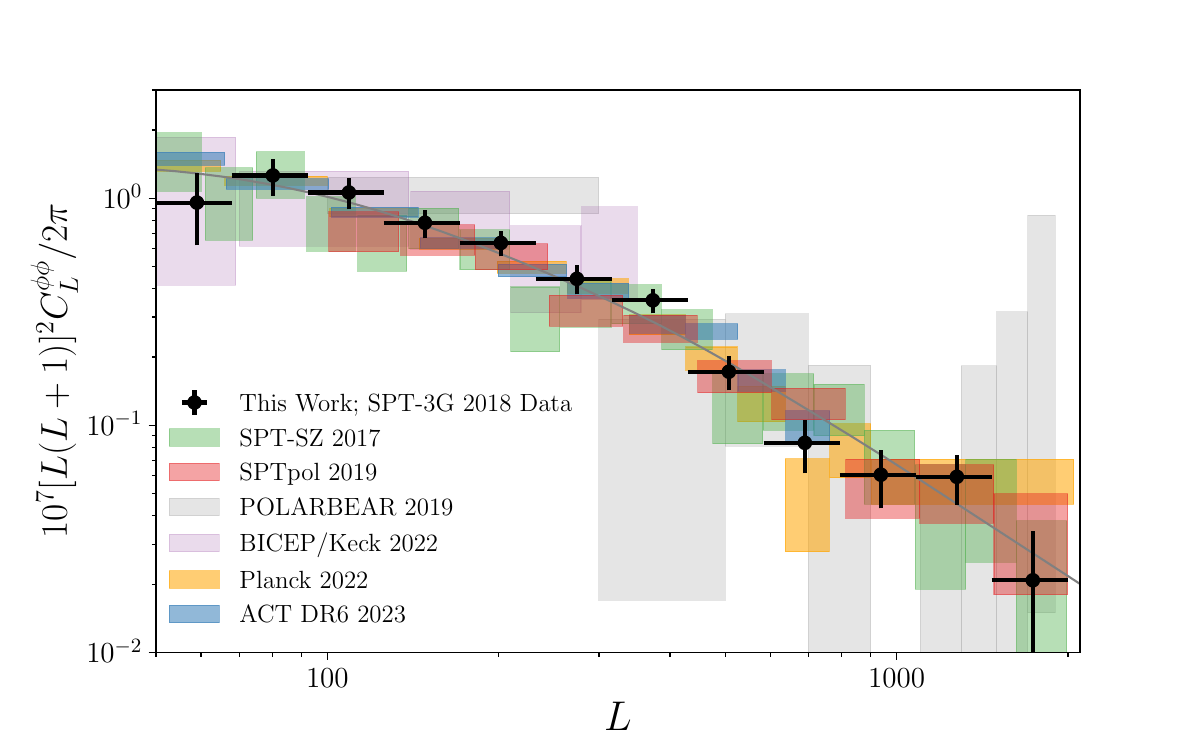}
	\caption{Lensing power spectrum measurements from this work, SPT-SZ (\citetalias{omori17}), \sptpol (\citetalias{wu19}),  POLARBEAR
	\citep{polarbear19b},
		BICEP/Keck \citep{ade23}, \planck \citep{carron22},
		 and \actpol \citep{qu23}.
		We also plot the lensing spectrum from the best-fit \LCDM{} model to the 2018 \planck  \texttt{TT,TE,EE+lowE+lensing}
		dataset (solid gray line)~\citep{planck18-6, planck18_archive}.}
	\label{fig:global_bp_log}
\end{figure*}

\subsection{Consistency and Null Tests}
\label{sect:consistency_tests}
To ensure the robustness of our analysis and identify potential systematic errors in the data, we conduct a suite of tests.
For each test, we modify one aspect of the analysis and recalculate the lensing power spectrum.
We then assess the consistency of the results by comparing the bandpower obtained from the modified analysis with that
obtained from the baseline analysis.
The bandpowers obtained from the different analysis options are summarized in Fig.~\ref{fig:clkk_mv_syst}.
We designed a suite of tests to specifically
 check the lensing reconstruction and confirm the pipeline's robustness
 against unmodeled nonidealities in the data.

We quantify the $\chi^2$ of the tests by comparing the difference data bandpower with the mean of 
the difference over 340 simulations (Set A in Sec. \ref{sect:bias_est_with_sims}), and using the 
variance of the per-bin simulation-difference spectrum to approximate the covariance diagonal, 
while neglecting the off-diagonal components, since the difference bandpowers are largely 
uncorrelated across different bins.
The $\chi^2$ of the systematic tests are
\beq
\label{eq:chisq}
\chi_{\mathrm{sys}}^2=\sum_b \frac{\left(\Delta \hat{C}_{b, \mathrm{data}}^{\phi \phi}-\left\langle\Delta \hat{C}_{b, \mathrm{sim}}^{\phi \phi}\right\rangle\right)^2}{\sigma_{b, \mathrm{sys}}^2}.
\eeq
Here $\left\langle\Delta \hat{C}_{b, \mathrm{sim}}^{\phi \phi}\right\rangle$ and $\sigma_{b, \mathrm{sys}}$
are estimated by performing the same analysis on the 340 simulations as on the data.
We compute the differences in overall lensing amplitude $\Delta A_{\mathrm{data}}$
between the alternate analysis choice and the baseline by subtracting the
two amplitudes and quantifying the $\chi^2$ using the distribution of amplitude differences for the simulations.
We also use Eq.~\ref{eq:chisq} to calculate the $\chi^2$ values for all our 340 simulations by replacing $\Delta \hat{C}_{b, \mathrm{data}}$ with individual simulation spectra differences.
The probability-to-exceed (PTE) is calculated as the fraction of simulation $\chi^2$ values that are higher than the $\chi^2$ value obtained from the data.
We summarize the data $\chi^2$ and PTE results in Tab.~\ref{tab:null_pte}.
To account for the look-elsewhere effect using the Bonferroni correction, we stipulate that
the PTE values should exceed the threshold of $5\%/N$, with $N=9$ here being the number of tests carried out~\citep{dunn61}. Here the different tests are being regarded as independent or weakly dependent.
As can be seen in Tab.~\ref{tab:null_pte}, all our PTE values are above $0.05/9 \approx 0.006$.  We
 conclude that we find no signs of significant systematic biases in these tests.

In Fig.~\ref{fig:clkk_mv_syst_deltasigma}, we present a visual overview of the consistency
 tests, illustrating the band power difference relative to the baseline result for each test.
The error bars represent the variations derived from the distributions of simulation-difference bandpowers.
In each $L$ bin, the data points and error bars are normalized by the 1$\sigma$ lensing spectrum uncertainties in that specific bin. We find that the difference band powers
 statistically meet expectations with no major systematic shifts for any test.  Below, we
 discuss the results of each consistency test in more detail.\\

{\it Varying $\ell_{\rm xmin}$, $\ell_{\rm min}$, and $\ell_{\rm max}$:}
The goals of these tests include checking the reconstruction with different amounts of CMB modes
and confirming the robustness of the results against contaminations.
Our baseline analysis excludes modes $\ell, \ell_{\mathrm{x}}<300$ at
95 GHz (and $\ell, \ell_{\mathrm{x}}<500$ at 150 GHz) and modes at $\ell>3000$ for both 95 and 150 GHz.
We increase $\ell_{\mathrm{min}}$ and $\ell_{\mathrm{xmin}}$
by 200 and reduce $\ell_{\mathrm{max}}$ by 500 compared to the baseline choices.
When varying $\ell_{\mathrm{max}}$, we find the lensing spectrum using
$\ell_{\mathrm{max}}=2500$ to be consistent with $\ell_{\mathrm{max}}=3000$
at a PTE of 0.09.
Also, the cumulative $S/N$ for lensing measurements does not go up significantly
beyond $\ell=3000$, so we set this as the
high-$\ell$ limit for our baseline analysis. Setting an $\ell_{\mathrm{max}}$ cut
also helps reduce the foreground bias, which we estimate in
Sec.~\ref{sect:astro_fgnds}.
For the lower $\ell$ and $\ell_x$ end,
we found $\ell_{\mathrm{min}} = \ell_{\mathrm{xmin}}=300$ for 95~GHz (and 500 for 150~GHz)
to be an optimal threshold that reduces low-frequency instrumental
and atmospheric noise leakage into other $\ell$ ranges (\citetalias{dutcher21})
while not removing too many modes. From Tab.~\ref{tab:null_pte},
the lensing spectra after changing the $\ell_{\mathrm{min}}$ or $\ell_{\mathrm{xmin}}$
are consistent with the baseline with PTEs of 0.30 and 0.24, respectively. \\

{\it Source Masking and Inpainting: }
We mask and inpaint extragalactic point sources and galaxy cluster imprints in the CMB map to reduce the bias they contribute (Sec.~\ref{subsect:mask_inpaint}).
We perform several tests to determine whether the masking or inpainting thresholds are sufficient.
Note that this is separate from the TOD-level masking discussed in
Sec. \ref{subsect:map_making}, which is validated with a separate
end-to-end analysis.
To explore the impact of masking, we tune the flux threshold 
of point source masking such that 30\% more
or 30\% fewer sources are masked compared to the baseline.
The modified cuts correspond to flux thresholds of 47~mJy and 80~mJy
compared to the baseline of 50~mJy at 150~GHz.
We also test the impact of inpainting for fainter sources by inpainting
50\% fewer sources than the baseline. The modified cuts correspond to source flux
thresholds of 6.0, 6.0, and 12.0~mJy at 95, 150, and 220~GHz and cluster
detection significance threshold of 9, compared to the baseline choices
described in Sec. \ref{subsect:mask_inpaint}. The PTEs for bandpower difference
relative to the baseline are 0.30 for 30\% more masking, 0.03 for 30\% less masking, 
and 0.01 for 50\% less inpainting. 
We note that the data points in Fig.~\ref{fig:clkk_mv_syst_deltasigma} do not move much relative
to the baseline results despite the relatively small PTEs for less masking and less inpainting, 
which indicate that the change in the bandpower is larger than expected
from the sims given the same analysis change.
Still, the bandpower PTEs are consistent
with expected changes from the simulations after we account for
the Bonferroni correction.
Furthermore, the lensing amplitude PTEs for less masking and less inpainting are 5\% and 90\%, respectively. 
We conclude that our results are robust to variations of masking and inpainting choices. \\

{\it Mask Apodization: }
Gradients in the masking edges can mimic lensing and be captured by the lensing estimator. 
The mean field subtraction discussed in Sec.~\ref{sect:lensing_estimation} should correct for this bias. 
To check whether the bias from mask  apodization is negligible, we change the cosine taper radius
for the boundary mask from the baseline of 30$'$ to 60$'$ and the source mask
from 10$'$ to 20$'$. We repeat the complete analysis process with
these changes and find the data difference to agree with the simulation difference distribution
with a PTE of 0.37. \\

{\it Transfer Function Variation: }
The goal of this test is to confirm our analysis is insensitive to variations in transfer functions estimated using
input maps with different power spectra.
We test the impact of the transfer function by estimating
the transfer functions using another set of Gaussian simulations following a power law spectrum
with a spectral index of $-1$ instead of the CMB power spectrum following the method described in Sec.~\ref{sect:simulations}.
The new transfer function has different mode coupling and shifts slightly
compared to the baseline transfer function due to a change
in the map spectra used to estimate it.
We expect the impact to be small because of the
response normalization discussed in Sec.~\ref{sect:lensing_estimation}.
Using a slightly varied transfer function should only result in a non-optimal weighting and
a small degradation of $S/N$.
The power spectra using the baseline transfer function
and the new one are consistent with a PTE of 0.29.
 \\

{\it  Curl Test:}
The lensing deflection field $\boldsymbol{d}$ can be decomposed into gradient and curl components:
$\boldsymbol{d}(\hat{\boldsymbol{n}})=\boldsymbol{\nabla} \phi(\hat{\boldsymbol{n}})+(\star \boldsymbol{\nabla}) \Omega(\hat{\boldsymbol{n}})$,
where $\phi$ is the lensing potential, $\Omega$ is the divergence-free or curl component,
and $\star $ is a 90$^{\circ}$ rotation operator \citep{namikawa12}.
We expect the curl component to be zero at our reconstruction noise levels
where higher-order and post-Born effects are negligible~\citep{robertson23}.
However, foregrounds or other systematic effects may introduce a non-zero curl signal to the data.
A curl estimation on the data can test for these signals.
The curl spectrum $C_L^{\Omega\Omega}$ is extracted using an estimator
analogous to the lensing estimator introduced in Sec.~\ref{sect:lensing_estimation}
but designed to respond to the curl component \citep{namikawa12}.
In addition, there are two key distinctions.
First, the theoretical input is set to a flat spectrum $C_L^{\Omega\Omega}=10^{-7}$,
which is used to uniformly weight the modes when binning,
as well as to establish a reference spectrum for the amplitude calculation.
Second, no MC response correction from simulations is applied to the reconstructions since the expected signal is zero.
Following the method for the lensing spectrum analysis, we correct for other biases to the curl spectrum, including the non-zero $N^1_L$ bias from the lensing trispectrum \citep{vanengelen12}.
We plot the curl spectrum for the MV combinration in Fig.~\ref{fig:clkk_mv_syst}, which is consistent with null at a PTE of 0.18. \\

\begin{figure}
	\centering
	\includegraphics[width=1\columnwidth]{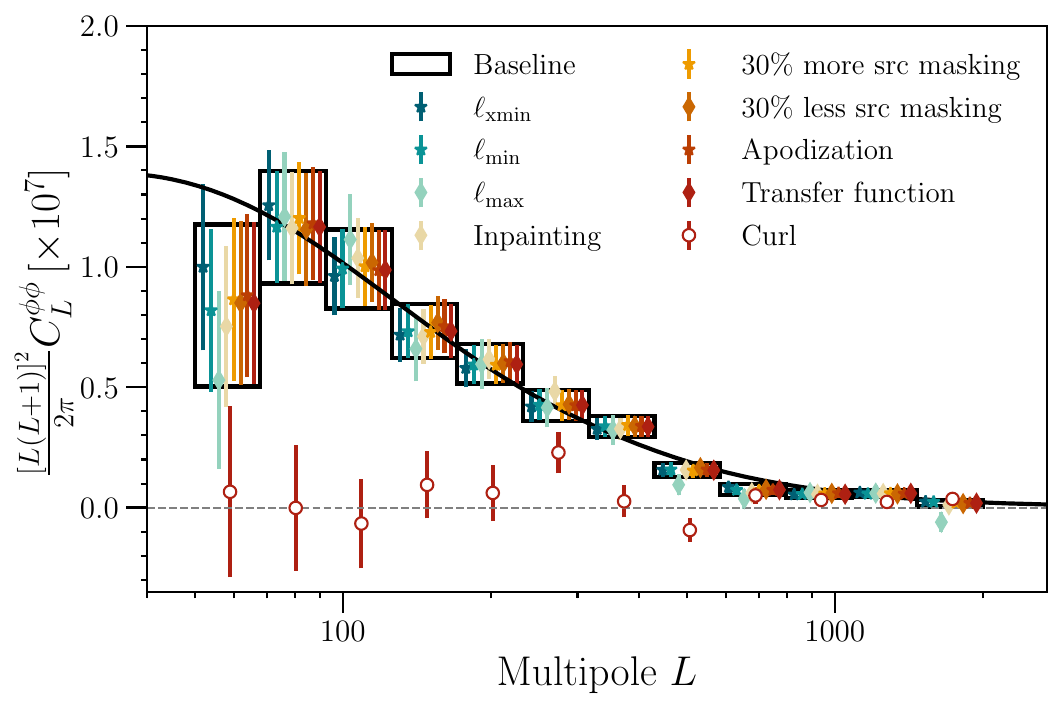}
	\caption{Results of the power spectrum consistency tests and curl null test for the minimum-variance lensing band
		powers formed by combining temperature data at 95 and 150 GHz. The band powers and errors for the baseline
		analysis are displayed as boxes. The band powers obtained from the different analysis choices are plotted with
		different colors and are in agreement with the baseline results. Note that we do not subtract the fiducial foreground bias template from these band powers.}
	\label{fig:clkk_mv_syst}
\end{figure}

\begin{figure}
	\centering
	\includegraphics[width=1\columnwidth]{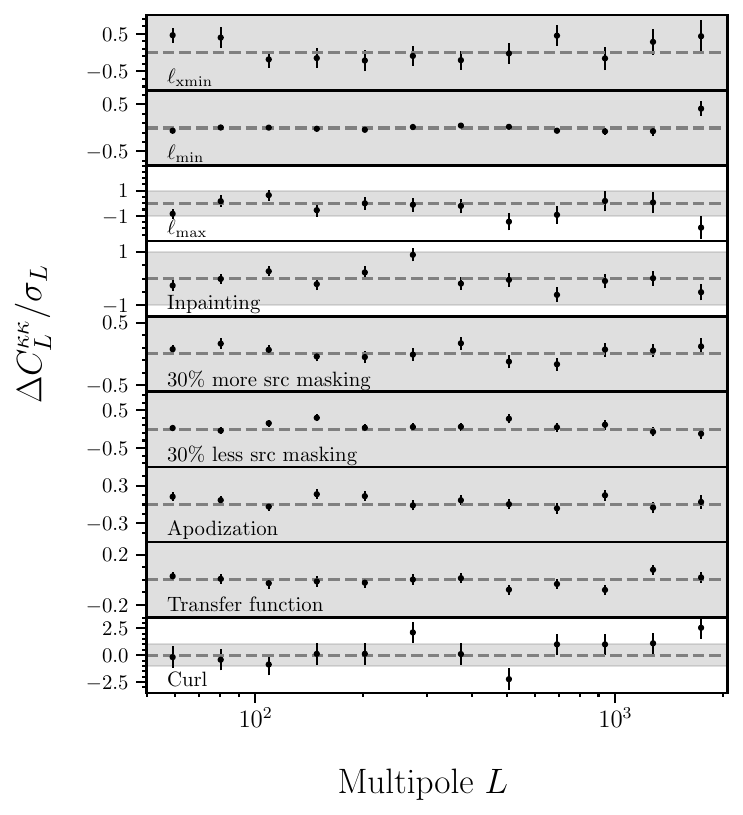}
	\caption{Comparison of difference bandpowers ($\Delta C_L^{\kappa\kappa}$) between the baseline analysis
		and those where we vary a given analysis setting, scaled by the uncertainties of the respective $\phi^{TT}$
		bandpowers. The error bars represent the standard deviations of the shifts for 340 simulations with the same analysis choice change.
		The shaded gray regions represent the 1$\sigma$ bands of the $\phi^{TT}$ estimators.}
	\label{fig:clkk_mv_syst_deltasigma}
\end{figure}

\begin{table}[t]
	\caption{Consistency Tests $\chi^2$ and PTEs}
   	\begin{adjustwidth}{0.5cm}{}
	\begin{tabular}{l |c c| c c }
		\hline\hline
		Test Name & $\chi^2_{\rm MV}$ & PTE  & $\Delta A_{\rm MV}$ & PTE\\
		&&& $\pm \sigma(\Delta A_{\rm MV})$&   \\ [0.5ex]
		\hline
		$\ell_{\rm xmin}$&15.1 & 0.24  & $-0.017 \pm 0.014$ & 0.32\\
		$\ell_{\rm min}$ & 13.8 & 0.30 & $\hphantom{-}0.003 \pm 0.003$  & 0.33  \\
		$\ell_{\rm max}$& 19.1 & 0.09 & $-0.032 \pm 0.043$ & 0.42  \\
		Less inpainting  & 27.7 & 0.01 & $\hphantom{-}0.001\pm 0.012$   & 0.90\\
		More masking   & 13.7 & 0.30 & $\hphantom{-}0.004\pm 0.006$   & 0.50\\
		Less masking   &  24.6 & 0.03 & $\hphantom{-}0.011\pm 0.006$   & 0.05\\
		Apod. mask      &  13.2 & 0.37 & $\hphantom{-}0.001\pm 0.005$  & 0.28 \\
		Transfer func.   &  14.4 & 0.29 & $-0.002 \pm 0.002$ & 0.54\\
		Curl                  &  16.0 & 0.18 & $-0.038 \pm 0.044$ & 0.40\\
		\hline
	\end{tabular}
   \end{adjustwidth}
   \centering
	\begin{tablenotes}[flushleft]
	\item	Summary of the $\chi^2$ for the difference band powers and
		the difference amplitudes as well as their corresponding PTEs for the systematics tests.
	\end{tablenotes}
	\label{tab:null_pte}
	\vspace{0.1cm}
\end{table}

\subsection{Systematic Uncertainties}
\label{sect:sys_uncertainties}
In this subsection, we assess the impact of uncertainties in the beam measurement and temperature calibrations on the lensing amplitude measurement.\\

{\it Beam uncertainty:}
To assess the impact of beam-related uncertainties, we introduce perturbations to the baseline beam profile using
the uncertainties ($\Delta B_{\ell}$) obtained from~\citetalias{dutcher21}.
These perturbations are applied by multiplying $1+\Delta B_{\ell}$ to the data map while keeping the simulations unchanged.
Subsequently, we divide both the data and simulations using the baseline beam,
which tests for a systematic 1$\sigma$ underestimation of the beam profile across the entire multipole range.
The resulting systematic shift in the amplitude of the lensing power spectrum is $\Delta A_{\rm beam} = 0.013$,
which is 22\% of the statistical uncertainty on $A_{\rm MV}$.\\

{\it Temperature calibration:}
We use a temperature calibration factor to calibrate the raw temperature maps via $T = T_{\rm raw} \times T_{\rm cal}$.
As outlined in Sec.~\ref{sect:tcal}, we calculate the uncertainty of the calibration factor
from Monte-Carlo using simulations with the same noise levels. The uncertainties for the four subfields are similar.
By taking their average, the uncertainties associated with
$T_{\rm cal}$, $\delta T_{\rm cal}$, are 0.3\% and 0.2\% for 95 and 150~GHz, respectively.
While keeping the simulated maps unchanged, we adjust the data maps by scaling them with
($1 + \delta T_{\rm cal}$) for the temperature map, and subsequently re-evaluate the data lensing amplitudes.
We then quantify the difference by conducting the baseline analysis with the temperature calibration of the data
maps shifted by 1$\sigma$, resulting in a $\Delta A_{\rm cal} = 0.010 $ for the MV
reconstruction, or about $0.16\sigma$ of the statistical uncertainties. \\

We report the quadrature sum of
the beam and temperature calibration uncertainties as the total systematic uncertainty
in Tab.~\ref{tab:amplitudes}.
The total systematic uncertainty is 0.016, which is smaller than the statistical uncertainty of 0.060.
This is smaller than \sptpol{}'s systematic uncertainty of  0.040 due to the absence of polarization calibration uncertainty, \sptpol's leading source of systematic uncertainty, in our case.  
 Our systematic uncertainty is similar to that of the temperature-only \sptsz{} measurements.

\section{Cosmological parameters}
\label{sect:params}
In this section, we present constraints from the 2018 \sptg lensing power spectrum on $\Lambda$CDM cosmological parameters as well as on a number of one- and two-parameter extensions.
As a reminder, we use the minimum-variance lensing band powers from temperature data introduced in Sec.~\ref{sect:mf_combo} to carry out the cosmological inference.

\subsection{Cosmological Inference Framework}
\label{sect:cosmo_framework}
Our baseline cosmology is a $\Lambda$CDM model with a single family of massive neutrinos having a total mass of $ \sum m_{\nu} =60$ meV.\footnote{The massless neutrino species contribute to the total effective number of relativistic species with $N_{\rm eff}=2.044$.}
The model is based on purely adiabatic scalar fluctuations and includes six parameters: the physical density of baryons ($\Omega_{\rm b}h^2$), the physical density of cold dark matter ($\Omega_{\rm c} h^2$), the (approximated) angular size of the sound horizon at recombination ($\theta_{\rm MC}$), the optical depth at reionization ($\tau$), the amplitude of curvature perturbations at $k=0.05$ Mpc$^{-1}$ ($A_{\rm s}$), and spectral index ($n_{\rm s}$) of the power law power spectrum of primordial scalar fluctuations.
We also quote constraints derived from these main six parameters, such as $\sigma_8$, the square root of the variance of the density field smoothed by a spherical top hat kernel with a radius of 8 Mpc/$h$, as calculated in linear perturbation theory \citep{peebles80},
and the Hubble constant $H_0$. We then examine a series of $\Lambda$CDM extensions including the sum of the neutrino masses $\sum m_{\rm \nu}$ and the spatial curvature $\Omega_{K}$.
The lensed CMB and CMB lensing potential power spectra are calculated with the \texttt{CAMB}\footnote{\url{https://camb.info} (\texttt{v1.4.1}). We use the Mead model \citep{mead20} to calculate the impact of non-linearities on the small-scale matter power spectrum $P_{\delta\delta}(k)$. The accuracy settings of \texttt{CAMB} are set to \texttt{lens\_potential\_accuracy=4; lens\_margin = 1250; AccuracyBoost = 1.0; lSampleBoost = 1.0; and lAccuracyBoost = 1.0}, which have been shown to be accurate enough for current sensitivities while enabling fast MCMC runs (see \citet{qu23,madhavacheril23}, and references therein).} Boltzmann code.
We sample the posterior space and infer cosmological parameter constraints using the Metropolis-Hastings sampler with adaptive covariance learning provided in the Markov Chain Monte Carlo (MCMC) \texttt{Cobaya}\footnote{\url{https://github.com/CobayaSampler/cobaya}}  package.

\subsubsection{Lensing Likelihood}
\label{sec:lens_like}
The CMB lensing log-likelihood is approximated to be Gaussian in the band powers of the measured lensing power spectrum
\begin{equation}
\label{eq:cmblike}
\begin{aligned}
& -2 \ln \mathcal{L}_\phi(\boldsymbol{\Theta})= \\
& \qquad \sum_{bb'}\left[\hat{C}_{b}^{\phi \phi}-C_{b}^{\phi \phi,\mathrm{th}}(\boldsymbol{\Theta})\right] \mathbb{C}_{bb'}^{-1}\left[\hat{C}_{b'}^{\phi \phi}-C_{b'}^{\phi \phi, \mathrm{th}}(\boldsymbol{\Theta})\right],
\end{aligned}
\end{equation}
where $C_{b}^{\phi \phi,\mathrm{th}}(\boldsymbol{\Theta})$ is the binned theory lensing spectrum evaluated at the position $\boldsymbol{\Theta}$ in the parameter space, as given by the Boltzmann solver.
When combining CMB lensing measurements with primary CMB data, we neglect correlations between the 2- and 4-point functions as these have been shown to not affect the cosmological inference at current noise levels \citep{schmittfull13,motloch17,peloton17, trendafilova23}.
Therefore, when jointly analyzing CMB lensing and primary CMB, we simply multiply the respective likelihoods.
As discussed in Sec.~\ref{sect:mf_combo}, the covariance matrix for the minimum-variance lensing binned spectrum is calculated using Monte Carlo simulations from the conditioned covariance matrices.
We rescale the inverse covariance matrix by a $\approx 4\%$ Hartlap correction factor~\citep{hartlap07}.

The lensing potential power spectrum estimate depends on cosmology through the response function $\mathcal{R}_L$ and on the $N^1_L$ bias as
\begin{equation}
C_L^{\phi \phi, \mathrm{th}} (\boldsymbol{\Theta})=\frac{\mathcal{R}_L^2(\boldsymbol{\Theta})}{\mathcal{R}_L^2(\boldsymbol{\Theta}_{\rm fid})} C_L^{\phi \phi} ({\boldsymbol{\Theta}})+ N_L^1(\boldsymbol{\Theta}) - N_L^1(\boldsymbol{\Theta}_{\rm fid}),
\end{equation}
where $\boldsymbol{\Theta}_{\rm fid}$ denotes the fiducial cosmology assumed to perform the lensing reconstruction.
Here, we follow \citet{planck15-15,sherwin17,simard18,bianchini20a} and take this cosmological dependence into account by linearly perturbing $\mathcal{R}_L$ and $N^1_L$ around  $\boldsymbol{\Theta}_{\rm fid}$, which amounts to calculating derivatives of the response function with respect to the primary CMB power spectra (in our case only $C_{\ell}^{TT}$) and of the $N^1_L$ bias with respect to the lensing potential spectrum $C_L^{\phi\phi}$.
These corrections are computed once for the fiducial cosmology and stored in the matrices $M_{bb'}^{TT}$ and $M_{bb'}^{\phi\phi}$,  respectively.
Including the contribution from residual foregrounds, the full prediction for the lensing potential power spectrum takes the following form
\beq
\label{eq:lincorr}
\begin{aligned}
C_b^{\phi \phi, \mathrm{th}}({\boldsymbol{\Theta}}) &= C_b^{\phi \phi}({\boldsymbol{\Theta}}) + A_{\rm fg} C_b^{\phi\phi,\rm fg} \\
&+ \sum_{x\in \{TT,\phi\phi\}} M^{x}_{bb'} \left(C_{b'}^{x}(\boldsymbol{\Theta})-C_{b'}^{x}(\boldsymbol{\Theta}_{\rm fid})\right),
\end{aligned}
\eeq
where summation over $b'$ is implied.
In Eq.~\ref{eq:lincorr}, $C_b^{\phi\phi,\rm fg}$ is the foreground contamination template introduced in Sec.~\ref{sect:astro_fgnds} and $A_{\rm fg}$ is the corresponding amplitude parameter on which we impose a uniform prior $A_{\rm fg} \sim \mathcal{U}\left(0,3\right)$.
The prior range is motivated by the approximately factor of three difference between the foreground biases estimated from  \texttt{AGORA} and from~\citet{vanengelen14}.
\footnote{The masking thresholds and noise levels are different in~\citet{vanengelen14} so we do not expect it to be representative of the foreground bias in our measurement. However, it is a reference point of how different simulations with different assumptions of astrophysics can produce factor of a few difference in the level of foregrounds biases.}

\subsubsection{Cosmological Datasets}
\label{sec:datasets}
In this work, we present parameter constraints obtained from a comprehensive analysis of three main classes of cosmological observables.
Our study incorporates the following key observables and survey data:
\begin{itemize}
\item{{\bf CMB lensing}: we use the 2018 \sptg lensing power spectrum measurement presented in this work along with the lensing bandpowers obtained from the analysis of the \planck CMB
PR4 (NPIPE) maps \citep{carron22}.
The \sptg{} and \planck{} datasets have different sensitivities to different components of the primary CMB due to their limited overlap in the observational footprints (3.6\% compared to 67\% of the sky), and different noise properties and angular resolution. These two datasets are therefore relatively independent. 
Additionally, we compare the new constraints from \sptg to the previous results from the \sptpol experiment presented in \citet{bianchini20a} and \sptsz{} in \citet{simard18}. }
The constraints in \citet{bianchini20a} and \citet{simard18} are based on the lensing measurements in \citetalias{wu19} and \citetalias{omori17}, respectively.
\item{{\bf BAO}: we utilize likelihoods obtained from spectroscopic galaxy surveys, including the BOSS (Baryon Oscillation Spectroscopic Survey) DR12~\citep{alam17}, SDSS MGS~\citep[Sloan Digital Sky Survey Main Galaxy Sample;][]{ross15}, 6dFGS~\citep[Six-degree Field Galaxy Survey;][]{beutler11}, and eBOSS DR16 Luminous Red Galaxy \citep[LRG,][]{alam21} surveys.
Incorporating information about the BAO scale into our analysis allows us to refine the parameter constraints in the $\Omega_{\rm m}$-$H_0$ plane and gain insights into the amplitude of the large-scale structure.}
\item{{\bf Primary CMB}: as our baseline early universe observable,\footnote{While CMB temperature and polarization power spectra mostly probe the early universe at $z \approx 1100$, we stress that secondary interactions like lensing, reionization, and the integrated Sachs-Wolfe effect confer sensitivity to the low-$z$ universe evolution.} we employ the power spectrum measurements of primary CMB temperature and polarization anisotropies from the \planck 2018 data release~\citep{planck18-5}.
Specifically, we use the low- and high-$\ell$ temperature and polarization likelihoods from PR3 maps.
In addition, we make use of the 2018 \sptg $TT/TE/EE$ intermediate and small-scale measurements from \citet{balkenhol23}. }
\end{itemize}

\begin{table}[t]
\centering
\caption{Cosmological parameters varied in this work and their respective priors. Parameters that are fixed are reported by a single number. $\mathcal{U}(a,b)$ denotes a uniform distribution between $[a,b]$, while $\mathcal{N}(\mu,\sigma^2)$ indicates a Gaussian distribution with mean $\mu$ and variance $\sigma^2$. In addition to these, we include and marginalize over the amplitude of the foreground bias template using a uniform prior $A_{\rm fg} \sim \mathcal{U}(0,3)$. Note that when adding primary CMB information from \sptg, we adopt a \planck{}-based Gaussian prior on  $\tau \sim \mathcal{N}(0.0540, 0.0074^2)$.}
\label{tab:priors}
\begin{tabular}{c|c|c}
Parameter               & Lensing only  (+ BAO)                 & Lensing +  CMB      \\
\hline
\hline
$\Omega_{\rm b} h^2$          & $\mathcal{N}(0.02233,0.00036^2)$ & $\mathcal{U}(0.005,0.1)$     \\
$\Omega_{\rm c} h^2$          & $\mathcal{U}(0.005,0.99)$      & $\mathcal{U}(0.001,0.99)$    \\
$H_0$ [km/s/Mpc]                & $\mathcal{U}(40,100)$          & $\mathcal{U}(40,100)$        \\
$\tau$                  & 0.055                          & $\mathcal{U}(0.01,0.8)$      \\
$n_{\rm s}$                   & $\mathcal{N}(0.96,0.02^2)$     & $\mathcal{U}(0.8,1.2)$       \\
$\ln (10^{10}A_{\rm s})$        & $\mathcal{U}(1.61,3.91)$       & $\mathcal{U}(1.61,3.91)$     \\
$\sum m_{\rm \nu}$ {[}eV{]} & 0.06                           & 0.06 or $\mathcal{U}(0,5)$   \\
$\Omega_{K}$              & 0                              & 0 or $\mathcal{U}(-0.3,0.3)$ \\
$A_L$                   & 1                              & 1 or $\mathcal{U}(0,10)$     \\
$A_L^{\phi\phi}$        & 1                              & 1 or $\mathcal{U}(0,10)$
\end{tabular}%
\end{table}

\subsection{Constraints from CMB Lensing Alone}
\label{sec:parameter_lensing_alone}
We start by considering constraints on \lcdm parameters from CMB lensing measurements alone, with a special focus on the amplitude of matter fluctuations.

\begin{figure*}
	\centering
	\includegraphics[width=1\textwidth]{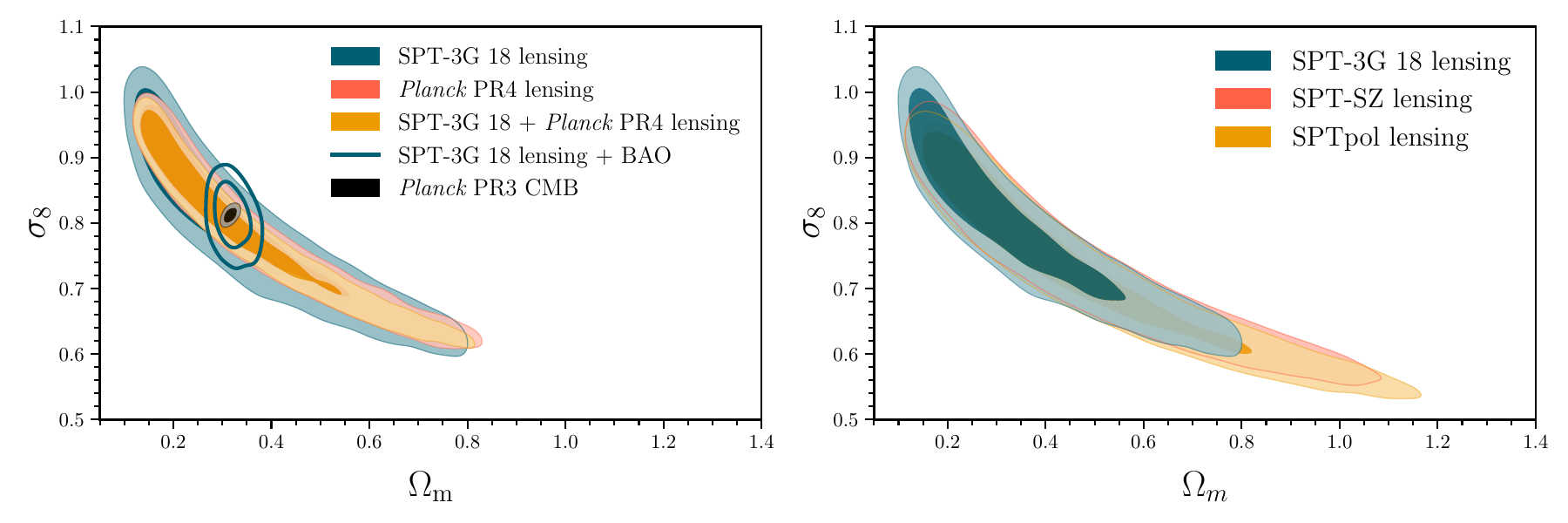}
	\caption{\textit{Left}: Constraints in the $\Omega_{\rm m}$-$\sigma_8$ plane from our \sptg{} CMB lensing measurements (filled blue contours). For comparison, we also include results from \planck{} PR4 lensing (red contours) as well as its combination with \sptg{} lensing data (yellow contours). The empty blue contours show the constraints combining our \sptg{} lensing likelihood with BAO data. The black filled contours representing the independent constraints derived from the \planck{} primary CMB power spectra are also found to be consistent with the CMB lensing measurements at lower redshifts. \textit{Right}: Comparison of $\Omega_{\rm m}$-$\sigma_8$ constraints across different SPT surveys.}
	\label{fig:s8_omegam}
\end{figure*}

The parameters that we vary in this analysis and their corresponding priors are listed in Tab.~\ref{tab:priors}.
Note in particular that we fix $\tau = 0.055$~\citep{planck18-6}, since CMB lensing is not directly sensitive to the optical depth, and that we impose a Gaussian prior on the baryon density $\Omega_{\rm b} h^2 =0.02233 \pm 0.00036$ based on recent element abundances and nucleosynthesis (BBN) modelling~\citep{cooke18,mossa20} as well as an informative albeit wide prior on $n_{\rm s}$ from \planck CMB anisotropies power spectra \citep{planck18-6}.
When analyzing \sptg CMB lensing measurements without primary CMB power spectra, we fix the linear corrections to the response function to the fiducial cosmology and only vary the ones related to the $N^1_L$ bias.

Within the base \lcdm model, constraints from CMB lensing measurements alone follow a narrow elongated tube in the 3D subspace spanned by $\sigma_8$-$H_0$-$\Omega_{\rm m}$.\footnote{See, e.g.,  \citet{pan14,planck15-15,madhavacheril23} for pedagogical discussions on the cosmological parameter dependence of the CMB lensing power spectrum.}
This is then projected as an elongated narrow region on the $\Omega_{\rm m}$-$\sigma_8$ plane, as shown in Fig.~\ref{fig:s8_omegam}.
The parameter combination optimally determined by CMB lensing measurements is $\sigma_8\Omega_{\rm m}^{0.25}$, which is constrained from \sptg data at the $\approx 4.4\%$ level:
\beq
\sigma_8\Omega_{\rm m}^{0.25} = 0.595\pm 0.026 \quad (\text{\sptg}).
\eeq
As a simple and quick way to estimate the agreement between measurements from two (independent) datasets, we calculate the differences in central parameter values normalized to the sum in quadrature of the uncertainties:  $(\mu_1-\mu_2)/\sqrt{\sigma_1^2+\sigma_2^2}$, where $\mu_i$ and $\sigma^2_i$ are the central values and variances of the two measurements, respectively.
With this definition, $\sigma_8\Omega_{\rm m}^{0.25}$ inferred from \sptg{} is only $0.13\sigma$ away from the \planck PR4 lensing value of $\sigma_8\Omega_{\rm m}^{0.25} = 0.599\pm 0.016$ and also in agreement at the 0.5$\sigma$ level with the value of $\sigma_8\Omega_{\rm m}^{0.25} = 0.609\pm 0.008$ based on the \planck CMB PR3 anisotropies.

In the right panel of Fig.~\ref{fig:s8_omegam}, we compare the \sptg results with those from the \sptsz{} analysis in \citetalias{omori17} and the \sptpol\ analysis in \citetalias{wu19}.\footnote{We have rerun the chains for the \sptsz{} and \sptpol{} surveys adopting the same priors and BAO dataset used in this analysis.}
It is important to note that apart from using three different cameras, these SPT measurements differ in other aspects as well.
Firstly, the sky coverage of these measurements are different. 
Secondly, while this measurement is similar to the temperature-only \sptsz{} analysis in that it relies solely on temperature data, the \citetalias{wu19} measurement used both temperature and polarization data, with the latter carrying more statistical weight in the final result.
Lastly, the \sptg and \sptsz{} lensing power spectra extend to lower multipoles ($L=50$) than the \sptpol\ measurement ($L=100$).
Hence, the \sptg and \sptsz{} measurements are more similar in that they both use temperature data
and cover a largely overlapping sky area, whereas \sptg and \sptpol provide relatively independent assessments of cosmology.

As can be seen from the right panel of Fig.~\ref{fig:s8_omegam}, the main difference between the \sptg{} dataset and its predecessors is that the bulk of the posterior mass moves to a region of the parameter space with lower $\Omega_{\rm m}$ and higher $\sigma_8$, excluding the high-$\Omega_{\rm m}$ tail present in the \sptsz{}/\sptpol{} datasets.
In particular, \sptg{} prefers a higher primordial scalar spectrum amplitude ($A_\mathrm{s}$), a parameter
 closely related to $\sigma_8$, with $\log(10^{10}A_s)= 3.22^{+0.27}_{-0.23}$ than
what is preferred by either \sptsz{}, at $2.76^{+0.31}_{-0.28}$, or \sptpol{}, at $2.65\pm 0.35$.
While bandpower fluctuations can lead to higher or lower preferred $\Omega_{\rm m}$ and $\sigma_8$ values, neither of which CMB lensing constrains particularly well, the combination
 $\sigma_8\Omega_{\rm m}^{0.25}=0.595\pm 0.026$ from \sptg is in good agreement with the values from \sptpol{} and from \sptsz{}.

We also note that our constraints are robust against details of the foreground treatment.
For example, when fixing the amplitude of the foreground contamination template $A_{\rm fg}$ to unity, the inferred constraint on the amplitude of structure becomes $\sigma_8 \Omega_\mathrm{m}^{0.25} = 0.589\pm 0.024$, only a 8\% reduction in uncertainty and less than $0.25\sigma$ shift in the central value.

We then proceed to combining the \sptg{} and \planck{} lensing measurements.
Given the independence (Sec.~\ref{sec:datasets}) and the consistency of the \sptg and \planck measurements,
their  combination then involves a straightforward multiplication of the lensing likelihoods associated with each dataset.
The constraint on $\sigma_8\Omega_{\rm m}^{0.25}=0.596\pm 0.014$ is improved by about 13\% with respect to \planck lensing-only statistical uncertainties.
A summary of the marginalized $\sigma_8\Omega_{\rm m}^{0.25}$ constraints across different datasets is provided in Fig.~\ref{fig:s8omegamp25} and the numerical values are reported in Tab.~\ref{tab:cmblens_bao}.

\begin{figure}
	\centering
	\includegraphics[width=1\columnwidth]{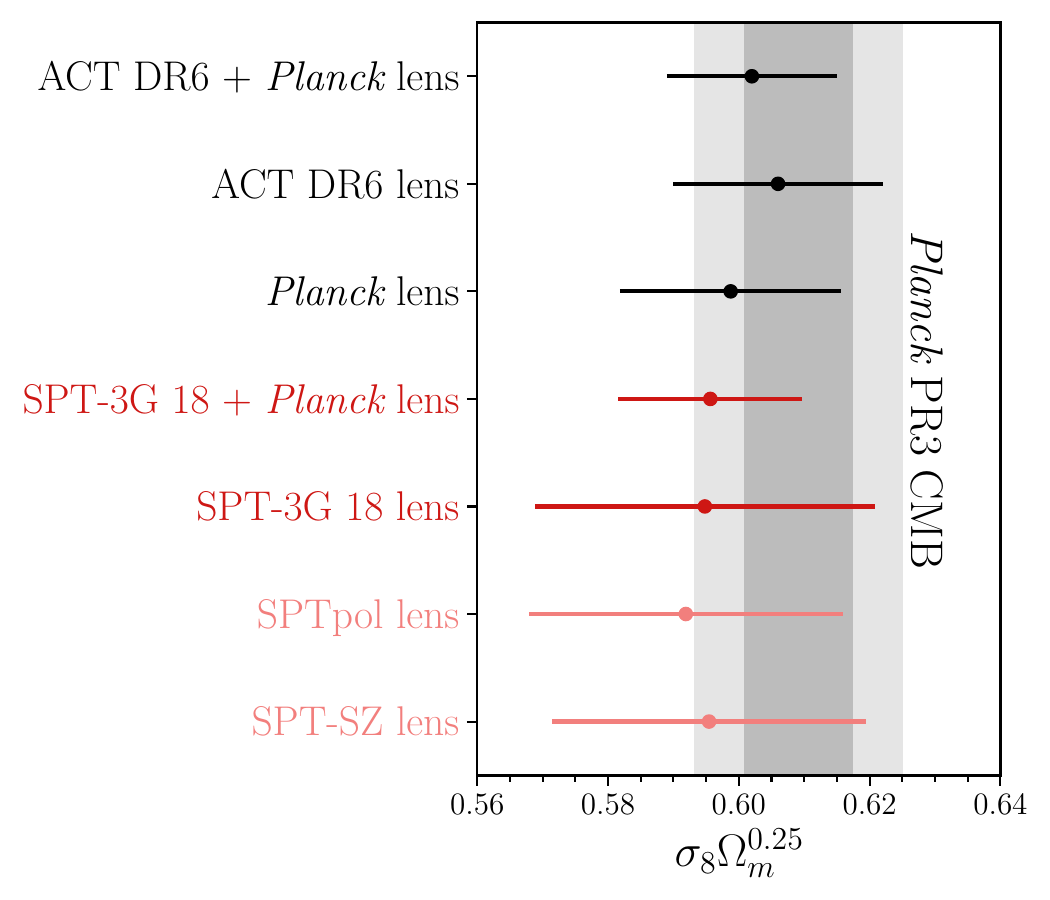}
\caption{Marginalized $\sigma_8 \Omega_{\rm m}^{0.25}$ posterior across different SPT, \planck, and ACT CMB lensing measurements. The shaded dark and light grey regions denote respectively the 1 and 2$\sigma$ statistical errors from \planck PR3 CMB temperature and polarization anisotropies.}
	\label{fig:s8omegamp25}
\end{figure}

\begin{table*}[htb]
\centering
\caption{Constraints on a subset of \lcdm parameters using the \planck and \sptg CMB lensing datasets alone, jointly analyzed, or combined with BAO information.  All intervals quoted in this table are 68\% intervals,  $H_0$ is in units of km s$^{-1}$Mpc$^{-1}$.}
\label{tab:cmblens_bao}
\begin{tabular}{cc|cccc}
\hline
\hline
 & \multicolumn{1}{c|}{Lensing} & \multicolumn{4}{c}{Lensing + BAO} \\
  \hline
 & $\sigma_8 \Omega_{\rm m}^{0.25}$ & $\sigma_8$   & $H_0$ & $\Omega_{\rm m}$     & $S_8$                 \\
\hline
\sptg  18            & $0.595\pm 0.026$ & $0.810 \pm 0.033$ &  $68.8^{+1.3}_{-1.6}$ & $0.320^{+0.021}_{-0.026}$ & $0.836 \pm 0.039$     \\
\planck          & $0.599 \pm 0.016$ & $0.815 \pm 0.016$ &  $68.4\pm 1.1        $ &  $0.313\pm 0.015$ & $0.833\pm 0.029$ \\
\sptg 18\,+\,\planck & $0.596 \pm 0.014$ & $0.810 \pm 0.014$ & $68.1\pm 1.0         $ & $0.309\pm 0.014$ & $0.822\pm 0.024$\\
\hline
\sptsz           & $0.597\pm 0.024$ & $0.789\pm 0.027$ & $71.0^{+1.7}_{-2.0}$ & $0.361^{+0.028}_{-0.033}$   & $0.865\pm 0.041$ \\
\sptpol           & $0.592\pm 0.024$ & $0.775\pm 0.023$ & $71.5^{+1.7}_{-2.1}$ & $0.369^{+0.029}_{-0.035}$  & $0.858\pm 0.037$   \\
\hline
\end{tabular}
\end{table*}

\subsection{Constraints from CMB Lensing and BAO}
\label{sec:parameter_lensing_bao}
Next, we turn our attention to the cosmological implications that arise from the inclusion of  BAO data with CMB lensing measurements.
In addition to providing constraints on $\sigma_8$ and $\Omega_{\rm m}$, the CMB lensing power spectrum is sensitive to the expansion rate $H_0$ due to its influence on the parameter combination $\sigma_8 \Omega_{\rm m}^{0.25} (\Omega_{\rm m} h^2 )^{-0.37}$~\citep{planck15-15,baxter21,madhavacheril23}.
Within \lcdm{}, the CMB lensing power spectrum can be written as an integral over the matter power spectrum $P_{\delta\delta}(k,z)$.
As a result, $C_L^{\phi\phi}$ is sensitive to the broadband shape of $P_{\delta\delta}(k,z)$, which is mostly controlled by the scale of matter-radiation equality $k_{\rm eq}\equiv a_{\rm eq}H_{\rm eq}\propto \Omega_{\rm m}h^2$ and the primordial amplitude $A_s$.
The shape and amplitude of the CMB lensing potential power spectrum are thus sensitive to a degenerate combination of the multipole corresponding to matter-radiation equality scale, $\ell_{\rm eq}\equiv k_{\rm eq}\chi_*\propto \Omega_{\rm m}^{0.6}h$, and $A_{\rm s}$  \citep[or $\sigma_8$, see, e.g.,][]{pan14,planck15-15,bianchini20a,madhavacheril23}.
The precise extent of the $A_{\rm s}$-$\ell_{\rm eq}$ parameter degeneracy is dictated by how accurately $C_L^{\phi\phi}$ is reconstructed and by the range of scales the lensing measurement probes.
Therefore, by effectively providing a handle on $\Omega_{\rm m}$, measurements of the projected BAO scale from low-$z$ galaxy surveys can break the $\sigma_8$-$\Omega_{\rm m}$-$H_0$ degeneracy and sharpen constraints on the individual parameters.

\begin{figure*}
	\centering
	\includegraphics[width=\textwidth]{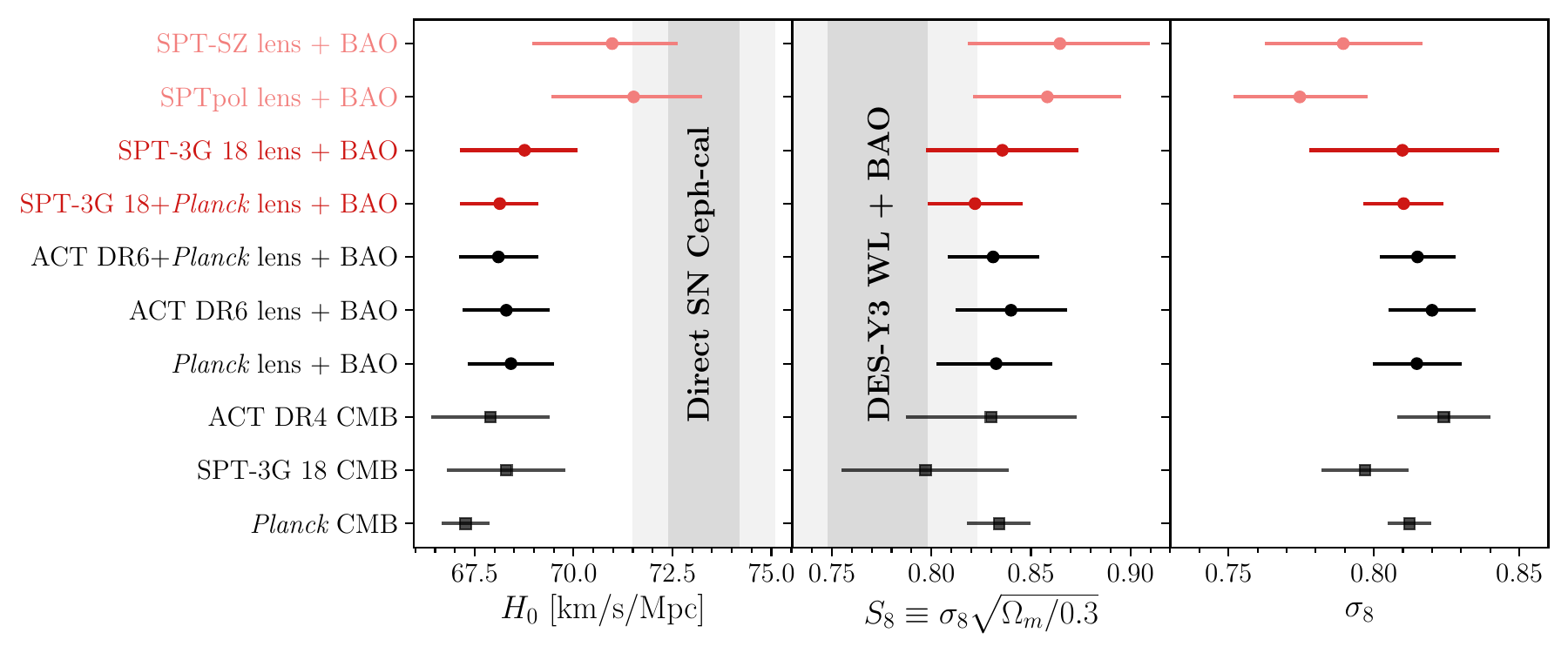}
	\caption{A comparison of the marginalized constraints on the Hubble constant $H_0$, $S_8\equiv \sigma_8\sqrt{\Omega_{\rm m}/0.3}$, and $\sigma_8$ (left to right) across different cosmological probes and surveys. The direct $H_0$ measurement is taken from~\citet{murakami23}, while the LSS-based constraint on $S_8$ is taken from the reanalysis of the DES-Y3 (+ BAO) data of~\citep{abbott22a} presented in~\citet{madhavacheril23}. 
	}
	\label{fig:H0_S8_sigma8_summary_plots}
\end{figure*}

In Tab.~\ref{tab:cmblens_bao}, we report constraints from CMB lensing and BAO on  $\sigma_8$, $\Omega_{\rm m}$, and $H_0$.
For our baseline \sptg measurement, we find the following constraints:
\beq
\label{eq:cmblens_bao}
\begin{rcases}
  H_0 &= 68.8^{+1.3}_{-1.6}\,\text{km}\,\text{s}^{-1}\,\text{Mpc}^{-1} \\
  \sigma_8 &= 0.810 \pm 0.033 \\
  \Omega_{\rm m} &= 0.320^{+0.021}_{-0.026}
\end{rcases}
\text{\sptg{} lensing + BAO.}
\eeq
The $\sigma_8$ and $\Omega_{\rm m}$ parameters are constrained at the $4\%$ and $7\%$ levels, respectively, and are both consistent (central values well within $1\sigma$) with the values inferred by the \planck{} primary CMB and lensing measurements.
The combination of \sptg CMB lensing, BAO, and BBN prior yields a $\approx 2\%$ constraint on the expansion rate $H_0$ with a central value of 68.8 km s$^{-1}$ Mpc$^{-1}$.
This value is consistent with other CMB lensing- and primary CMB-based constraints of the Hubble constant within the \lcdm{} model (see Fig.~\ref{fig:H0_S8_sigma8_summary_plots}), as well as the TRGB-calibrated local distance ladder measurement of~\citet{freedman19}.
When compared to recent direct $H_0$ constraints from Cepheid-calibrated \texttt{SH0ES} supernovae (SN) measurements \citep{murakami23}, we find the two estimates to be different at 2.6$\sigma$ significance.

We explore the sensitivity of our results to the low-$L$ bins and foreground marginalization.
We compare our baseline result (Eq.~\ref{eq:cmblens_bao}) to ones without the lowest $L$ bins because
the magnitude of the MF bias surpasses that of the signal for scales below $L \lesssim 100$ (see Fig.~\ref{fig:clkk_fg_bias}),
which might lead one to question the robustness of the baseline result given potential inaccuracies on the MF estimate.
When discarding the first two bins of the measured \sptg lensing power spectrum (i.e. throwing away information below $L < 92$), the parameter constraints become:
\begin{equation}
\left.
\begin{array}{l}{H_0 = 68.9 \pm 1.6}  \text{ km s$^{-1}$ Mpc$^{-1}$} \\
{\sigma_{8}\,\,= 0.809\pm 0.033} \\ {\Omega_{\rm m}=0.322^{+0.027}_{-0.030}}
\end{array}\right\}
\begin{array}{l}{\text {\sptg{} lensing}}
\\ {\text{($L_{\rm min}=92$) + BAO }}.
\end{array}
\end{equation}
While the uncertainties on some of the parameters slightly deteriorate, the inferred central values are largely unchanged.
The parameter that is affected most is $\Omega_{\rm m}$, whose 1$\sigma$ uncertainty is degraded from $0.024 \to 0.028$, followed by $\sigma(H_0) = 1.5 \to 1.6$.
We also find foreground marginalization does not dramatically affect the cosmological inference.
We test this by setting $A_{\rm fg} = 1$ and repeating the MCMC analysis.
The parameter that is affected most by the change is $\sigma_8$ for which we find an updated 1$\sigma$ constraint of $\sigma_8 = 0.805^{+0.029}_{-0.026}$.
As can be seen, the shift in the central value is 0.15$\sigma$ and the $A_{\rm fg}$ marginalization degrades the sensitivity by about 17\%.

The constraints in Eqs.~\ref{eq:cmblens_bao} are sharpened when lensing information from \planck{} PR4 lensing is included:
\begin{equation}
\left.
\begin{array}{l}{H_0 = 68.1 \pm 1.0}  \text{ km s$^{-1}$ Mpc$^{-1}$} \\
{\sigma_{8}\,\,= 0.810 \pm 0.014} \\ {\Omega_{\rm m}=0.309\pm 0.014}
\end{array}\right\}
\begin{array}{l}{\text {\sptg{} lensing  }}
\\ {\text{+\planck{} lensing  }}
\\ {\text{+ \rm BAO  }}.
\end{array}
\end{equation}

There are several intriguing differences between these results and those from the previous \sptpol{} and \sptsz{} lensing+BAO analyses.
Firstly, as shown in Tab.~\ref{tab:cmblens_bao}, the precision of constraints on $\Omega_{\rm m}$ and $H_0$ obtained from \sptg{} improves by approximately 30\%  when compared to the corresponding values from \sptpol{} and \sptsz{}.
The $\sigma_8$ uncertainties are degraded by $20-30\%$ compared to the other two SPT results, which is shown above to be largely due to foreground marginalization.
Secondly, the central values for the $H_0$ and  $\Omega_{\rm m}$ obtained through \sptg{} lensing shift towards lower values than those from \sptpol{} and \sptsz{}, while $\sigma_8$ increases slightly, mirroring the shifts of $\Omega_{\rm m}$ and $\sigma_8$ from lensing-alone constraints in Sec.~\ref{sec:parameter_lensing_alone}.
The improvement in the precision of $\Omega_{\rm m}$ and $H_0$ is in part explained by the decrease in the $\sigma_8$-$H_0$-$\Omega_{\rm m}$ posterior volume in \sptg compared to \sptsz/\sptpol (because of the lower noise in \sptg compared to \sptsz and the larger area/lower $L_{\rm min}$ compared to \sptpol).
The shift to a lower central $H_0$ value can be understood as follows.
The \sptg $\sigma_8$-$H_0$-$\Omega_{\rm m}$ posterior subspace is shifted to lower $\Omega_{\rm m}$-$H_0$ (and higher $A_{\rm s}$ and $\sigma_8$) relative to those from \sptsz and \sptpol.
Intersected by the BAO contours, which are positively correlated in the $\Omega_{\rm m}$-$H_0$ plane and favor a high value of $H_0 = 72.7^{+2.2}_{-2.9}$, gives the resulting $\Omega_{\rm m}$-$H_0$-$\sigma_8$ combinations, with lower $H_0$~\cite[e.g.][]{wu20}.
In other words, the high $H_0$ preference from BAO-alone constraints is more effectively pulled down by \sptg's combination of a smaller and shifted posterior subspace. 
While a thorough analysis of the consistency across the various SPT lensing measurements
requires common simulations to properly account for the correlations and is beyond the scope of this work, we note that reasonable shifts in the tilt and amplitude, as well as the magnitude of the uncertainties, within the SPT-3G measurement drive $\lesssim 1\sigma$ shifts in the parameter space.
As an example, fixing the amplitude of the foreground template to $A_{\rm fg}=0$ (i.e. neglecting residual foreground contamination in the lensing reconstruction) suppresses the lensing power by about 5\% and slightly perturbs the spectrum tilt across the whole $L\in[50,2000]$ range.
As a consequence, the corresponding central values of the relevant parameters from \sptg{} move towards the \sptsz{}/\sptpol{} constraints (but not completely) and become $\Omega_{\rm m} = 0.330^{+0.024}_{-0.028}$, $\sigma_8 = 0.783\pm 0.029$, and $ H_0 = 69.3^{+1.5}_{-1.6}$.

In order to compare our constraints with the findings from optical weak lensing surveys, we also provide the inference on $S_8$, which is defined as $\sigma_8\sqrt{\Omega_{\rm m}/0.3}$.
This parameter combination is known to be most accurately estimated from galaxy shear measurements and has recently been the subject of intense scrutiny due to $2-3 \sigma$ tensions between the primary CMB and galaxy lensing constraints \citep[e.g.,][]{abdalla22}.
From \sptg{} lensing data in combination with BAO scale information, we find a $4.7\%$ determination of the parameter at the level of
\beq
S_8 =  0.836 \pm 0.039 \quad \text{(\sptg{} lensing + BAO)},
\eeq
or $S_8 =  0.822 \pm 0.024$ when jointly analyzed with \planck{} lensing data.
Note that these statistical uncertainties are smaller than the typical errors from current galaxy lensing surveys such as the Dark Energy Survey\footnote{\url{https://www.darkenergysurvey.org}} \citep[DES-Y3,][]{abbott22a,amon22,secco22},  the Hyper Suprime-Cam Subaru Strategic Program\footnote{\url{https://hsc-release.mtk.nao.ac.jp/}}  \citep[HSC-Y3,][]{li23,dalal23}, and the Kilo Degree Survey\footnote{\url{https://kids.strw.leidenuniv.nl}} \citep[KiDS-1000,][]{asgari21}.
The $S_8$ value favored by \sptg lensing + BAO is approximately 1.7$\sigma$, 1.6$\sigma$, and 1.8$\sigma$ higher than that from DES, HSC-Y3, and KiDS, respectively.
We note that the precision of our $S_8$ constraint from CMB lensing and BAO is slightly better than that achieved using primary CMB information from \sptg, $S_8=0.797 \pm 0.042$ \citep{balkenhol23}, with the central value from CMB lensing + BAO being about 0.7$\sigma$ higher.

A visual comparison of the marginalized constraints on the Hubble constant $H_0$, $S_8$, and $\sigma_8$ across different probes and surveys is provided in Fig.~\ref{fig:H0_S8_sigma8_summary_plots}.
We choose two representative values in the literature to highlight direct $H_0$ and LSS $S_8$ measurements. Our results are in excellent agreement with the ACT DR6 and \planck{} lensing constraints.
For $H_0$, we show the result from \citet{murakami23}, $H_0=73.29\pm0.90$, which currently maximizes the tension with the indirect $H_0$ \planck{}-based estimates at the $ 5.7\sigma$ level, while for the low-$z$ amplitude of structure we pick the galaxy-galaxy lensing and galaxy clustering (the so-called $3 \times 2$ pt) measurements from DES-Y3 \citep{abbott22a} in combination with BAO data, as presented in \citet{madhavacheril23}.
For completeness, we also point the reader to the recent results from HSC \citep{li23,dalal23}, KiDS \citep{asgari21}, and the joint analysis of DES and KiDS data \citep{abbott23b}.

\subsection{Constraints on \lcdm{} extensions}

In the following sections, we study the consistency between the amount of lensing favored by direct and indirect lensing measurements, as well as the impact on the sum of the neutrino masses $\sum m_{\rm \nu}$ and curvature density $\Omega_{K}$ constraints when including \sptg $C_L^{\phi\phi}$ bandpowers in the cosmological inference.
We adopt the priors listed in the right column of Tab.~\ref{tab:priors} and include both the response function and $N^1_L$ linear corrections to the lensing likelihood, as discussed in Sec.~\ref{sec:lens_like}.
Note that when replacing \planck{} primary CMB data with \sptg{} $TT/TE/EE$ measurements from \citet{balkenhol23}, we impose a \planck{}-based Gaussian prior on the optical depth to reionization of  $\tau = 0.0540 \pm 0.0074$ (which is primarily constrained by a feature in the $TE$ and $EE$ spectra at $\ell < 10$).

\subsubsection{Lensing Amplitudes}
\label{sec:lensing_amp}
Lensing imprints in CMB maps can either be reconstructed directly, for example by exploiting the induced higher-order correlations between Fourier modes to measure $C_L^{\phi\phi}$ (the main focus of this paper), or indirectly through, e.g., the smearing of the acoustic peaks in the primary CMB spectra.
In this section, we compare direct and indirect CMB lensing measurements and ask ourselves 
two questions: 1) is the amplitude of the reconstructed lensing power spectrum consistent with the fiducial \planck{} \lcdm{} cosmology and 2) does the amount of lensing preferred by the primary CMB smoothing agree with the one suggested by direct lensing measurement?

We have already partially answered the former question in Sec.~\ref{sec:results_spectra}, where we have defined a lensing amplitude parameter $A_L^{\phi\phi}$ as a weighted rescaling of the measured $\hat{C}_L^{\phi\phi}$ with respect to a binned fiducial power spectrum.
The lensing amplitudes in Sec.~\ref{sec:results_spectra} are directly corrected for the estimated foreground bias template, assuming its amplitude is exact and fixed to $A_{\rm fg}=1$.
Here we revisit the lensing amplitude calculation using MCMC by marginalizing over uncertainties in the foreground cleaning.
We first fix the lensing power spectrum to the assumed fiducial cosmology, $C_L^{\phi\phi}(\boldsymbol{\Theta}_{\rm fid})$, and then rescale it by an overall amplitude parameter $C_L^{\phi\phi}(\boldsymbol{\Theta}_{\rm fid}) \to A_L^{\phi\phi}C_L^{\phi\phi}(\boldsymbol{\Theta}_{\rm fid})$.
In this run we have two free parameters, $A_L^{\phi\phi}$ and $A_{\rm fg}$ (defined in Eq.~\ref{eq:lincorr}), and we include $N^1_L$ bias correction in Eq.~\ref{eq:lincorr}.
When using the uniform prior on $A_{\rm fg} \sim \mathcal{U}\left(0,3\right)$, we obtain a lensing power spectrum amplitude normalized to the \planck{} 2018 \texttt{TTTEEE+lowE+lensing} fiducial cosmology of
\beq
A_L^{\phi\phi} = 1.063 \pm 0.090,
\eeq
which is consistent with unity within $\approx 0.7\sigma$.
This value is about 4\% higher than was found in Sec.~\ref{sec:results_spectra}  and its associated uncertainty is degraded by a factor of 1.5.\footnote{If we instead impose a less conservative Gaussian prior $1.0 \pm 0.3$ on the amplitude of the foreground bias template, where the dispersion is taken from the scatter in the template amplitudes of 16 cutouts of the \texttt{AGORA} simulations as discussed in Sec.~\ref{sect:astro_fgnds}, we find $A_L^{\phi\phi} = 1.014 \pm 0.076$.}

The second question is perhaps more interesting since the amount of lensing determined from the smoothing of the acoustic peaks in the \planck{} temperature and polarization power spectra is larger, at the $2$ to 3$\sigma$ level, than what is predicted by \lcdm{} and when compared to direct CMB lensing measurements \citep[e.g.,][]{planck18-6,motloch20,lemos23}.
To complicate things further, the magnitude and significance of this anomaly vary between different \planck{} data releases and likelihood versions~\citep[e.g.,][]{rosenberg22}, and ground-based CMB surveys such as SPT and ACT do not observe a similar amount of 2-pt function smearing~\citep{aiola20,balkenhol23}.
To check whether the \sptg{} lensing power spectrum is consistent with the lensing information in the primary CMB 2-point function and with \lcdm{} predictions, we follow \citet{calabrese08} and~\citet{planck18-6}.
In addition to $A_L^{\phi\phi}$, we introduce a non-physical parameter $A_L$ which scales the lensing power spectrum that affects both the lens reconstruction power and the smoothing of the acoustic peaks.

\begin{figure}
	\centering
	\includegraphics[width=\columnwidth]{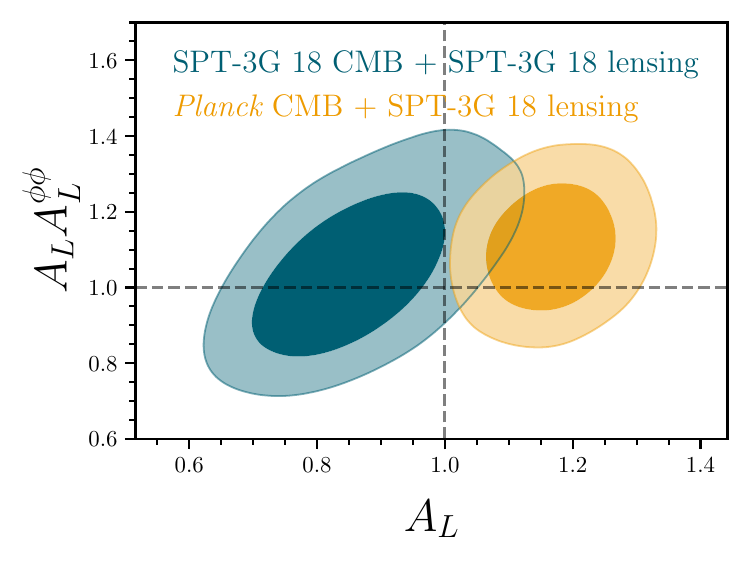}
	\caption{Consistency check between direct and indirect CMB lensing measurements. The amount of lensing directly inferred from \sptg{} data ($A_L \times A_L^{\phi\phi}$) is consistent with what is predicted by the best-fit \lcdm{} cosmology as determined from primary CMB when the smearing effects ($A_L$) are marginalized over. This holds for both \planck{} PR3 primary CMB (orange contours) as well as 2018 \sptg{} data (blue contours).}
	\label{fig:lens_amp}
\end{figure}

In Fig.~\ref{fig:lens_amp}, we show the constraints on the lensing amplitude parameters from the joint analysis of \sptg{} lensing with either \planck{} or \sptg{} primary CMB anisotropies.
In the figure, the combination $A_L \times A_L^{\phi\phi}$ quantifies the overall amplitude of the measured lensing power with respect to \lcdm{} expectations, when the inferred \lcdm{} parameters have been marginalized over the observed level of peak smearing.
As can be seen, while features in the \planck{} and \sptg{} 2-point CMB spectra drive the preference for either $A_L > 1$ or $< 1$, the amplitude of the lensing power spectrum is consistent with both \planck{} and \sptg{} primary CMB \lcdm{} predictions when the sensitivity to the peak smoothing effect is artificially removed:
\begin{equation}
	\begin{aligned}
		A_L \times A_L^{\phi\phi} &= 1.11\pm 0.11 \,(\text{\planck{} CMB + \sptg{} lens}), \\
		A_L \times A_L^{\phi\phi} &= 1.04^{+0.13}_{-0.16}  \, (\text{\sptg{} CMB + \sptg{} lens}).
	\end{aligned}
\end{equation}
The \sptg{} lensing reconstruction shows no evidence for an unusually high or low amount of lensing relative to that predicted by the best-fit \lcdm{} parameters as determined from primary CMB data when the peak smoothing effect has been marginalized over.

\subsubsection{Neutrino Mass}
\label{sec:neutrinos}
By being sensitive to matter clustering at intermediate redshifts and mostly linear scales, CMB lensing measurements can also provide insights on the neutrino sector~\citep[e.g.,][]{smith09,abazajian15b,pan15}, one of the most elusive constituents of the standard model of particles.
Observations of neutrino flavor oscillations have established that neutrinos are massive particles and that the three known mass-eigenstates are not completely degenerate~\citep{fukuda98, ahmad02}.
However, their absolute mass scale is still unknown.
The mass-squared differences measured in oscillation experiments put a lower bound on the sum of the neutrino masses $\sum m_{\rm \nu} > 60$ meV, for a normal hierarchy, or greater than 100 meV, for the inverted hierarchy.
While tritium beta decay end-point experiments like KATRIN have put constraints on the effective electron anti-neutrino mass at the level of $\sum m_{\rm \nu} < 800$ meV~\citep[90\% C.L.,][]{aker22}, and are expected to improve by a factor of four,  cosmological measurements already have a stronger (albeit model dependent) sensitivity  and have the potential to make a mass measurement in the next decade or so~\citep{planck18-6,abazajian19,gerbino22}.
In particular, neutrinos become non-relativistic at low redshifts and contribute to both the matter density parameter and the expansion rate but not to the matter clustering on scales below their free-streaming length.
Consequently, in a universe where neutrinos possess mass, the growth of structures below the free-streaming length will be suppressed compared to a universe where neutrinos are massless.

\begin{figure}
	\centering
	\includegraphics[width=\columnwidth]{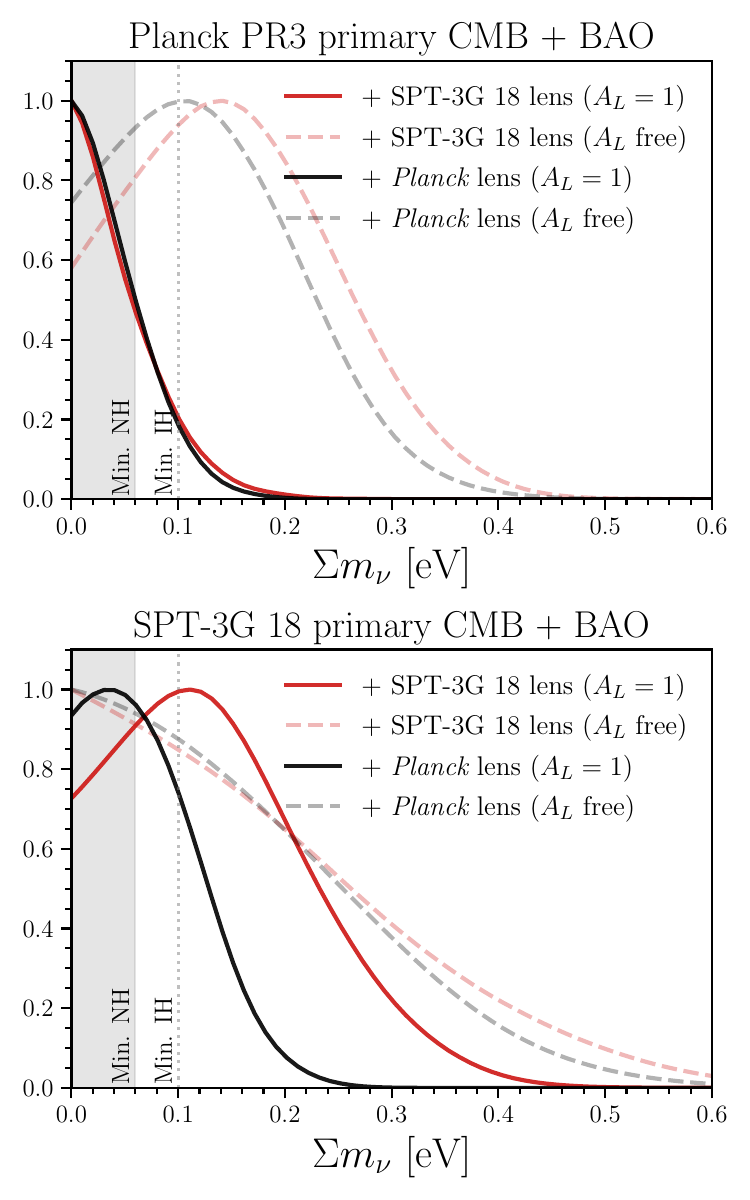}
	\caption{Marginalized constraints on the sum of the neutrino masses $\sum m_{\rm \nu}$ when BAO and either \planck (upper panel) or \sptg (lower panel) CMB temperature and polarization power spectra are folded into the cosmological inference. In each panel, the red and black lines show the effect of including direct lensing measurements from \sptg and \planck, respectively. The corresponding light dashed lines show instead the constraint when we remove the lensing information in the primary CMB, i.e. we allow $A_L$ to vary. The expectations for minimal masses based on oscillation measurements in the normal (NH) and inverted (IH) hierarchies are denoted by the grey shaded band and the dotted vertical line, respectively.}
	\label{fig:neutrinos}
\end{figure}

In Fig.~\ref{fig:neutrinos}, we show the marginalized posterior on the sum of the neutrino masses in a \lcdm{} + $\sum m_{\rm \nu}$ cosmology when combining primary CMB, BAO, and CMB lensing datasets (solid lines).
When freeing $\sum m_{\rm \nu}$, we follow the argument of \citet{lesgourgues06b} and consider a degenerate combination of three equally massive neutrinos, as was done in the \planck{} papers \citep{planck18-6}.
The top panel shows the inclusion of \planck{} primary CMB data, while the bottom one shows the constraints obtained using \sptg{} CMB temperature and polarization spectra as our early universe probe.
In each panel, the red and black lines denote respectively the addition of either \sptg{} or \planck{} lensing datasets.
In addition, we also show the corresponding constraints in a \lcdm{} + $\sum m_{\rm \nu} + A_L$ cosmology, i.e. when we marginalize over the lensing information in the primary CMB (semi-transparent dashed lines, same color coding as above).

The 95\% C.L. constraint on the sum of the neutrino masses from \planck{} CMB + BAO is $\sum m_{\rm \nu} < 0.13$ eV, which remains
\begin{equation}
	\begin{split}
&\sum m_{\rm \nu} < 0.13\,  \text{eV}\\
&\quad (\text{\planck{} CMB + BAO + \sptg{} lens}, \, 2\sigma),
\end{split}
\end{equation}
when \sptg{} lensing information is added. This constraint with  \sptg{} lensing is similar to that with \planck{} lensing at $\sum m_{\rm \nu}< 0.11$ eV.
While the inclusion of CMB lensing does not seem to improve or change the constraint significantly, we recall that the neutrino constraints based on  \planck{} PR3 primary CMB are known to be artificially tight because of the anomalous lensing smoothing amplitude (see Sec.~\ref{sec:lensing_amp}).
We therefore repeat the exercise when allowing $A_L$ to vary.
As expected, the $2\sigma$ upper limits combined with \sptg{} and \planck{} lensing are relaxed to $\sum m_{\rm \nu}< 0.32$~eV and $\sum m_{\rm \nu}< 0.28$~eV, respectively.
In both cases, the posteriors peak at neutrino masses around $\sum m_{\rm \nu} $ around $ 0.1-0.2$~eV.

To investigate the neutrino constraint sensitivity to primary CMB and potential lensing anomalies, we replace \planck{} PR3 primary CMB with the latest \sptg{} temperature and polarization power spectra.
Although not as constraining as the \planck{} dataset, \sptg{} can provide a useful consistency check.
A joint analysis of \sptg{} CMB + BAO reveals a 2$\sigma$ constraint on the neutrino mass of $\sum m_{\rm \nu} < 0.43$ eV, which becomes
\begin{equation}
	\begin{split}
		&\sum m_{\rm \nu} < 0.30 \, \text{eV} \quad \\
		&(\text{\sptg{} CMB + BAO + \sptg{} lens}, \, 2\sigma),
	\end{split}
\end{equation}
with the inclusion of \sptg{} lensing.
The corresponding constraint using \planck{} lensing is comparatively tighter, $\sum m_{\rm \nu} < 0.17$ eV.
Both posteriors peak at values $\sum m_{\rm \nu}>0$.
The lensing parameter $A_L$ from \sptg{} primary CMB is about 1$\sigma$ low with respect to \lcdm{} expectations (see Fig.~\ref{fig:lens_amp}).
When $A_L$ is allowed to vary, the posteriors peak at zero mass and broaden so that the constraints become  $\sum m_{\rm \nu} < 0.45$ eV (for \sptg{} lensing) and  $\sum m_{\rm \nu} < 0.32$ eV (for \planck{} lensing).

\subsubsection{Curvature}
CMB lensing measurements also allow us to test the mean spatial curvature of the universe, which is predicted to be close to flat in the majority of inflationary models.
Curvature constraints from primary CMB spectra are largely driven by the lensing smoothing on the acoustic peaks, which partially breaks the geometrical degeneracy \citep{stompor99}.
Therefore, a direct measurement of the lensing amplitude can further resolve the degeneracy and sharpen constraints on $\Omega_{K}$.
Once again, \planck{} temperature and polarization anisotropies alone present us with a curious feature, showing a $2-3\sigma$ preference for non-flat models~\citep[e.g.,][]{planck18-6,handley21,divalentino19}.
This is driven by the $A_L$-$\Omega_{K}$ degeneracy and the presence of an enhanced lensing amplitude in the \planck{} 2018 data release.
The preference for a negative curvature weakens in the analysis of \planck{} \texttt{NPIPE} maps \citep{rosenberg22} and is not present in either SPT~\citep{balkenhol21} or ACT data~\citep{aiola20}.
Moreover, the addition of BAO data strongly breaks the geometric degeneracy and eliminates the preference for non-zero $\Omega_{K}$.

\begin{figure}
	\centering
	\includegraphics[width=\columnwidth]{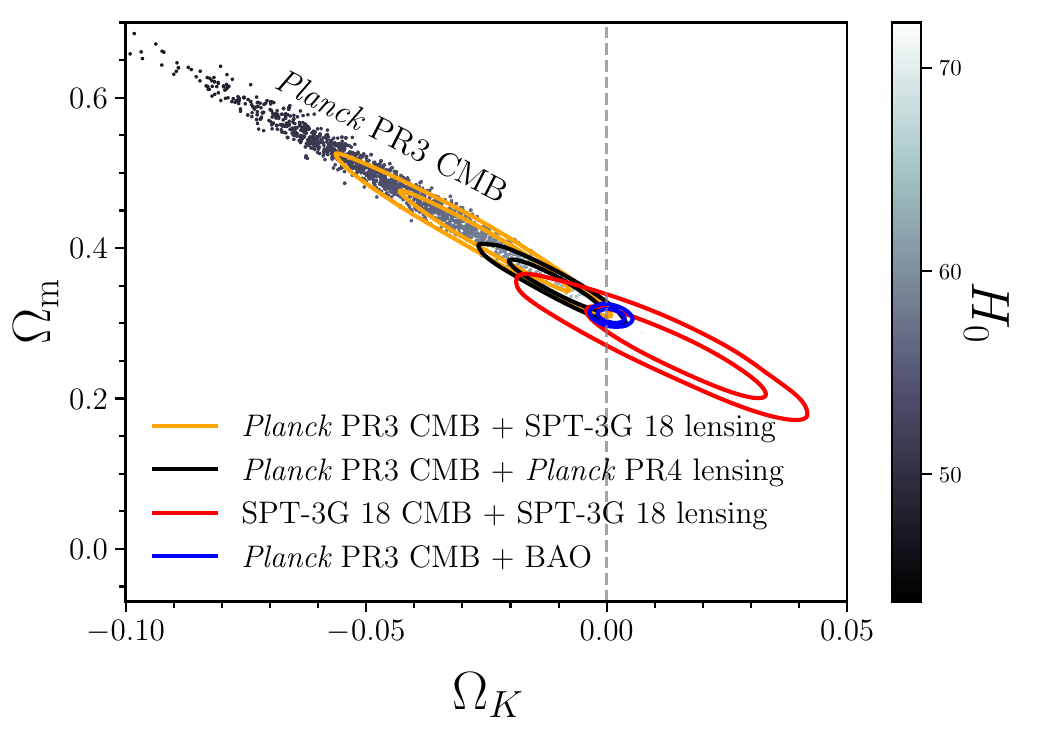}
	\caption{Constraints on curvature and  the matter density parameter obtained from the analysis of \planck primary CMB, with scattered points color-coded according to their corresponding Hubble constant values. The orange and black solid lines show the constraints inferred by adding either the \sptg{} or \planck{} lensing datasets, respectively.  The blue solid lines denote instead the constraints from \planck{} primary CMB combined with BAO data, and indicate consistency with a flat geometry. The constraints obtained from both 2018 SPT-3G primary CMB and lensing are highlighted in red.}
	\label{fig:curvature}
\end{figure}

The geometrical degeneracy is clearly visible in the colored scatter points in Fig.~\ref{fig:curvature}, which show the allowed region in the $\Omega_{K}$-$\Omega_{\rm m}$ parameter space using only primary CMB information from \planck{}, for which we find $-0.078 < \Omega_{K} <  -0.010$ (95\% C.L.).
The inclusion of lensing information from \sptg{} shrinks the allowed 2$\sigma$ region to
\beq
\Omega_{K} = -0.026^{+0.022}_{-0.024}\,(\text{\planck{} CMB + \sptg{} lens},\, 2\sigma),
\eeq
which is still $\approx 2\sigma$ away from $\Omega_{K} = 0$.\footnote{We note that in this run, the posterior of the inferred foreground contamination parameter $A_{\rm fg}$ tends to shift to the upper limit of the prior boundary.}
For comparison, when we instead include \planck{} lensing, the 2$\sigma$ constraint becomes $\Omega_{K} = -0.011^{+0.012}_{-0.012}$.
Using only SPT data for both primary CMB and lensing, the curvature is constrained to
\beq
\Omega_{K} = 0.014^{+0.023}_{-0.026}\,(\text{\sptg{} CMB + \sptg{} lens},\, 2\sigma),
\eeq
consistent with a flat universe.
Finally, we can turn the limit on the spatial curvature into a constraint on the cosmological constant density parameter $\Omega_\Lambda = 1 - \Omega_{\rm m}-\Omega_{K}$ as done in the first indirect evidence for dark energy using only CMB data \citep{sherwin11}.
Combining \sptg{} CMB lensing with either \planck{} or \sptg{} primary CMB, we obtain the following $1\sigma$ bounds:
\begin{equation}
	\begin{aligned}
		\Omega_\Lambda &= 0.613^{+0.035}_{-0.029}\, (\text{\planck{} CMB + \sptg{} lens}), \\
                 \Omega_\Lambda &= 0.722^{+0.031}_{-0.026}\, (\text{\sptg{}\,CMB + \sptg{}\,lens}).
	\end{aligned}
\end{equation}
Again, the constraint from \planck{} CMB and SPT-3G CMB lensing is shifted due to the anomalous peak smoothing in \planck{}.

\section{Conclusions}
\label{sect:conclusions}

We have presented the first analysis of CMB lensing using data from \sptg{}, the current camera on the South Pole Telescope.
With the 95 and 150~GHz temperature data from the \sptg{} 2018 data set, we produced high-fidelity convergence mass maps over 1500~deg$^2$ of the Southern sky, with $S/N$ per mode
greater than unity for modes at $L\lesssim 70$ to 100 depending on the frequency combination.
We combined the cross-frequency convergence maps and reported a minimum-variance CMB lensing
power spectrum over the multipole range $50 \le L \le 2000$.
We constrained the lensing amplitude to be $A^{\rm MV}=1.020\pm0.060$, a $5.9\%$ precision measurement.
The systematic uncertainty, stemming from beam and map calibration uncertainties, is $\pm0.016$, about 27\% of the statistical uncertainty.
When marginalizing over a foreground bias template, we found  $A^{\phi\phi}_L=1.063\pm0.090$.
We conducted a set of consistency checks and null tests on our results and found no
evidence of significant bias in our measurement.

We have discussed the cosmological implications of the \sptg{} 2018 data lensing measurements within
 the \lcdm{} model and in a number of 1- and 2-parameter extensions.
Our lensing amplitude agrees with those from \actpol and \planck with or without foreground marginalization.
While foreground marginalization has minimal impact on cosmological constraints---foreground bias is small compared to the uncertainty---it is included for all cosmological parameter constraints.
With \sptg{} 2018 lensing alone, we constrained $\sigma_8 \Omega_{\rm m}^{0.25}$ to be $0.595 \pm 0.026$---a precision of 4.4\%.
This result is broadly consistent with other CMB lensing measurements, including previous  \sptsz{}, \sptpol{}, \actpol{}, and \planck{} results, as well as with \planck{}'s primary CMB anisotropy measurements. 
When complementing this with BAO data from galaxy surveys, we were able to constrain the matter
density and structure amplitude parameters to $\Omega_{\rm m}=0.320^{+0.021}_{-0.026}$ and
$\sigma_8 = 0.810 \pm 0.033$, consistent with \lcdm{} expectations based on \planck{} primary CMB.
There is some tension between CMB and low redshift measurements of $H_0$ and $S_8$. When combined with BAO, our measurement is consistent with the cosmology
inferred from \planck primary CMB measurements.
Specifically, the parameter combination $S_8=\sigma_8\sqrt{\Omega_{\rm m}/0.3}$ was
determined to be $0.836 \pm 0.039$, about $1.6-1.8\sigma$ higher than the values inferred from optical surveys such as DES, HSC, and KiDS.
The Hubble constant result of $H_0 =68.8^{+1.3}_{-1.6}$\text{ km s$^{-1}$ Mpc$^{-1}$} agrees with other
early-universe estimates and is approximately $2.6\sigma$ lower
than estimates based on Cepheid-calibrated local distance ladder measurements.

The degree of peak smoothing in the temperature power spectrum of \planck{}, which is 2-3$\sigma$ higher than what is expected from the \lcdm{} model, has notable effects on estimates for cosmological parameters like $\Sigma m_{\rm \nu}$ and $\Omega_{K}$.
To isolate the impact from the excess smoothing, we either marginalized over $A_L$ when combining with \planck's primary spectrum
or combined with primary spectrum measurements from the \sptg{} 2018 data.
First, we demonstrated that \sptg 2018 lensing amplitude is consistent with the \lcdm expectation when effects of
peak smoothing of either \planck or \sptg 2018 primary CMB were marginalized over.
In combination with the \planck and BAO datasets, the sum of neutrino masses is
$\Sigma m_{\rm \nu}< 0.13$~eV and $\Sigma m_{\rm \nu}< 0.32$~eV (95\% C.L) without
and with $A_L$ marginalized over.
Alternatively, when using \sptg 2018 CMB data in place of \planck, we obtained $\Sigma m_{\rm \nu}< 0.30$~eV (95\% C. L.; $A_L$=1).
Using primary CMB from \planck in combination
with SPT-3G lensing drives a $2\sigma$ preference for a non-flat universe and favors a lower
$\Omega_{\rm \Lambda}$ of $0.613^{+0.059}_{-0.067}$.
The $2\sigma$ preference disappears when replacing the primary CMB from \planck with
that from \sptg{} 2018 data, yielding $\Omega_{K}=0.014^{+0.023}_{-0.026}$ (95\% C. L.)
and $\Omega_{\rm \Lambda}$ of $0.722^{+0.054}_{-0.059}$.

While data for this analysis comes from only half of the 2018 observing season with
a half-functioning focal plane resulting in higher per-year map noise, map depths
using data from subsequent observing seasons are matching expectations~\citep{sobrin22}.
Our lensing measurement is signal dominated up to
$L\simeq 100$; upcoming analyses using two years of data on the 1500 deg$^2$ field will expand the signal-dominated region out to multipoles of a few hundred,
resulting in a $\approx$2\% measurement of the lensing amplitude.
At these low map noise levels, polarization measurements will contribute the
majority of $S/N$ in future \sptg lensing analyses, and employing optimal methods~\citep[e.g.,][]{millea22,legrand22,bianchini23} will further this dataset's science reach.
High-fidelity lensing maps generated by \sptg
will be critical in removing the lensing contamination in the search of
 primordial gravitational waves, parametrized
by the tensor-to-scalar ratio $r$, when jointly analyzed with data from BICEP Array~\citep{hui18}.
The most stringent constraint on $r$ to date from~\citep{bicep2keck21}
is already limited by lensing.
With \sptg's full-survey lensing map, we expect to improve BICEP Array's
$r$ uncertainty by a factor of about 2.5, reaching  $\sigma(r)$ of about 0.003.

\begin{acknowledgements}
We would like to thank Mathew Madhavacheril and Frank Qu for clarifications regarding the ACT DR6 lensing analysis.
The South Pole Telescope program is supported by the National Science Foundation (NSF) through the Grant No. OPP-1852617.
Partial support is also provided by the Kavli Institute of Cosmological Physics at the University of Chicago.
Argonne National Laboratory's work was supported by the U.S. Department of Energy, Office of High Energy Physics, under Contract No. DE-AC02-06CH11357.
Work at Fermi National Accelerator Laboratory, a DOE-OS, HEP User Facility managed by the Fermi Research Alliance, LLC, was supported under Contract No. DE-AC02-07CH11359.
The Cardiff authors acknowledge support from the UK Science and Technologies Facilities Council (STFC).
The IAP authors acknowledge support from the Centre National d'\'{E}tudes Spatiales (CNES). 
This project has received funding from the European Research Council (ERC) under the European Union’s
 Horizon 2020 research and innovation programme (grant agreement No 101001897).
The Melbourne authors acknowledge support from the Australian Research Council's Discovery Project scheme (No. DP210102386).
The McGill authors acknowledge funding from the Natural Sciences and Engineering Research Council of Canada, Canadian Institute for Advanced Research, and the Fonds de recherche du Qu\'ebec Nature et technologies.
The SLAC authors acknowledge support by the Department of Energy, Contract DE-AC02-76SF00515.
The UCLA and MSU authors acknowledge support from NSF AST-1716965 and CSSI-1835865.
M.A. and J.V. acknowledge support from the Center for AstroPhysical Surveys at the
National Center for Supercomputing Applications in Urbana, IL.
K.F. acknowledges support from the Department of Energy Office of Science Graduate Student Research (SCGSR) Program.
Z.P. was supported by Laboratory Directed Research and Development (LDRD) funding
from Argonne National Laboratory, provided by the Director, Office of Science, of the U.S. Department of Energy
under Contract No. LDRD-2021-0186.
J.V. acknowledges support from the Sloan Foundation.
W.L.K.W is supported in part by the Department of Energy, Laboratory Directed Research and Development program and as part of the Panofsky Fellowship program at SLAC National Accelerator Laboratory, under contract DE-AC02-76SF00515.
This research was done using resources provided by the Open Science Grid \citep{pordes07, sfiligoi09}, which is supported by the NSF Award No. 1148698, and the U.S. Department of Energy's Office of Science.
Some of the computing for this project was performed on the Sherlock cluster. We would like to thank Stanford University and the Stanford Research Computing Center for providing computational resources and support that contributed to these research results.

\end{acknowledgements}

\bibliography{spt.bib}

\begin{thebibliography}{}
\expandafter\ifx\csname natexlab\endcsname\relax\def\natexlab#1{#1}\fi
\providecommand{\url}[1]{\href{#1}{#1}}
\providecommand{\dodoi}[1]{doi:~\href{http://doi.org/#1}{\nolinkurl{#1}}}
\providecommand{\doeprint}[1]{\href{http://ascl.net/#1}{\nolinkurl{http://ascl.net/#1}}}
\providecommand{\doarXiv}[1]{\href{https://arxiv.org/abs/#1}{\nolinkurl{https://arxiv.org/abs/#1}}}

\bibitem[{{Lewis} \& {Challinor}(2006)}]{lewis06}
{Lewis}, A., \& {Challinor}, A.
\newblock {Weak gravitational lensing of the CMB}. 2006, \physrep, 429, 1,
  \dodoi{10.1016/j.physrep.2006.03.002}

\bibitem[{{Bianchini} {et~al.}(2020){Bianchini}, {Wu}, {Ade}, {Anderson},
  {Austermann}, {Avva}, {Beall}, {Bender}, {Benson}, {Bleem}, {Carlstrom},
  {Chang}, {Chaubal}, {Chiang}, {Citron}, {Moran}, {Crawford}, {Crites}, {de
  Haan}, {Dobbs}, {Everett}, {Gallicchio}, {George}, {Gilbert}, {Gupta},
  {Halverson}, {Harrington}, {Henning}, {Hilton}, {Holder}, {Holzapfel},
  {Hrubes}, {Huang}, {Hubmayr}, {Irwin}, {Knox}, {Lee}, {Li}, {Lowitz},
  {Manzotti}, {McMahon}, {Meyer}, {Millea}, {Mocanu}, {Montgomery}, {Nadolski},
  {Natoli}, {Nibarger}, {Noble}, {Novosad}, {Omori}, {Padin}, {Patil}, {Pryke},
  {Reichardt}, {Ruhl}, {Saliwanchik}, {Sayre}, {Schaffer}, {Sievers}, {Simard},
  {Smecher}, {Stark}, {Story}, {Tucker}, {Vanderlinde}, {Veach}, {Vieira},
  {Wang}, {Whitehorn}, \& {Yefremenko}}]{bianchini20a}
{Bianchini}, F., {Wu}, W.~L.~K., {Ade}, P.~A.~R., {et~al.}
\newblock {Constraints on Cosmological Parameters from the 500 deg$^{2}$ SPTPOL
  Lensing Power Spectrum}. 2020, \apj, 888, 119,
  \dodoi{10.3847/1538-4357/ab6082}

\bibitem[{{Qu} {et~al.}(2023){Qu}, {Sherwin}, {Madhavacheril}, {Han},
  {Crowley}, {Abril-Cabezas}, {Ade}, {Aiola}, {Alford}, {Amiri}, {Amodeo},
  {An}, {Atkins}, {Austermann}, {Battaglia}, {Battistelli}, {Beall}, {Bean},
  {Beringue}, {Bhandarkar}, {Biermann}, {Bolliet}, {Bond}, {Cai}, {Calabrese},
  {Calafut}, {Capalbo}, {Carrero}, {Carron}, {Challinor}, {Chesmore}, {Cho},
  {Choi}, {Clark}, {C{\'o}rdova Rosado}, {Cothard}, {Coughlin}, {Coulton},
  {Dalal}, {Darwish}, {Devlin}, {Dicker}, {Doze}, {Duell}, {Duff},
  {Duivenvoorden}, {Dunkley}, {D{\"u}nner}, {Fanfani}, {Fankhanel}, {Farren},
  {Ferraro}, {Freundt}, {Fuzia}, {Gallardo}, {Garrido}, {Gluscevic}, {Golec},
  {Guan}, {Halpern}, {Harrison}, {Hasselfield}, {Healy}, {Henderson},
  {Hensley}, {Herv{\'\i}as-Caimapo}, {Hill}, {Hilton}, {Hilton}, {Hincks},
  {Hlo{\v{z}}ek}, {Ho}, {Huber}, {Hubmayr}, {Huffenberger}, {Hughes}, {Irwin},
  {Isopi}, {Jense}, {Keller}, {Kim}, {Knowles}, {Koopman}, {Kosowsky},
  {Kramer}, {Kusiak}, {La Posta}, {Lague}, {Lakey}, {Lee}, {Li}, {Li}, {Limon},
  {Lokken}, {Louis}, {Lungu}, {MacCrann}, {MacInnis}, {Maldonado}, {Maldonado},
  {Mallaby-Kay}, {Marques}, {McMahon}, {Mehta}, {Menanteau}, {Moodley},
  {Morris}, {Mroczkowski}, {Naess}, {Namikawa}, {Nati}, {Newburgh}, {Nicola},
  {Niemack}, {Nolta}, {Orlowski-Scherer}, {Page}, {Pandey}, {Partridge},
  {Prince}, {Puddu}, {Radiconi}, {Robertson}, {Rojas}, {Sakuma}, {Salatino},
  {Schaan}, {Schmitt}, {Sehgal}, {Shaikh}, {Sierra}, {Sievers}, {Sif{\'o}n},
  {Simon}, {Sonka}, {Spergel}, {Staggs}, {Storer}, {Switzer}, {Tampier},
  {Thornton}, {Trac}, {Treu}, {Tucker}, {Ulluom}, {Vale}, {Van Engelen}, {Van
  Lanen}, {van Marrewijk}, {Vargas}, {Vavagiakis}, {Wagoner}, {Wang}, {Wenzl},
  {Wollack}, {Xu}, {Zago}, \& {Zhang}}]{qu23}
{Qu}, F.~J., {Sherwin}, B.~D., {Madhavacheril}, M.~S., {et~al.}
\newblock {The Atacama Cosmology Telescope: A Measurement of the DR6 CMB
  Lensing Power Spectrum and its Implications for Structure Growth}. 2023,
  arXiv e-prints, arXiv:2304.05202, \dodoi{10.48550/arXiv.2304.05202}

\bibitem[{Kaplinghat {et~al.}(2003)Kaplinghat, Knox, \& Song}]{kaplinghat03}
Kaplinghat, M., Knox, L., \& Song, Y.-S.
\newblock Determining Neutrino Mass from the Cosmic Microwave Background Alone.
  2003, Phys. Rev. Lett., 91, 241301, \dodoi{10.1103/PhysRevLett.91.241301}

\bibitem[{{Lesgourgues} {et~al.}(2006){Lesgourgues}, {Perotto}, {Pastor}, \&
  {Piat}}]{lesgourgues06}
{Lesgourgues}, J., {Perotto}, L., {Pastor}, S., \& {Piat}, M.
\newblock {Probing neutrino masses with CMB lensing extraction}. 2006, \prd,
  73, 045021, \dodoi{10.1103/PhysRevD.73.045021}

\bibitem[{{Lesgourgues} \& {Pastor}(2006)}]{lesgourgues06b}
{Lesgourgues}, J., \& {Pastor}, S.
\newblock {Massive neutrinos and cosmology}. 2006, Physics Reports, 429, 307,
  \dodoi{10.1016/j.physrep.2006.04.001}

\bibitem[{{de Putter} {et~al.}(2009){de Putter}, {Zahn}, \&
  {Linder}}]{deputter09}
{de Putter}, R., {Zahn}, O., \& {Linder}, E.~V.
\newblock {CMB lensing constraints on neutrinos and dark energy}. 2009, \prd,
  79, 065033, \dodoi{10.1103/PhysRevD.79.065033}

\bibitem[{{Pan} \& {Knox}(2015)}]{pan15}
{Pan}, Z., \& {Knox}, L.
\newblock {Constraints on neutrino mass from cosmic microwave background and
  large-scale structure}. 2015, \mnras, 454, 3200,
  \dodoi{10.1093/mnras/stv2164}

\bibitem[{Sherwin {et~al.}(2011)Sherwin, Dunkley, Das, Appel, Bond, Carvalho,
  Devlin, D\"unner, Essinger-Hileman, Fowler, Hajian, Halpern, Hasselfield,
  Hincks, Hlozek, Hughes, Irwin, Klein, Kosowsky, Marriage, Marsden, Moodley,
  Menanteau, Niemack, Nolta, Page, Parker, Reese, Schmitt, Sehgal, Sievers,
  Spergel, Staggs, Swetz, Switzer, Thornton, Visnjic, \& Wollack}]{sherwin11}
Sherwin, B.~D., Dunkley, J., Das, S., {et~al.}
\newblock Evidence for Dark Energy from the Cosmic Microwave Background Alone
  Using the Atacama Cosmology Telescope Lensing Measurements. 2011, Phys. Rev.
  Lett., 107, 021302, \dodoi{10.1103/PhysRevLett.107.021302}

\bibitem[{{Calabrese} {et~al.}(2009){Calabrese}, {Cooray}, {Martinelli},
  {Melchiorri}, {Pagano}, {Slosar}, \& {Smoot}}]{calabrese09}
{Calabrese}, E., {Cooray}, A., {Martinelli}, M., {Melchiorri}, A., {Pagano},
  L., {Slosar}, A., \& {Smoot}, G.~F.
\newblock {CMB lensing constraints on dark energy and modified gravity
  scenarios}. 2009, \prd, 80, 103516, \dodoi{10.1103/PhysRevD.80.103516}

\bibitem[{{Namikawa} {et~al.}(2018){Namikawa}, {Bouchet}, \&
  {Taruya}}]{namikawa18}
{Namikawa}, T., {Bouchet}, F.~R., \& {Taruya}, A.
\newblock {CMB lensing bispectrum as a probe of modified gravity theories}.
  2018, \prd, 98, 043530, \dodoi{10.1103/PhysRevD.98.043530}

\bibitem[{Singh {et~al.}(2018)Singh, Alam, Mandelbaum, Seljak,
  Rodriguez-Torres, \& Ho}]{singh18}
Singh, S., Alam, S., Mandelbaum, R., Seljak, U., Rodriguez-Torres, S., \& Ho,
  S.
\newblock {Probing gravity with a joint analysis of galaxy and CMB lensing and
  SDSS spectroscopy}. 2018, Monthly Notices of the Royal Astronomical Society,
  482, 785, \dodoi{10.1093/mnras/sty2681}

\bibitem[{Zhang {et~al.}(2020)Zhang, Pullen, Alam, Singh, Burtin, Chuang, Hou,
  Lyke, Myers, Neveux, Ross, Rossi, \& Zhao}]{zhang20}
Zhang, Y., Pullen, A.~R., Alam, S., Singh, S., Burtin, E., Chuang, C.-H., Hou,
  J., Lyke, B.~W., Myers, A.~D., Neveux, R., Ross, A.~J., Rossi, G., \& Zhao,
  C.
\newblock {Testing general relativity on cosmological scales at redshift
  z$\sim$1.5 with quasar and CMB lensing}. 2020, Monthly Notices of the Royal
  Astronomical Society, 501, 1013, \dodoi{10.1093/mnras/staa3672}

\bibitem[{{Das} {et~al.}(2011){Das}, {Sherwin}, {Aguirre}, {Appel}, {Bond},
  {Carvalho}, {Devlin}, {Dunkley}, {D{\"u}nner}, {Essinger-Hileman}, {Fowler},
  {Hajian}, {Halpern}, {Hasselfield}, {Hincks}, {Hlozek}, {Huffenberger},
  {Hughes}, {Irwin}, {Klein}, {Kosowsky}, {Lupton}, {Marriage}, {Marsden},
  {Menanteau}, {Moodley}, {Niemack}, {Nolta}, {Page}, {Parker}, {Reese},
  {Schmitt}, {Sehgal}, {Sievers}, {Spergel}, {Staggs}, {Swetz}, {Switzer},
  {Thornton}, {Visnjic}, \& {Wollack}}]{das11}
{Das}, S., {Sherwin}, B.~D., {Aguirre}, P., {et~al.}
\newblock {Detection of the Power Spectrum of Cosmic Microwave Background
  Lensing by the Atacama Cosmology Telescope}. 2011, Physical Review Letters,
  107, 021301, \dodoi{10.1103/PhysRevLett.107.021301}

\bibitem[{{Sherwin} {et~al.}(2017){Sherwin}, {van Engelen}, {Sehgal},
  {Madhavacheril}, {Addison}, {Aiola}, {Allison}, {Battaglia}, {Becker},
  {Beall}, {Bond}, {Calabrese}, {Datta}, {Devlin}, {D{\"u}nner}, {Dunkley},
  {Fox}, {Gallardo}, {Halpern}, {Hasselfield}, {Henderson}, {Hill}, {Hilton},
  {Hubmayr}, {Hughes}, {Hincks}, {Hlozek}, {Huffenberger}, {Koopman},
  {Kosowsky}, {Louis}, {Maurin}, {McMahon}, {Moodley}, {Naess}, {Nati},
  {Newburgh}, {Niemack}, {Page}, {Sievers}, {Spergel}, {Staggs}, {Thornton},
  {Van Lanen}, {Vavagiakis}, \& {Wollack}}]{sherwin17}
{Sherwin}, B.~D., {van Engelen}, A., {Sehgal}, N., {et~al.}
\newblock {Two-season Atacama Cosmology Telescope polarimeter lensing power
  spectrum}. 2017, \prd, 95, 123529, \dodoi{10.1103/PhysRevD.95.123529}

\bibitem[{Darwish {et~al.}(2020)Darwish, Madhavacheril, Sherwin, Aiola,
  Battaglia, Beall, Becker, Bond, Calabrese, Choi, Devlin, Dunkley, D{\"u}nner,
  Ferraro, Fox, Gallardo, Guan, Halpern, Han, Hasselfield, Hill, Hilton,
  Hilton, Hincks, Patty~Ho, Hubmayr, Hughes, Koopman, Kosowsky, Van~Lanen,
  Louis, Lungu, MacInnis, Maurin, McMahon, Moodley, Naess, Namikawa, Nati,
  Newburgh, Nibarger, Niemack, Page, Partridge, Qu, Robertson, Schillaci,
  Schmitt, Sehgal, Sif{\'o}n, Spergel, Staggs, Storer, van Engelen, \&
  Wollack}]{darwish20}
Darwish, O., Madhavacheril, M.~S., Sherwin, B.~D., {et~al.}
\newblock {The Atacama Cosmology Telescope: a CMB lensing mass map over 2100
  square degrees of sky and its cross-correlation with BOSS-CMASS galaxies}.
  2020, Monthly Notices of the Royal Astronomical Society, 500, 2250,
  \dodoi{10.1093/mnras/staa3438}

\bibitem[{{BICEP2 Collaboration} {et~al.}(2016){BICEP2 Collaboration}, {Keck
  Array Collaboration}, {Ade}, {Ahmed}, {Aikin}, {Alexander}, {Barkats},
  {Benton}, {Bischoff}, {Bock}, {Bowens-Rubin}, {Brevik}, {Buder}, {Bullock},
  {Buza}, {Connors}, {Crill}, {Duband}, {Dvorkin}, {Filippini}, {Fliescher},
  {Grayson}, {Halpern}, {Harrison}, {Hildebrandt}, {Hilton}, {Hui}, {Irwin},
  {Kang}, {Karkare}, {Karpel}, {Kaufman}, {Keating}, {Kefeli}, {Kernasovskiy},
  {Kovac}, {Kuo}, {Leitch}, {Lueker}, {Megerian}, {Namikawa}, {Netterfield},
  {Nguyen}, {O'Brient}, {Ogburn}, {Orlando}, {Pryke}, {Richter}, {Schwarz},
  {Sheehy}, {Staniszewski}, {Steinbach}, {Sudiwala}, {Teply}, {Thompson},
  {Tolan}, {Tucker}, {Turner}, {Vieregg}, {Weber}, {Wiebe}, {Willmert}, {Wong},
  {Wu}, \& {Yoon}}]{bicep2keck16b}
{BICEP2 Collaboration}, {Keck Array Collaboration}, {Ade}, P.~A.~R., {et~al.}
\newblock {BICEP2/Keck Array VIII: Measurement of Gravitational Lensing from
  Large-scale B-mode Polarization}. 2016, \apj, 833, 228,
  \dodoi{10.3847/1538-4357/833/2/228}

\bibitem[{Ade {et~al.}(2023)Ade, Ahmed, Amiri, Barkats, Thakur, Bischoff, Beck,
  Bock, Boenish, Bullock, Buza, IV, Connors, Cornelison, Crumrine, Cukierman,
  Denison, Dierickx, Duband, Eiben, Fatigoni, Filippini, Fliescher,
  Giannakopoulos, Goeckner-Wald, Goldfinger, Grayson, Grimes, Hall, Halal,
  Halpern, Hand, Harrison, Henderson, Hildebrandt, Hubmayr, Hui, Irwin, Kang,
  Karkare, Karpel, Kefeli, Kernasovskiy, Kovac, Kuo, Lau, Leitch, Lennox,
  Megerian, Minutolo, Moncelsi, Nakato, Namikawa, Nguyen, O'Brient, IV,
  Palladino, Petroff, Prouve, Pryke, Racine, Reintsema, Richter, Schillaci,
  Schwarz, Schmitt, Sheehy, Singari, Soliman, Germaine, Steinbach, Sudiwala,
  Teply, Thompson, Tolan, Tucker, Turner, Umilt{\`a}, Verg{\`e}s, Vieregg,
  Wandui, Weber, Wiebe, Willmert, Wong, Wu, Yang, Yoon, Young, Yu, Zeng, Zhang,
  Zhang, \& Collaboration}]{ade23}
Ade, P. A.~R., Ahmed, Z., Amiri, M., {et~al.}
\newblock BICEP/Keck. XVII. Line-of-sight Distortion Analysis: Estimates of
  Gravitational Lensing, Anisotropic Cosmic Birefringence, Patchy Reionization,
  and Systematic Errors. 2023, The Astrophysical Journal, 949, 43,
  \dodoi{10.3847/1538-4357/acc85c}

\bibitem[{{Planck Collaboration} {et~al.}(2014){Planck Collaboration}, {Ade},
  {Aghanim}, {Armitage-Caplan}, {Arnaud}, {Ashdown}, {Atrio-Barandela},
  {Aumont}, {Baccigalupi}, {Banday}, \& et~al.}]{planck13-17}
{Planck Collaboration}, {Ade}, P.~A.~R., {Aghanim}, N., {Armitage-Caplan}, C.,
  {Arnaud}, M., {Ashdown}, M., {Atrio-Barandela}, F., {Aumont}, J.,
  {Baccigalupi}, C., {Banday}, A.~J., \& et~al.
\newblock {Planck 2013 results. XVII. Gravitational lensing by large-scale
  structure}. 2014, \aap, 571, A17, \dodoi{10.1051/0004-6361/201321543}

\bibitem[{{Planck Collaboration} {et~al.}(2016{\natexlab{a}}){Planck
  Collaboration}, {Ade}, {Aghanim}, {Arnaud}, {Ashdown}, {Aumont},
  {Baccigalupi}, {Banday}, {Barreiro}, {Bartlett}, \& et~al.}]{planck15-15}
{Planck Collaboration}, {Ade}, P.~A.~R., {Aghanim}, N., {Arnaud}, M.,
  {Ashdown}, M., {Aumont}, J., {Baccigalupi}, C., {Banday}, A.~J., {Barreiro},
  R.~B., {Bartlett}, J.~G., \& et~al.
\newblock {Planck 2015 results. XV. Gravitational lensing}. 2016{\natexlab{a}},
  \aap, 594, A15, \dodoi{10.1051/0004-6361/201525941}

\bibitem[{{Planck Collaboration} {et~al.}(2020{\natexlab{a}}){Planck
  Collaboration}, {Aghanim}, {Akrami}, {Ashdown}, {Aumont}, {Baccigalupi},
  {Ballardini}, {Banday}, {Barreiro}, {Bartolo}, {Basak}, {Benabed}, {Bernard},
  {Bersanelli}, {Bielewicz}, {Bock}, {Bond}, {Borrill}, {Bouchet}, {Boulanger},
  {Bucher}, {Burigana}, {Calabrese}, {Cardoso}, {Carron}, {Challinor},
  {Chiang}, {Colombo}, {Combet}, {Crill}, {Cuttaia}, {de Bernardis}, {de
  Zotti}, {Delabrouille}, {Di Valentino}, {Diego}, {Dor{\'e}}, {Douspis},
  {Ducout}, {Dupac}, {Efstathiou}, {Elsner}, {En{\ss}lin}, {Eriksen},
  {Fantaye}, {Fernandez-Cobos}, {Finelli}, {Forastieri}, {Frailis}, {Fraisse},
  {Franceschi}, {Frolov}, {Galeotta}, {Galli}, {Ganga}, {G{\'e}nova-Santos},
  {Gerbino}, {Ghosh}, {Gonz{\'a}lez-Nuevo}, {G{\'o}rski}, {Gratton},
  {Gruppuso}, {Gudmundsson}, {Hamann}, {Handley}, {Hansen}, {Herranz}, {Hivon},
  {Huang}, {Jaffe}, {Jones}, {Karakci}, {Keih{\"a}nen}, {Keskitalo}, {Kiiveri},
  {Kim}, {Knox}, {Krachmalnicoff}, {Kunz}, {Kurki-Suonio}, {Lagache},
  {Lamarre}, {Lasenby}, {Lattanzi}, {Lawrence}, {Le Jeune}, {Levrier}, {Lewis},
  {Liguori}, {Lilje}, {Lindholm}, {L{\'o}pez-Caniego}, {Lubin}, {Ma},
  {Mac{\'\i}as-P{\'e}rez}, {Maggio}, {Maino}, {Mand olesi}, {Mangilli},
  {Marcos-Caballero}, {Maris}, {Martin}, {Mart{\'\i}nez-Gonz{\'a}lez},
  {Matarrese}, {Mauri}, {McEwen}, {Melchiorri}, {Mennella}, {Migliaccio},
  {Miville-Desch{\^e}nes}, {Molinari}, {Moneti}, {Montier}, {Morgante}, {Moss},
  {Natoli}, {Pagano}, {Paoletti}, {Partridge}, {Patanchon}, {Perrotta},
  {Pettorino}, {Piacentini}, {Polastri}, {Polenta}, {Puget}, {Rachen},
  {Reinecke}, {Remazeilles}, {Renzi}, {Rocha}, {Rosset}, {Roudier},
  {Rubi{\~n}o-Mart{\'\i}n}, {Ruiz-Granados}, {Salvati}, {Sandri}, {Savelainen},
  {Scott}, {Sirignano}, {Sunyaev}, {Suur-Uski}, {Tauber}, {Tavagnacco},
  {Tenti}, {Toffolatti}, {Tomasi}, {Trombetti}, {Valiviita}, {Van Tent},
  {Vielva}, {Villa}, {Vittorio}, {Wandelt}, {Wehus}, {White}, {White},
  {Zacchei}, \& {Zonca}}]{planck18-8}
{Planck Collaboration}, {Aghanim}, N., {Akrami}, Y., {et~al.}
\newblock {Planck 2018 results. VIII. Gravitational lensing}.
  2020{\natexlab{a}}, \aap, 641, A8, \dodoi{10.1051/0004-6361/201833886}

\bibitem[{Carron {et~al.}(2022)Carron, Mirmelstein, \& Lewis}]{carron22}
Carron, J., Mirmelstein, M., \& Lewis, A.
\newblock CMB lensing from Planck PR4 maps. 2022, Journal of Cosmology and
  Astroparticle Physics, 2022, 039, \dodoi{10.1088/1475-7516/2022/09/039}

\bibitem[{{POLARBEAR Collaboration} {et~al.}(2014){POLARBEAR Collaboration},
  {Ade}, {Akiba}, {Anthony}, {Arnold}, {Atlas}, {Barron}, {Boettger},
  {Borrill}, {Chapman}, {Chinone}, {Dobbs}, {Elleflot}, {Errard}, {Fabbian},
  {Feng}, {Flanigan}, {Gilbert}, {Grainger}, {Halverson}, {Hasegawa},
  {Hattori}, {Hazumi}, {Holzapfel}, {Hori}, {Howard}, {Hyland}, {Inoue},
  {Jaehnig}, {Jaffe}, {Keating}, {Kermish}, {Keskitalo}, {Kisner}, {Le Jeune},
  {Lee}, {Linder}, {Leitch}, {Lungu}, {Matsuda}, {Matsumura}, {Meng}, {Miller},
  {Morii}, {Moyerman}, {Myers}, {Navaroli}, {Nishino}, {Paar}, {Peloton},
  {Quealy}, {Rebeiz}, {Reichardt}, {Richards}, {Ross}, {Schanning}, {Schenck},
  {Sherwin}, {Shimizu}, {Shimmin}, {Shimon}, {Siritanasak}, {Smecher},
  {Spieler}, {Stebor}, {Steinbach}, {Stompor}, {Suzuki}, {Takakura}, {Tomaru},
  {Wilson}, {Yadav}, \& {Zahn}}]{polarbear14c}
{POLARBEAR Collaboration}, {Ade}, P.~A.~R., {Akiba}, Y., {et~al.}
\newblock {Measurement of the Cosmic Microwave Background Polarization Lensing
  Power Spectrum with the POLARBEAR experiment}. 2014, Physical Review Letters,
  021301.
\newblock \doarXiv{1312.6646}

\bibitem[{{Andr'es Osvaldo Aguilar Fa'undez} {et~al.}(2019){Andr'es Osvaldo
  Aguilar Fa'undez}, {Arnold}, {Baccigalupi}, {Barron}, {Beck}, {Beckman},
  {Bianchini}, {El Bouhargani}, {Carron}, {Cheung}, {Chinone}, {Elleflot},
  {Errard}, {Fabbian}, {Feng}, {Fujino}, {Goeckner-Wald}, {Hamada}, {Hasegawa},
  {Hazumi}, {Hill}, {Hirose}, {Jeong}, {Katayama}, {Keating}, {Kikuchi},
  {Kusaka}, {Lee}, {Leon}, {Linder}, {Lowry}, {Matsuda}, {Matsumura}, {Minami},
  {Navaroli}, {Nishino}, {Phuong Pham}, {Poletti}, {Puglisi}, {Reichardt},
  {Segawa}, {Sherwin}, {Silva-Feaver}, {Siritanasak}, {Stompor}, {Suzuki},
  {Tajima}, {Takatori}, {Tanabe}, {Teply}, \& {Tsai}}]{polarbear19b}
{Andr'es Osvaldo Aguilar Fa'undez}, M., {Arnold}, K., {Baccigalupi}, C.,
  {et~al.}
\newblock {Measurement of the Cosmic Microwave Background Polarization Lensing
  Power Spectrum from Two Years of POLARBEAR Data}. 2019, arXiv e-prints,
  arXiv:1911.10980.
\newblock \doarXiv{1911.10980}

\bibitem[{{van Engelen} {et~al.}(2012){van Engelen}, {Keisler}, {Zahn}, {Aird},
  {Benson}, {Bleem}, {Carlstrom}, {Chang}, {Cho}, {Crawford}, {Crites}, {de
  Haan}, {Dobbs}, {Dudley}, {George}, {Halverson}, {Holder}, {Holzapfel},
  {Hoover}, {Hou}, {Hrubes}, {Joy}, {Knox}, {Lee}, {Leitch}, {Lueker},
  {Luong-Van}, {McMahon}, {Mehl}, {Meyer}, {Millea}, {Mohr}, {Montroy},
  {Natoli}, {Padin}, {Plagge}, {Pryke}, {Reichardt}, {Ruhl}, {Sayre},
  {Schaffer}, {Shaw}, {Shirokoff}, {Spieler}, {Staniszewski}, {Stark}, {Story},
  {Vanderlinde}, {Vieira}, \& {Williamson}}]{vanengelen12}
{van Engelen}, A., {Keisler}, R., {Zahn}, O., {et~al.}
\newblock {A Measurement of Gravitational Lensing of the Microwave Background
  Using South Pole Telescope Data}. 2012, \apj, 756, 142,
  \dodoi{10.1088/0004-637X/756/2/142}

\bibitem[{{Story} {et~al.}(2015){Story}, {Hanson}, {Ade}, {Aird}, {Austermann},
  {Beall}, {Bender}, {Benson}, {Bleem}, {Carlstrom}, {Chang}, {Chiang}, {Cho},
  {Citron}, {Crawford}, {Crites}, {de Haan}, {Dobbs}, {Everett}, {Gallicchio},
  {Gao}, {George}, {Gilbert}, {Halverson}, {Harrington}, {Henning}, {Hilton},
  {Holder}, {Holzapfel}, {Hoover}, {Hou}, {Hrubes}, {Huang}, {Hubmayr},
  {Irwin}, {Keisler}, {Knox}, {Lee}, {Leitch}, {Li}, {Liang}, {Luong-Van},
  {McMahon}, {Mehl}, {Meyer}, {Mocanu}, {Montroy}, {Natoli}, {Nibarger},
  {Novosad}, {Padin}, {Pryke}, {Reichardt}, {Ruhl}, {Saliwanchik}, {Sayre},
  {Schaffer}, {Smecher}, {Stark}, {Tucker}, {Vanderlinde}, {Vieira}, {Wang},
  {Whitehorn}, {Yefremenko}, \& {Zahn}}]{story15}
{Story}, K.~T., {Hanson}, D., {Ade}, P.~A.~R., {et~al.}
\newblock {A Measurement of the Cosmic Microwave Background Gravitational
  Lensing Potential from 100 Square Degrees of SPTpol Data}. 2015, \apj, 810,
  50, \dodoi{10.1088/0004-637X/810/1/50}

\bibitem[{{Omori} {et~al.}(2017){Omori}, {Chown}, {Simard}, {Story}, {Aylor},
  {Baxter}, {Benson}, {Bleem}, {Carlstrom}, {Chang}, {Cho}, {Crawford},
  {Crites}, {de Haan}, {Dobbs}, {Everett}, {George}, {Halverson}, {Harrington},
  {Holder}, {Hou}, {Holzapfel}, {Hrubes}, {Knox}, {Lee}, {Leitch}, {Luong-Van},
  {Manzotti}, {Marrone}, {McMahon}, {Meyer}, {Mocanu}, {Mohr}, {Natoli},
  {Padin}, {Pryke}, {Reichardt}, {Ruhl}, {Sayre}, {Schaffer}, {Shirokoff},
  {Staniszewski}, {Stark}, {Vanderlinde}, {Vieira}, {Williamson}, \&
  {Zahn}}]{omori17}
{Omori}, Y., {Chown}, R., {Simard}, G., {et~al.}
\newblock {A 2500 deg$^{2}$ CMB Lensing Map from Combined South Pole Telescope
  and Planck Data}. 2017, \apj, 849, 124, \dodoi{10.3847/1538-4357/aa8d1d}

\bibitem[{{Wu} {et~al.}(2019){Wu}, {Mocanu}, {Ade}, {Anderson}, {Austermann},
  {Avva}, {Beall}, {Bender}, {Benson}, {Bianchini}, {Bleem}, {Carlstrom},
  {Chang}, {Chiang}, {Citron}, {Corbett Moran}, {Crawford}, {Crites}, {de
  Haan}, {Dobbs}, {Everett}, {Gallicchio}, {George}, {Gilbert}, {Gupta},
  {Halverson}, {Harrington}, {Henning}, {Hilton}, {Holder}, {Holzapfel}, {Hou},
  {Hrubes}, {Huang}, {Hubmayr}, {Irwin}, {Knox}, {Lee}, {Li}, {Lowitz},
  {Manzotti}, {McMahon}, {Meyer}, {Millea}, {Montgomery}, {Nadolski}, {Natoli},
  {Nibarger}, {Noble}, {Novosad}, {Omori}, {Padin}, {Patil}, {Pryke},
  {Reichardt}, {Ruhl}, {Saliwanchik}, {Sayre}, {Schaffer}, {Sievers}, {Simard},
  {Smecher}, {Stark}, {Story}, {Tucker}, {Vanderlinde}, {Veach}, {Vieira},
  {Wang}, {Whitehorn}, \& {Yefremenko}}]{wu19}
{Wu}, W.~L.~K., {Mocanu}, L.~M., {Ade}, P.~A.~R., {et~al.}
\newblock {A Measurement of the Cosmic Microwave Background Lensing Potential
  and Power Spectrum from 500 deg$^{2}$ of SPTpol Temperature and Polarization
  Data}. 2019, \apj, 884, 70, \dodoi{10.3847/1538-4357/ab4186}

\bibitem[{Millea {et~al.}(2021)Millea, Daley, Chou, Anderes, Ade, Anderson,
  Austermann, Avva, Beall, Bender, Benson, Bianchini, Bleem, Carlstrom, Chang,
  Chaubal, Chiang, Citron, Moran, Crawford, Crites, de~Haan, Dobbs, Everett,
  Gallicchio, George, Goeckner-Wald, Guns, Gupta, Halverson, Henning, Hilton,
  Holder, Holzapfel, Hrubes, Huang, Hubmayr, Irwin, Knox, Lee, Li, Lowitz,
  McMahon, Meyer, Mocanu, Montgomery, Natoli, Nibarger, Noble, Novosad, Omori,
  Padin, Patil, Pryke, Reichardt, Ruhl, Saliwanchik, Schaffer, Sievers,
  Smecher, Stark, Thorne, Tucker, Veach, Vieira, Wang, Whitehorn, Wu, \&
  Yefremenko}]{millea21}
Millea, M., Daley, C.~M., Chou, T.-L., {et~al.}
\newblock Optimal Cosmic Microwave Background Lensing Reconstruction and
  Parameter Estimation with SPTpol Data. 2021, The Astrophysical Journal, 922,
  259, \dodoi{10.3847/1538-4357/ac02bb}

\bibitem[{Omori {et~al.}(2023)Omori, Baxter, Chang, Friedrich, Alarcon, Alves,
  Amon, Andrade-Oliveira, Bechtol, Becker, Bernstein, Blazek, Bleem, Camacho,
  Campos, Carnero~Rosell, Carrasco~Kind, Cawthon, Chen, Choi, Cordero,
  Crawford, Crocce, Davis, DeRose, Dodelson, Doux, Drlica-Wagner, Eckert,
  Eifler, Elsner, Elvin-Poole, Everett, Fang, Fert\'e, Fosalba, Gatti,
  Giannini, Gruen, Gruendl, Harrison, Herner, Huang, Huff, Huterer, Jarvis,
  Krause, Kuropatkin, Leget, Lemos, Liddle, MacCrann, McCullough, Muir, Myles,
  Navarro-Alsina, Pandey, Park, Porredon, Prat, Raveri, Rollins, Roodman,
  Rosenfeld, Ross, Rykoff, S\'anchez, Sanchez, Secco, Sevilla-Noarbe, Sheldon,
  Shin, Troxel, Tutusaus, Varga, Weaverdyck, Wechsler, Wu, Yanny, Yin, Zhang,
  Zuntz, Abbott, Aguena, Allam, Annis, Bacon, Benson, Bertin, Bocquet, Brooks,
  Burke, Carlstrom, Carretero, Chang, Chown, Costanzi, da~Costa, Crites,
  Pereira, de~Haan, De~Vicente, Desai, Diehl, Dobbs, Doel, Everett, Ferrero,
  Flaugher, Friedel, Frieman, Garc\'{\i}a-Bellido, Gaztanaga, George,
  Giannantonio, Halverson, Hinton, Holder, Hollowood, Holzapfel, Honscheid,
  Hrubes, James, Knox, Kuehn, Lahav, Lee, Lima, Luong-Van, March, McMahon,
  Melchior, Menanteau, Meyer, Miquel, Mocanu, Mohr, Morgan, Natoli, Padin,
  Palmese, Paz-Chinch\'on, Pieres, Plazas~Malag\'on, Pryke, Reichardt, Romer,
  Ruhl, Sanchez, Schaffer, Schubnell, Serrano, Shirokoff, Smith, Staniszewski,
  Stark, Suchyta, Tarle, Thomas, To, Vieira, Weller, \& Williamson}]{omori23}
Omori, Y., Baxter, E.~J., Chang, C., {et~al.}
\newblock Joint analysis of Dark Energy Survey Year 3 data and CMB lensing from
  SPT and Planck. I. Construction of CMB lensing maps and modeling choices.
  2023, Phys. Rev. D, 107, 023529, \dodoi{10.1103/PhysRevD.107.023529}

\bibitem[{{Dutcher} {et~al.}(2021){Dutcher}, {Balkenhol}, {Ade}, {Ahmed},
  {Anderes}, {Anderson}, {Archipley}, {Avva}, {Aylor}, {Barry}, {Basu Thakur},
  {Benabed}, {Bender}, {Benson}, {Bianchini}, {Bleem}, {Bouchet}, {Bryant},
  {Byrum}, {Carlstrom}, {Carter}, {Cecil}, {Chang}, {Chaubal}, {Chen}, {Cho},
  {Chou}, {Cliche}, {Crawford}, {Cukierman}, {Daley}, {de Haan}, {Denison},
  {Dibert}, {Ding}, {Dobbs}, {Everett}, {Feng}, {Ferguson}, {Foster}, {Fu},
  {Galli}, {Gambrel}, {Gardner}, {Goeckner-Wald}, {Gualtieri}, {Guns}, {Gupta},
  {Guyser}, {Halverson}, {Harke-Hosemann}, {Harrington}, {Henning}, {Hilton},
  {Hivon}, {Holder}, {Holzapfel}, {Hood}, {Howe}, {Huang}, {Irwin}, {Jeong},
  {Jonas}, {Jones}, {Khaire}, {Knox}, {Kofman}, {Korman}, {Kubik}, {Kuhlmann},
  {Kuo}, {Lee}, {Leitch}, {Lowitz}, {Lu}, {Meyer}, {Michalik}, {Millea},
  {Montgomery}, {Nadolski}, {Natoli}, {Nguyen}, {Noble}, {Novosad}, {Omori},
  {Padin}, {Pan}, {Paschos}, {Pearson}, {Posada}, {Prabhu}, {Quan},
  {Raghunathan}, {Rahlin}, {Reichardt}, {Riebel}, {Riedel}, {Rouble}, {Ruhl},
  {Sayre}, {Schiappucci}, {Shirokoff}, {Smecher}, {Sobrin}, {Stark}, {Stephen},
  {Story}, {Suzuki}, {Thompson}, {Thorne}, {Tucker}, {Umilta}, {Vale},
  {Vanderlinde}, {Vieira}, {Wang}, {Whitehorn}, {Wu}, {Yefremenko}, {Yoon},
  {Young}, \& {SPT-3G Collaboration}}]{dutcher21}
{Dutcher}, D., {Balkenhol}, L., {Ade}, P.~A.~R., {et~al.}
\newblock {Measurements of the E -mode polarization and temperature-E -mode
  correlation of the CMB from SPT-3G 2018 data}. 2021, \prd, 104, 022003,
  \dodoi{10.1103/PhysRevD.104.022003}

\bibitem[{{Balkenhol} {et~al.}(2021){Balkenhol}, {Dutcher}, {Ade}, {Ahmed},
  {Anderes}, {Anderson}, {Archipley}, {Avva}, {Aylor}, {Barry}, {Basu Thakur},
  {Benabed}, {Bender}, {Benson}, {Bianchini}, {Bleem}, {Bouchet}, {Bryant},
  {Byrum}, {Carlstrom}, {Carter}, {Cecil}, {Chang}, {Chaubal}, {Chen}, {Cho},
  {Chou}, {Cliche}, {Crawford}, {Cukierman}, {Daley}, {de Haan}, {Denison},
  {Dibert}, {Ding}, {Dobbs}, {Everett}, {Feng}, {Ferguson}, {Foster}, {Fu},
  {Galli}, {Gambrel}, {Gardner}, {Goeckner-Wald}, {Gualtieri}, {Guns}, {Gupta},
  {Guyser}, {Halverson}, {Harke-Hosemann}, {Harrington}, {Henning}, {Hilton},
  {Hivon}, {Holder}, {Holzapfel}, {Hood}, {Howe}, {Huang}, {Irwin}, {Jeong},
  {Jonas}, {Jones}, {Khaire}, {Knox}, {Kofman}, {Korman}, {Kubik}, {Kuhlmann},
  {Kuo}, {Lee}, {Leitch}, {Lowitz}, {Lu}, {Meyer}, {Michalik}, {Millea},
  {Montgomery}, {Nadolski}, {Natoli}, {Nguyen}, {Noble}, {Novosad}, {Omori},
  {Padin}, {Pan}, {Paschos}, {Pearson}, {Posada}, {Prabhu}, {Quan}, {Rahlin},
  {Reichardt}, {Riebel}, {Riedel}, {Rouble}, {Ruhl}, {Sayre}, {Schiappucci},
  {Shirokoff}, {Smecher}, {Sobrin}, {Stark}, {Stephen}, {Story}, {Suzuki},
  {Thompson}, {Thorne}, {Tucker}, {Umilta}, {Vale}, {Vanderlinde}, {Vieira},
  {Wang}, {Whitehorn}, {Wu}, {Yefremenko}, {Yoon}, {Young}, \& {SPT-3G
  Collaboration}}]{balkenhol21}
{Balkenhol}, L., {Dutcher}, D., {Ade}, P.~A.~R., {et~al.}
\newblock {Constraints on {\ensuremath{\Lambda}} CDM extensions from the SPT-3G
  2018 E E and T E power spectra}. 2021, \prd, 104, 083509,
  \dodoi{10.1103/PhysRevD.104.083509}

\bibitem[{Balkenhol {et~al.}(2023)Balkenhol, Dutcher, Spurio~Mancini, Doussot,
  Benabed, Galli, Ade, Anderson, Ansarinejad, Archipley, Bender, Benson,
  Bianchini, Bleem, Bouchet, Bryant, Camphuis, Carlstrom, Cecil, Chang,
  Chaubal, Chichura, Chou, Coerver, Crawford, Cukierman, Daley, de~Haan,
  Dibert, Dobbs, Everett, Feng, Ferguson, Foster, Gambrel, Gardner,
  Goeckner-Wald, Gualtieri, Guidi, Guns, Halverson, Hivon, Holder, Holzapfel,
  Hood, Huang, Knox, Korman, Kuo, Lee, Lowitz, Lu, Millea, Montgomery, Nakato,
  Natoli, Noble, Novosad, Omori, Padin, Pan, Paschos, Prabhu, Quan, Rahimi,
  Rahlin, Reichardt, Rouble, Ruhl, Schiappucci, Smecher, Sobrin, Stark,
  Stephen, Suzuki, Tandoi, Thompson, Thorne, Tucker, Umilta, Vieira, Wang,
  Whitehorn, Wu, Yefremenko, Young, \& Zebrowski}]{balkenhol23}
Balkenhol, L., Dutcher, D., Spurio~Mancini, A., {et~al.}
\newblock Measurement of the CMB temperature power spectrum and constraints on
  cosmology from the SPT-3G 2018 $TT$, $TE$, and $EE$ dataset. 2023, Phys. Rev.
  D, 108, 023510, \dodoi{10.1103/PhysRevD.108.023510}

\bibitem[{{Planck Collaboration} {et~al.}(2020{\natexlab{b}}){Planck
  Collaboration}, {Aghanim}, {Akrami}, {Ashdown}, {Aumont}, {Baccigalupi},
  {Ballardini}, {Banday}, {Barreiro}, {Bartolo}, {Basak}, {Battye}, {Benabed},
  {Bernard}, {Bersanelli}, {Bielewicz}, {Bock}, {Bond}, {Borrill}, {Bouchet},
  {Boulanger}, {Bucher}, {Burigana}, {Butler}, {Calabrese}, {Cardoso},
  {Carron}, {Challinor}, {Chiang}, {Chluba}, {Colombo}, {Combet}, {Contreras},
  {Crill}, {Cuttaia}, {de Bernardis}, {de Zotti}, {Delabrouille}, {Delouis},
  {Di Valentino}, {Diego}, {Dor{\'e}}, {Douspis}, {Ducout}, {Dupac}, {Dusini},
  {Efstathiou}, {Elsner}, {En{\ss}lin}, {Eriksen}, {Fantaye}, {Farhang},
  {Fergusson}, {Fernandez-Cobos}, {Finelli}, {Forastieri}, {Frailis},
  {Fraisse}, {Franceschi}, {Frolov}, {Galeotta}, {Galli}, {Ganga},
  {G{\'e}nova-Santos}, {Gerbino}, {Ghosh}, {Gonz{\'a}lez-Nuevo}, {G{\'o}rski},
  {Gratton}, {Gruppuso}, {Gudmundsson}, {Hamann}, {Handley}, {Hansen},
  {Herranz}, {Hildebrandt}, {Hivon}, {Huang}, {Jaffe}, {Jones}, {Karakci},
  {Keih{\"a}nen}, {Keskitalo}, {Kiiveri}, {Kim}, {Kisner}, {Knox},
  {Krachmalnicoff}, {Kunz}, {Kurki-Suonio}, {Lagache}, {Lamarre}, {Lasenby},
  {Lattanzi}, {Lawrence}, {Le Jeune}, {Lemos}, {Lesgourgues}, {Levrier},
  {Lewis}, {Liguori}, {Lilje}, {Lilley}, {Lindholm}, {L{\'o}pez-Caniego},
  {Lubin}, {Ma}, {Mac{\'\i}as-P{\'e}rez}, {Maggio}, {Maino}, {Mandolesi},
  {Mangilli}, {Marcos-Caballero}, {Maris}, {Martin}, {Martinelli},
  {Mart{\'\i}nez-Gonz{\'a}lez}, {Matarrese}, {Mauri}, {McEwen}, {Meinhold},
  {Melchiorri}, {Mennella}, {Migliaccio}, {Millea}, {Mitra},
  {Miville-Desch{\^e}nes}, {Molinari}, {Montier}, {Morgante}, {Moss}, {Natoli},
  {N{\o}rgaard-Nielsen}, {Pagano}, {Paoletti}, {Partridge}, {Patanchon},
  {Peiris}, {Perrotta}, {Pettorino}, {Piacentini}, {Polastri}, {Polenta},
  {Puget}, {Rachen}, {Reinecke}, {Remazeilles}, {Renzi}, {Rocha}, {Rosset},
  {Roudier}, {Rubi{\~n}o-Mart{\'\i}n}, {Ruiz-Granados}, {Salvati}, {Sandri},
  {Savelainen}, {Scott}, {Shellard}, {Sirignano}, {Sirri}, {Spencer},
  {Sunyaev}, {Suur-Uski}, {Tauber}, {Tavagnacco}, {Tenti}, {Toffolatti},
  {Tomasi}, {Trombetti}, {Valenziano}, {Valiviita}, {Van Tent}, {Vibert},
  {Vielva}, {Villa}, {Vittorio}, {Wand elt}, {Wehus}, {White}, {White},
  {Zacchei}, \& {Zonca}}]{planck18-6}
{Planck Collaboration}, {Aghanim}, N., {Akrami}, Y., {et~al.}
\newblock {Planck 2018 results. VI. Cosmological parameters}.
  2020{\natexlab{b}}, \aap, 641, A6, \dodoi{10.1051/0004-6361/201833910}

\bibitem[{Freedman(2021)}]{freedman21}
Freedman, W.~L.
\newblock Measurements of the Hubble Constant: Tensions in Perspective*. 2021,
  The Astrophysical Journal, 919, 16, \dodoi{10.3847/1538-4357/ac0e95}

\bibitem[{{Riess} {et~al.}(2022){Riess}, {Yuan}, {Macri}, {Scolnic}, {Brout},
  {Casertano}, {Jones}, {Murakami}, {Anand}, {Breuval}, {Brink}, {Filippenko},
  {Hoffmann}, {Jha}, {D'arcy Kenworthy}, {Mackenty}, {Stahl}, \&
  {Zheng}}]{riess22}
{Riess}, A.~G., {Yuan}, W., {Macri}, L.~M., {et~al.}
\newblock {A Comprehensive Measurement of the Local Value of the Hubble
  Constant with 1 km s$^{-1}$ Mpc$^{-1}$ Uncertainty from the Hubble Space
  Telescope and the SH0ES Team}. 2022, \apjl, 934, L7,
  \dodoi{10.3847/2041-8213/ac5c5b}

\bibitem[{{Murakami} {et~al.}(2023){Murakami}, {Riess}, {Stahl}, {Kenworthy},
  {Pluck}, {Macoretta}, {Brout}, {Jones}, {Scolnic}, \&
  {Filippenko}}]{murakami23}
{Murakami}, Y.~S., {Riess}, A.~G., {Stahl}, B.~E., {Kenworthy}, W.~D., {Pluck},
  D.-M.~A., {Macoretta}, A., {Brout}, D., {Jones}, D.~O., {Scolnic}, D.~M., \&
  {Filippenko}, A.~V.
\newblock {Leveraging SN Ia spectroscopic similarity to improve the measurement
  of $H_0$}. 2023, arXiv e-prints, arXiv:2306.00070,
  \dodoi{10.48550/arXiv.2306.00070}

\bibitem[{{Asgari, Marika} {et~al.}(2021){Asgari, Marika}, {Lin, Chieh-An},
  {Joachimi, Benjamin}, {Giblin, Benjamin}, {Heymans, Catherine}, {Hildebrandt,
  Hendrik}, {Kannawadi, Arun}, {St\"olzner, Benjamin}, {Tr\"oster, Tilman},
  {van den Busch, Jan Luca}, {Wright, Angus H.}, {Bilicki, Maciej}, {Blake,
  Chris}, {de Jong, Jelte}, {Dvornik, Andrej}, {Erben, Thomas}, {Getman,
  Fedor}, {Hoekstra, Henk}, {K\"ohlinger, Fabian}, {Kuijken, Konrad}, {Miller,
  Lance}, {Radovich, Mario}, {Schneider, Peter}, {Shan, HuanYuan}, \&
  {Valentijn, Edwin}}]{asgari21}
{Asgari, Marika}, {Lin, Chieh-An}, {Joachimi, Benjamin}, {et~al.}
\newblock KiDS-1000 cosmology: Cosmic shear constraints and comparison between
  two point statistics. 2021, A\&A, 645, A104,
  \dodoi{10.1051/0004-6361/202039070}

\bibitem[{{Joudaki, S.} {et~al.}(2020){Joudaki, S.}, {Hildebrandt, H.},
  {Traykova, D.}, {Chisari, N. E.}, {Heymans, C.}, {Kannawadi, A.}, {Kuijken,
  K.}, {Wright, A. H.}, {Asgari, M.}, {Erben, T.}, {Hoekstra, H.}, {Joachimi,
  B.}, {Miller, L.}, {Tr\"oster, T.}, \& {van den Busch, J. L.}}]{joudaki19}
{Joudaki, S.}, {Hildebrandt, H.}, {Traykova, D.}, {Chisari, N. E.}, {Heymans,
  C.}, {Kannawadi, A.}, {Kuijken, K.}, {Wright, A. H.}, {Asgari, M.}, {Erben,
  T.}, {Hoekstra, H.}, {Joachimi, B.}, {Miller, L.}, {Tr\"oster, T.}, \& {van
  den Busch, J. L.}
\newblock KiDS+VIKING-450 and DES-Y1 combined: Cosmology with cosmic shear.
  2020, A\&A, 638, L1, \dodoi{10.1051/0004-6361/201936154}

\bibitem[{{Abbott} {et~al.}(2022){Abbott}, {Aguena}, {Alarcon}, {Allam},
  {Alves}, {Amon}, {Andrade-Oliveira}, {Annis}, {Avila}, {Bacon}, {Baxter},
  {Bechtol}, {Becker}, {Bernstein}, {Bhargava}, {Birrer}, {Blazek},
  {Brandao-Souza}, {Bridle}, {Brooks}, {Buckley-Geer}, {Burke}, {Camacho},
  {Campos}, {Carnero Rosell}, {Carrasco Kind}, {Carretero}, {Castander},
  {Cawthon}, {Chang}, {Chen}, {Chen}, {Choi}, {Conselice}, {Cordero},
  {Costanzi}, {Crocce}, {da Costa}, {da Silva Pereira}, {Davis}, {Davis}, {De
  Vicente}, {DeRose}, {Desai}, {Di Valentino}, {Diehl}, {Dietrich}, {Dodelson},
  {Doel}, {Doux}, {Drlica-Wagner}, {Eckert}, {Eifler}, {Elsner}, {Elvin-Poole},
  {Everett}, {Evrard}, {Fang}, {Farahi}, {Fernandez}, {Ferrero}, {Fert{\'e}},
  {Fosalba}, {Friedrich}, {Frieman}, {Garc{\'\i}a-Bellido}, {Gatti},
  {Gaztanaga}, {Gerdes}, {Giannantonio}, {Giannini}, {Gruen}, {Gruendl},
  {Gschwend}, {Gutierrez}, {Harrison}, {Hartley}, {Herner}, {Hinton},
  {Hollowood}, {Honscheid}, {Hoyle}, {Huff}, {Huterer}, {Jain}, {James},
  {Jarvis}, {Jeffrey}, {Jeltema}, {Kovacs}, {Krause}, {Kron}, {Kuehn},
  {Kuropatkin}, {Lahav}, {Leget}, {Lemos}, {Liddle}, {Lidman}, {Lima}, {Lin},
  {MacCrann}, {Maia}, {Marshall}, {Martini}, {McCullough}, {Melchior},
  {Mena-Fern{\'a}ndez}, {Menanteau}, {Miquel}, {Mohr}, {Morgan}, {Muir},
  {Myles}, {Nadathur}, {Navarro-Alsina}, {Nichol}, {Ogando}, {Omori},
  {Palmese}, {Pandey}, {Park}, {Paz-Chinch{\'o}n}, {Petravick}, {Pieres},
  {Plazas Malag{\'o}n}, {Porredon}, {Prat}, {Raveri}, {Rodriguez-Monroy},
  {Rollins}, {Romer}, {Roodman}, {Rosenfeld}, {Ross}, {Rykoff}, {Samuroff},
  {S{\'a}nchez}, {Sanchez}, {Sanchez}, {Sanchez Cid}, {Scarpine}, {Schubnell},
  {Scolnic}, {Secco}, {Serrano}, {Sevilla-Noarbe}, {Sheldon}, {Shin}, {Smith},
  {Soares-Santos}, {Suchyta}, {Swanson}, {Tabbutt}, {Tarle}, {Thomas}, {To},
  {Troja}, {Troxel}, {Tucker}, {Tutusaus}, {Varga}, {Walker}, {Weaverdyck},
  {Wechsler}, {Weller}, {Yanny}, {Yin}, {Zhang}, {Zuntz}, \& {DES
  Collaboration}}]{abbott22a}
{Abbott}, T.~M.~C., {Aguena}, M., {Alarcon}, A., {et~al.}
\newblock {Dark Energy Survey Year 3 results: Cosmological constraints from
  galaxy clustering and weak lensing}. 2022, \prd, 105, 023520,
  \dodoi{10.1103/PhysRevD.105.023520}

\bibitem[{{Carlstrom} {et~al.}(2011){Carlstrom}, {Ade}, {Aird}, {Benson},
  {Bleem}, {Busetti}, {Chang}, {Chauvin}, {Cho}, {Crawford}, {Crites}, {Dobbs},
  {Halverson}, {Heimsath}, {Holzapfel}, {Hrubes}, {Joy}, {Keisler}, {Lanting},
  {Lee}, {Leitch}, {Leong}, {Lu}, {Lueker}, {Luong-van}, {McMahon}, {Mehl},
  {Meyer}, {Mohr}, {Montroy}, {Padin}, {Plagge}, {Pryke}, {Ruhl}, {Schaffer},
  {Schwan}, {Shirokoff}, {Spieler}, {Staniszewski}, {Stark}, {Tucker},
  {Vanderlinde}, {Vieira}, \& {Williamson}}]{carlstrom11}
{Carlstrom}, J.~E., {Ade}, P.~A.~R., {Aird}, K.~A., {et~al.}
\newblock {The 10 Meter South Pole Telescope}. 2011, \pasp, 123, 568,
  \dodoi{10.1086/659879}

\bibitem[{{Sobrin} {et~al.}(2022){Sobrin}, {Anderson}, {Bender}, {Benson},
  {Dutcher}, {Foster}, {Goeckner-Wald}, {Montgomery}, {Nadolski}, {Rahlin},
  {Ade}, {Ahmed}, {Anderes}, {Archipley}, {Austermann}, {Avva}, {Aylor},
  {Balkenhol}, {Barry}, {Thakur}, {Benabed}, {Bianchini}, {Bleem}, {Bouchet},
  {Bryant}, {Byrum}, {Carlstrom}, {Carter}, {Cecil}, {Chang}, {Chaubal},
  {Chen}, {Cho}, {Chou}, {Cliche}, {Crawford}, {Cukierman}, {Daley}, {Haan},
  {Denison}, {Dibert}, {Ding}, {Dobbs}, {Everett}, {Feng}, {Ferguson}, {Fu},
  {Galli}, {Gambrel}, {Gardner}, {Gualtieri}, {Guns}, {Gupta}, {Guyser},
  {Halverson}, {Harke-Hosemann}, {Harrington}, {Henning}, {Hilton}, {Hivon},
  {Holder}, {Holzapfel}, {Hood}, {Howe}, {Huang}, {Irwin}, {Jeong}, {Jonas},
  {Jones}, {Khaire}, {Knox}, {Kofman}, {Korman}, {Kubik}, {Kuhlmann}, {Kuo},
  {Lee}, {Leitch}, {Lowitz}, {Lu}, {Meyer}, {Michalik}, {Millea}, {Natoli},
  {Nguyen}, {Noble}, {Novosad}, {Omori}, {Padin}, {Pan}, {Paschos}, {Pearson},
  {Posada}, {Prabhu}, {Quan}, {Reichardt}, {Riebel}, {Riedel}, {Rouble},
  {Ruhl}, {Saliwanchik}, {Sayre}, {Schiappucci}, {Shirokoff}, {Smecher},
  {Stark}, {Stephen}, {Story}, {Suzuki}, {Tandoi}, {Thompson}, {Thorne},
  {Tucker}, {Umilta}, {Vale}, {Vanderlinde}, {Vieira}, {Wang}, {Whitehorn},
  {Wu}, {Yefremenko}, {Yoon}, \& {Young}}]{sobrin22}
{Sobrin}, J.~A., {Anderson}, A.~J., {Bender}, A.~N., {et~al.}
\newblock {The Design and Integrated Performance of SPT-3G}. 2022, \apjs, 258,
  42, \dodoi{10.3847/1538-4365/ac374f}

\bibitem[{{Keisler} {et~al.}(2011){Keisler}, {Reichardt}, {Aird}, {Benson},
  {Bleem}, {Carlstrom}, {Chang}, {Cho}, {Crawford}, {Crites}, {de Haan},
  {Dobbs}, {Dudley}, {George}, {Halverson}, {Holder}, {Holzapfel}, {Hoover},
  {Hou}, {Hrubes}, {Joy}, {Knox}, {Lee}, {Leitch}, {Lueker}, {Luong-Van},
  {McMahon}, {Mehl}, {Meyer}, {Millea}, {Mohr}, {Montroy}, {Natoli}, {Padin},
  {Plagge}, {Pryke}, {Ruhl}, {Schaffer}, {Shaw}, {Shirokoff}, {Spieler},
  {Staniszewski}, {Stark}, {Story}, {van Engelen}, {Vanderlinde}, {Vieira},
  {Williamson}, \& {Zahn}}]{keisler11}
{Keisler}, R., {Reichardt}, C.~L., {Aird}, K.~A., {et~al.}
\newblock {A Measurement of the Damping Tail of the Cosmic Microwave Background
  Power Spectrum with the South Pole Telescope}. 2011, \apj, 743, 28,
  \dodoi{10.1088/0004-637X/743/1/28}

\bibitem[{{Hoffman} \& {Ribak}(1991)}]{hoffman91}
{Hoffman}, Y., \& {Ribak}, E.
\newblock {Constrained realizations of Gaussian fields - A simple algorithm}.
  1991, \apjl, 380, L5, \dodoi{10.1086/186160}

\bibitem[{{Benoit-L{\'e}vy} {et~al.}(2013){Benoit-L{\'e}vy}, {D{\'e}chelette},
  {Benabed}, {Cardoso}, {Hanson}, \& {Prunet}}]{benoitlevy13}
{Benoit-L{\'e}vy}, A., {D{\'e}chelette}, T., {Benabed}, K., {Cardoso}, J.-F.,
  {Hanson}, D., \& {Prunet}, S.
\newblock {Full-sky CMB lensing reconstruction in presence of sky-cuts}. 2013,
  \aap, 555, A37, \dodoi{10.1051/0004-6361/201321048}

\bibitem[{Raghunathan {et~al.}(2019)Raghunathan, Holder, Bartlett, Patil,
  Reichardt, \& Whitehorn}]{raghunathan19c}
Raghunathan, S., Holder, G.~P., Bartlett, J.~G., Patil, S., Reichardt, C.~L.,
  \& Whitehorn, N.
\newblock An inpainting approach to tackle the kinematic and thermal SZ induced
  biases in CMB-cluster lensing estimators. 2019, Journal of Cosmology and
  Astroparticle Physics, 2019, 037, \dodoi{10.1088/1475-7516/2019/11/037}

\bibitem[{{Bleem} {et~al.}(2015){Bleem}, {Stalder}, {de Haan}, {Aird}, {Allen},
  {Applegate}, {Ashby}, {Bautz}, {Bayliss}, {Benson}, {Bocquet}, {Brodwin},
  {Carlstrom}, {Chang}, {Chiu}, {Cho}, {Clocchiatti}, {Crawford}, {Crites},
  {Desai}, {Dietrich}, {Dobbs}, {Foley}, {Forman}, {George}, {Gladders},
  {Gonzalez}, {Halverson}, {Hennig}, {Hoekstra}, {Holder}, {Holzapfel},
  {Hrubes}, {Jones}, {Keisler}, {Knox}, {Lee}, {Leitch}, {Liu}, {Lueker},
  {Luong-Van}, {Mantz}, {Marrone}, {McDonald}, {McMahon}, {Meyer}, {Mocanu},
  {Mohr}, {Murray}, {Padin}, {Pryke}, {Reichardt}, {Rest}, {Ruel}, {Ruhl},
  {Saliwanchik}, {Saro}, {Sayre}, {Schaffer}, {Schrabback}, {Shirokoff},
  {Song}, {Spieler}, {Stanford}, {Staniszewski}, {Stark}, {Story}, {Stubbs},
  {Vanderlinde}, {Vieira}, {Vikhlinin}, {Williamson}, {Zahn}, \&
  {Zenteno}}]{bleem15b}
{Bleem}, L.~E., {Stalder}, B., {de Haan}, T., {et~al.}
\newblock {Galaxy Clusters Discovered via the Sunyaev-Zel'dovich Effect in the
  2500-Square-Degree SPT-SZ Survey}. 2015, \apjs, 216, 27,
  \dodoi{10.1088/0067-0049/216/2/27}

\bibitem[{Lembo {et~al.}(2022)Lembo, Fabbian, Carron, \& Lewis}]{lembo22}
Lembo, M., Fabbian, G., Carron, J., \& Lewis, A.
\newblock CMB lensing reconstruction biases from masking extragalactic sources.
  2022, Phys. Rev. D, 106, 023525, \dodoi{10.1103/PhysRevD.106.023525}

\bibitem[{{Planck Collaboration}(2018)}]{planck18_archive}
{Planck Collaboration}. 2018, {Planck Legacy Archive}.
\newblock \url{https://www.cosmos.esa.int/web/planck/pla}

\bibitem[{{Lewis} \& {Challinor}(2011)}]{lewis11b}
{Lewis}, A., \& {Challinor}, A. 2011, {CAMB: Code for Anisotropies in the
  Microwave Background}, Astrophysics Source Code Library, record
  ascl:1102.026.
\newblock \doeprint{1102.026}

\bibitem[{{G{\'o}rski} {et~al.}(2005){G{\'o}rski}, {Hivon}, {Banday},
  {Wandelt}, {Hansen}, {Reinecke}, \& {Bartelmann}}]{gorski05}
{G{\'o}rski}, K.~M., {Hivon}, E., {Banday}, A.~J., {Wandelt}, B.~D., {Hansen},
  F.~K., {Reinecke}, M., \& {Bartelmann}, M.
\newblock {HEALPix: A Framework for High-Resolution Discretization and Fast
  Analysis of Data Distributed on the Sphere}. 2005, \apj, 622, 759,
  \dodoi{10.1086/427976}

\bibitem[{{Lewis}(2005)}]{lewis05}
{Lewis}, A.
\newblock {Lensed CMB simulation and parameter estimation}. 2005, \prd, 71,
  083008, \dodoi{10.1103/PhysRevD.71.083008}

\bibitem[{{Reichardt} {et~al.}(2021){Reichardt}, {Patil}, {Ade}, {Anderson},
  {Austermann}, {Avva}, {Baxter}, {Beall}, {Bender}, {Benson}, {Bianchini},
  {Bleem}, {Carlstrom}, {Chang}, {Chaubal}, {Chiang}, {Chou}, {Citron},
  {Moran}, {Crawford}, {Crites}, {de Haan}, {Dobbs}, {Everett}, {Gallicchio},
  {George}, {Gilbert}, {Gupta}, {Halverson}, {Harrington}, {Henning}, {Hilton},
  {Holder}, {Holzapfel}, {Hrubes}, {Huang}, {Hubmayr}, {Irwin}, {Knox}, {Lee},
  {Li}, {Lowitz}, {Luong-Van}, {McMahon}, {Mehl}, {Meyer}, {Millea}, {Mocanu},
  {Mohr}, {Montgomery}, {Nadolski}, {Natoli}, {Nibarger}, {Noble}, {Novosad},
  {Omori}, {Padin}, {Pryke}, {Ruhl}, {Saliwanchik}, {Sayre}, {Schaffer},
  {Shirokoff}, {Sievers}, {Smecher}, {Spieler}, {Staniszewski}, {Stark},
  {Tucker}, {Vanderlinde}, {Veach}, {Vieira}, {Wang}, {Whitehorn},
  {Williamson}, {Wu}, \& {Yefremenko}}]{reichardt21}
{Reichardt}, C.~L., {Patil}, S., {Ade}, P.~A.~R., {et~al.}
\newblock {An Improved Measurement of the Secondary Cosmic Microwave Background
  Anisotropies from the SPT-SZ + SPTpol Surveys}. 2021, \apj, 908, 199,
  \dodoi{10.3847/1538-4357/abd407}

\bibitem[{{Viero} {et~al.}(2013){Viero}, {Wang}, {Zemcov}, {Addison},
  {Amblard}, {Arumugam}, {Aussel}, {B{\'e}thermin}, {Bock}, {Boselli}, {Buat},
  {Burgarella}, {Casey}, {Clements}, {Conley}, {Conversi}, {Cooray}, {De
  Zotti}, {Dowell}, {Farrah}, {Franceschini}, {Glenn}, {Griffin},
  {Hatziminaoglou}, {Heinis}, {Ibar}, {Ivison}, {Lagache}, {Levenson},
  {Marchetti}, {Marsden}, {Nguyen}, {O'Halloran}, {Oliver}, {Omont}, {Page},
  {Papageorgiou}, {Pearson}, {P{\'e}rez-Fournon}, {Pohlen}, {Rigopoulou},
  {Roseboom}, {Rowan-Robinson}, {Schulz}, {Scott}, {Seymour}, {Shupe}, {Smith},
  {Symeonidis}, {Vaccari}, {Valtchanov}, {Vieira}, {Wardlow}, \&
  {Xu}}]{viero13a}
{Viero}, M.~P., {Wang}, L., {Zemcov}, M., {et~al.}
\newblock {HerMES: Cosmic Infrared Background Anisotropies and the Clustering
  of Dusty Star-forming Galaxies}. 2013, \apj, 772, 77,
  \dodoi{10.1088/0004-637X/772/1/77}

\bibitem[{{Shaw} {et~al.}(2010){Shaw}, {Nagai}, {Bhattacharya}, \&
  {Lau}}]{shaw10}
{Shaw}, L.~D., {Nagai}, D., {Bhattacharya}, S., \& {Lau}, E.~T.
\newblock {Impact of Cluster Physics on the Sunyaev-Zel'dovich Power Spectrum}.
  2010, \apj, 725, 1452, \dodoi{10.1088/0004-637X/725/2/1452}

\bibitem[{{Shaw} {et~al.}(2012){Shaw}, {Rudd}, \& {Nagai}}]{shaw12}
{Shaw}, L.~D., {Rudd}, D.~H., \& {Nagai}, D.
\newblock {Deconstructing the Kinetic SZ Power Spectrum}. 2012, \apj, 756, 15,
  \dodoi{10.1088/0004-637X/756/1/15}

\bibitem[{{Zahn} {et~al.}(2012){Zahn}, {Reichardt}, {Shaw}, {Lidz}, {Aird},
  {Benson}, {Bleem}, {Carlstrom}, {Chang}, {Cho}, {Crawford}, {Crites}, {de
  Haan}, {Dobbs}, {Dor{\'e}}, {Dudley}, {George}, {Halverson}, {Holder},
  {Holzapfel}, {Hoover}, {Hou}, {Hrubes}, {Joy}, {Keisler}, {Knox}, {Lee},
  {Leitch}, {Lueker}, {Luong-Van}, {McMahon}, {Mehl}, {Meyer}, {Millea},
  {Mohr}, {Montroy}, {Natoli}, {Padin}, {Plagge}, {Pryke}, {Ruhl}, {Schaffer},
  {Shirokoff}, {Spieler}, {Staniszewski}, {Stark}, {Story}, {van Engelen},
  {Vanderlinde}, {Vieira}, \& {Williamson}}]{zahn12}
{Zahn}, O., {Reichardt}, C.~L., {Shaw}, L., {et~al.}
\newblock {Cosmic Microwave Background Constraints on the Duration and Timing
  of Reionization from the South Pole Telescope}. 2012, \apj, 756, 65,
  \dodoi{10.1088/0004-637X/756/1/65}

\bibitem[{{van Engelen} {et~al.}(2014){van Engelen}, {Bhattacharya}, {Sehgal},
  {Holder}, {Zahn}, \& {Nagai}}]{vanengelen14}
{van Engelen}, A., {Bhattacharya}, S., {Sehgal}, N., {Holder}, G.~P., {Zahn},
  O., \& {Nagai}, D.
\newblock {CMB Lensing Power Spectrum Biases from Galaxies and Clusters Using
  High-angular Resolution Temperature Maps}. 2014, \apj, 786, 13,
  \dodoi{10.1088/0004-637X/786/1/13}

\bibitem[{{Omori}(2022)}]{omori22}
{Omori}, Y.
\newblock {Agora: Multi-Component Simulation for Cross-Survey Science}. 2022,
  arXiv e-prints, arXiv:2212.07420, \dodoi{10.48550/arXiv.2212.07420}

\bibitem[{{Everett} {et~al.}(2020){Everett}, {Zhang}, {Crawford}, {Vieira},
  {Aravena}, {Archipley}, {Austermann}, {Benson}, {Bleem}, {Carlstrom},
  {Chang}, {Chapman}, {Crites}, {de Haan}, {Dobbs}, {George}, {Halverson},
  {Harrington}, {Holder}, {Holzapfel}, {Hrubes}, {Knox}, {Lee}, {Luong-Van},
  {Mangian}, {Marrone}, {McMahon}, {Meyer}, {Mocanu}, {Mohr}, {Natoli},
  {Padin}, {Pryke}, {Reichardt}, {Reuter}, {Ruhl}, {Sayre}, {Schaffer},
  {Shirokoff}, {Spilker}, {Stalder}, {Staniszewski}, {Stark}, {Story},
  {Switzer}, {Vanderlinde}, {Wei{\ss}}, \& {Williamson}}]{everett20}
{Everett}, W.~B., {Zhang}, L., {Crawford}, T.~M., {et~al.}
\newblock {Millimeter-wave Point Sources from the 2500 Square Degree SPT-SZ
  Survey: Catalog and Population Statistics}. 2020, \apj, 900, 55,
  \dodoi{10.3847/1538-4357/ab9df7}

\bibitem[{Cavaliere \& Fusco-Femiano(1978)}]{cavaliere78}
Cavaliere, A., \& Fusco-Femiano, R.
\newblock The Distribution of Hot Gas in Clusters of Galaxies. 1978, \aap, 70,
  677

\bibitem[{{Hu} \& {Okamoto}(2002)}]{hu02a}
{Hu}, W., \& {Okamoto}, T.
\newblock {Mass Reconstruction with Cosmic Microwave Background Polarization}.
  2002, \apj, 574, 566, \dodoi{10.1086/341110}

\bibitem[{Fabbian {et~al.}(2019)Fabbian, Lewis, \& Beck}]{fabbian19b}
Fabbian, G., Lewis, A., \& Beck, D.
\newblock CMB lensing reconstruction biases in cross-correlation with
  large-scale structure probes. 2019, Journal of Cosmology and Astroparticle
  Physics, 2019, 057, \dodoi{10.1088/1475-7516/2019/10/057}

\bibitem[{{Namikawa} {et~al.}(2013){Namikawa}, {Hanson}, \&
  {Takahashi}}]{namikawa13}
{Namikawa}, T., {Hanson}, D., \& {Takahashi}, R.
\newblock {Bias-hardened CMB lensing}. 2013, \mnras, 431, 609,
  \dodoi{10.1093/mnras/stt195}

\bibitem[{{Mocanu} {et~al.}(2019){Mocanu}, {Crawford}, {Aylor}, {Benson},
  {Bleem}, {Carlstrom}, {Chang}, {Cho}, {Chown}, {Crites}, {de Haan}, {Dobbs},
  {Everett}, {George}, {Halverson}, {Harrington}, {Henning}, {Holder},
  {Holzapfel}, {Hou}, {Hrubes}, {Knox}, {Lee}, {Luong-Van}, {Marrone},
  {McMahon}, {Meyer}, {Millea}, {Mohr}, {Natoli}, {Omori}, {Padin}, {Pryke},
  {Reichardt}, {Ruhl}, {Sayre}, {Schaffer}, {Shirokoff}, {Staniszewski},
  {Stark}, {Story}, {Vanderlinde}, {Vieira}, {Williamson}, \& {Wu}}]{mocanu19}
{Mocanu}, L.~M., {Crawford}, T.~M., {Aylor}, K., {et~al.}
\newblock {Consistency of cosmic microwave background temperature measurements
  in three frequency bands in the 2500-square-degree SPT-SZ survey}. 2019,
  \jcap, 2019, 038, \dodoi{10.1088/1475-7516/2019/07/038}

\bibitem[{{Planck Collaboration} {et~al.}(2016{\natexlab{b}}){Planck
  Collaboration}, {Aghanim}, {Arnaud}, {Ashdown}, {Aumont}, {Baccigalupi},
  {Banday}, {Barreiro}, {Bartlett}, {Bartolo}, \& et~al.}]{planck15-11}
{Planck Collaboration}, {Aghanim}, N., {Arnaud}, M., {Ashdown}, M., {Aumont},
  J., {Baccigalupi}, C., {Banday}, A.~J., {Barreiro}, R.~B., {Bartlett}, J.~G.,
  {Bartolo}, N., \& et~al.
\newblock {Planck 2015 results. XI. CMB power spectra, likelihoods, and
  robustness of parameters}. 2016{\natexlab{b}}, \aap, 594, A11,
  \dodoi{10.1051/0004-6361/201526926}

\bibitem[{{Dunn}(1961)}]{dunn61}
{Dunn}, O.~J.
\newblock Multiple Comparisons Among Means. 1961, American Statistical
  Association, 52

\bibitem[{{Namikawa} {et~al.}(2012){Namikawa}, {Yamauchi}, \&
  {Taruya}}]{namikawa12}
{Namikawa}, T., {Yamauchi}, D., \& {Taruya}, A.
\newblock {Full-sky lensing reconstruction of gradient and curl modes from CMB
  maps}. 2012, \jcap, 1, 007, \dodoi{10.1088/1475-7516/2012/01/007}

\bibitem[{{Robertson} \& {Lewis}(2023)}]{robertson23}
{Robertson}, M., \& {Lewis}, A.
\newblock {How to detect lensing rotation}. 2023, arXiv e-prints,
  arXiv:2303.13313, \dodoi{10.48550/arXiv.2303.13313}

\bibitem[{Peebles(1980)}]{peebles80}
Peebles, P. 1980, The Large Scale Structure of the Universe (Princeton:
  Princeton University Press)

\bibitem[{Mead {et~al.}(2020)Mead, Tr{\"o}ster, Heymans, Van~Waerbeke, \&
  McCarthy}]{mead20}
Mead, A.~J., Tr{\"o}ster, T., Heymans, C., Van~Waerbeke, L., \& McCarthy, I.~G.
\newblock A hydrodynamical halo model for weak-lensing cross correlations.
  2020, Astronomy \& Astrophysics, 641, A130,
  \dodoi{10.1051/0004-6361/202038308}

\bibitem[{{Madhavacheril} {et~al.}(2023){Madhavacheril}, {Qu}, {Sherwin},
  {MacCrann}, {Li}, {Abril-Cabezas}, {Ade}, {Aiola}, {Alford}, {Amiri},
  {Amodeo}, {An}, {Atkins}, {Austermann}, {Battaglia}, {Battistelli}, {Beall},
  {Bean}, {Beringue}, {Bhandarkar}, {Biermann}, {Bolliet}, {Bond}, {Cai},
  {Calabrese}, {Calafut}, {Capalbo}, {Carrero}, {Challinor}, {Chesmore}, {Cho},
  {Choi}, {Clark}, {C{\'o}rdova Rosado}, {Cothard}, {Coughlin}, {Coulton},
  {Crowley}, {Dalal}, {Darwish}, {Devlin}, {Dicker}, {Doze}, {Duell}, {Duff},
  {Duivenvoorden}, {Dunkley}, {D{\"u}nner}, {Fanfani}, {Fankhanel}, {Farren},
  {Ferraro}, {Freundt}, {Fuzia}, {Gallardo}, {Garrido}, {Givans}, {Gluscevic},
  {Golec}, {Guan}, {Hall}, {Halpern}, {Han}, {Harrison}, {Hasselfield},
  {Healy}, {Henderson}, {Hensley}, {Herv{\'\i}as-Caimapo}, {Hill}, {Hilton},
  {Hilton}, {Hincks}, {Hlo{\v{z}}ek}, {Ho}, {Huber}, {Hubmayr}, {Huffenberger},
  {Hughes}, {Irwin}, {Isopi}, {Jense}, {Keller}, {Kim}, {Knowles}, {Koopman},
  {Kosowsky}, {Kramer}, {Kusiak}, {La Posta}, {Lague}, {Lakey}, {Lee}, {Li},
  {Limon}, {Lokken}, {Louis}, {Lungu}, {MacInnis}, {Maldonado}, {Maldonado},
  {Mallaby-Kay}, {Marques}, {McMahon}, {Mehta}, {Menanteau}, {Moodley},
  {Morris}, {Mroczkowski}, {Naess}, {Namikawa}, {Nati}, {Newburgh}, {Nicola},
  {Niemack}, {Nolta}, {Orlowski-Scherer}, {Page}, {Pandey}, {Partridge},
  {Prince}, {Puddu}, {Radiconi}, {Robertson}, {Rojas}, {Sakuma}, {Salatino},
  {Schaan}, {Schmitt}, {Sehgal}, {Shaikh}, {Sierra}, {Sievers}, {Sif{\'o}n},
  {Simon}, {Sonka}, {Spergel}, {Staggs}, {Storer}, {Switzer}, {Tampier},
  {Thornton}, {Trac}, {Treu}, {Tucker}, {Ulluom}, {Vale}, {Van Engelen}, {Van
  Lanen}, {van Marrewijk}, {Vargas}, {Vavagiakis}, {Wagoner}, {Wang}, {Wenzl},
  {Wollack}, {Xu}, {Zago}, \& {Zhang}}]{madhavacheril23}
{Madhavacheril}, M.~S., {Qu}, F.~J., {Sherwin}, B.~D., {et~al.}
\newblock {The Atacama Cosmology Telescope: DR6 Gravitational Lensing Map and
  Cosmological Parameters}. 2023, arXiv e-prints, arXiv:2304.05203,
  \dodoi{10.48550/arXiv.2304.05203}

\bibitem[{{Schmittfull} {et~al.}(2013){Schmittfull}, {Challinor}, {Hanson}, \&
  {Lewis}}]{schmittfull13}
{Schmittfull}, M.~M., {Challinor}, A., {Hanson}, D., \& {Lewis}, A.
\newblock {Joint analysis of CMB temperature and lensing-reconstruction power
  spectra}. 2013, \prd, 88, 063012, \dodoi{10.1103/PhysRevD.88.063012}

\bibitem[{Motloch {et~al.}(2017)Motloch, Hu, \& Benoit-L\'evy}]{motloch17}
Motloch, P., Hu, W., \& Benoit-L\'evy, A.
\newblock CMB lens sample covariance and consistency relations. 2017, Phys.
  Rev. D, 95, 043518, \dodoi{10.1103/PhysRevD.95.043518}

\bibitem[{{Peloton} {et~al.}(2017){Peloton}, {Schmittfull}, {Lewis}, {Carron},
  \& {Zahn}}]{peloton17}
{Peloton}, J., {Schmittfull}, M., {Lewis}, A., {Carron}, J., \& {Zahn}, O.
\newblock {Full covariance of CMB and lensing reconstruction power spectra}.
  2017, \prd, 95, 043508, \dodoi{10.1103/PhysRevD.95.043508}

\bibitem[{Trendafilova(2023)}]{trendafilova23}
Trendafilova, C.
\newblock The impact of cross-covariances between the CMB and reconstructed
  lensing power. 2023, in prep

\bibitem[{{Hartlap} {et~al.}(2007){Hartlap}, {Simon}, \&
  {Schneider}}]{hartlap07}
{Hartlap}, J., {Simon}, P., \& {Schneider}, P.
\newblock {Why your model parameter confidences might be too optimistic.
  Unbiased estimation of the inverse covariance matrix}. 2007, \aap, 464, 399,
  \dodoi{10.1051/0004-6361:20066170}

\bibitem[{{Simard} {et~al.}(2018){Simard}, {Omori}, {Aylor}, {Baxter},
  {Benson}, {Bleem}, {Carlstrom}, {Chang}, {Cho}, {Chown}, {Crawford},
  {Crites}, {de Haan}, {Dobbs}, {Everett}, {George}, {Halverson}, {Harrington},
  {Henning}, {Holder}, {Hou}, {Holzapfel}, {Hrubes}, {Knox}, {Lee}, {Leitch},
  {Luong-Van}, {Manzotti}, {McMahon}, {Meyer}, {Mocanu}, {Mohr}, {Natoli},
  {Padin}, {Pryke}, {Reichardt}, {Ruhl}, {Sayre}, {Schaffer}, {Shirokoff},
  {Staniszewski}, {Stark}, {Story}, {Vand erlinde}, {Vieira}, {Williamson}, \&
  {Wu}}]{simard18}
{Simard}, G., {Omori}, Y., {Aylor}, K., {et~al.}
\newblock {Constraints on Cosmological Parameters from the Angular Power
  Spectrum of a Combined 2500 deg$^{2}$ SPT-SZ and Planck Gravitational Lensing
  Map}. 2018, \apj, 860, 137, \dodoi{10.3847/1538-4357/aac264}

\bibitem[{{Alam} {et~al.}(2017){Alam}, {Ata}, {Bailey}, {Beutler}, {Bizyaev},
  {Blazek}, {Bolton}, {Brownstein}, {Burden}, {Chuang}, {Comparat}, {Cuesta},
  {Dawson}, {Eisenstein}, {Escoffier}, {Gil-Mar{\'{\i}}n}, {Grieb}, {Hand},
  {Ho}, {Kinemuchi}, {Kirkby}, {Kitaura}, {Malanushenko}, {Malanushenko},
  {Maraston}, {McBride}, {Nichol}, {Olmstead}, {Oravetz}, {Padmanabhan},
  {Palanque-Delabrouille}, {Pan}, {Pellejero-Ibanez}, {Percival}, {Petitjean},
  {Prada}, {Price-Whelan}, {Reid}, {Rodr{\'{\i}}guez-Torres}, {Roe}, {Ross},
  {Ross}, {Rossi}, {Rubi{\~n}o-Mart{\'{\i}}n}, {Saito}, {Salazar-Albornoz},
  {Samushia}, {S{\'a}nchez}, {Satpathy}, {Schlegel}, {Schneider},
  {Sc{\'o}ccola}, {Seo}, {Sheldon}, {Simmons}, {Slosar}, {Strauss}, {Swanson},
  {Thomas}, {Tinker}, {Tojeiro}, {Maga{\~n}a}, {Vazquez}, {Verde}, {Wake},
  {Wang}, {Weinberg}, {White}, {Wood-Vasey}, {Y{\`e}che}, {Zehavi}, {Zhai}, \&
  {Zhao}}]{alam17}
{Alam}, S., {Ata}, M., {Bailey}, S., {et~al.}
\newblock {The clustering of galaxies in the completed SDSS-III Baryon
  Oscillation Spectroscopic Survey: cosmological analysis of the DR12 galaxy
  sample}. 2017, \mnras, 470, 2617, \dodoi{10.1093/mnras/stx721}

\bibitem[{{Ross} {et~al.}(2015){Ross}, {Samushia}, {Howlett}, {Percival},
  {Burden}, \& {Manera}}]{ross15}
{Ross}, A.~J., {Samushia}, L., {Howlett}, C., {Percival}, W.~J., {Burden}, A.,
  \& {Manera}, M.
\newblock {The clustering of the SDSS DR7 main Galaxy sample - I. A 4 per cent
  distance measure at z = 0.15}. 2015, \mnras, 449, 835,
  \dodoi{10.1093/mnras/stv154}

\bibitem[{{Beutler} {et~al.}(2011){Beutler}, {Blake}, {Colless}, {Jones},
  {Staveley-Smith}, {Campbell}, {Parker}, {Saunders}, \& {Watson}}]{beutler11}
{Beutler}, F., {Blake}, C., {Colless}, M., {Jones}, D.~H., {Staveley-Smith},
  L., {Campbell}, L., {Parker}, Q., {Saunders}, W., \& {Watson}, F.
\newblock {The 6dF Galaxy Survey: baryon acoustic oscillations and the local
  Hubble constant}. 2011, \mnras, 416, 3017,
  \dodoi{10.1111/j.1365-2966.2011.19250.x}

\bibitem[{Alam {et~al.}(2021)Alam, Aubert, Avila, Balland, Bautista, Bershady,
  Bizyaev, Blanton, Bolton, Bovy, Brinkmann, Brownstein, Burtin, Chabanier,
  Chapman, Choi, Chuang, Comparat, Cousinou, Cuceu, Dawson, de~la Torre,
  de~Mattia, Agathe, des Bourboux, Escoffier, Etourneau, Farr, Font-Ribera,
  Frinchaboy, Fromenteau, Gil-Mar\'{\i}n, Le~Goff, Gonzalez-Morales,
  Gonzalez-Perez, Grabowski, Guy, Hawken, Hou, Kong, Parker, Klaene, Kneib,
  Lin, Long, Lyke, de~la Macorra, Martini, Masters, Mohammad, Moon, Mueller,
  Mu\~noz Guti\'errez, Myers, Nadathur, Neveux, Newman, Noterdaeme, Oravetz,
  Oravetz, Palanque-Delabrouille, Pan, Paviot, Percival, P\'erez-R\`afols,
  Petitjean, Pieri, Prakash, Raichoor, Ravoux, Rezaie, Rich, Ross, Rossi,
  Ruggeri, Ruhlmann-Kleider, S\'anchez, S\'anchez, S\'anchez-Gallego, Sayres,
  Schneider, Seo, Shafieloo, Slosar, Smith, Stermer, Tamone, Tinker, Tojeiro,
  Vargas-Maga\~na, Variu, Wang, Weaver, Weijmans, Y\`eche, Zarrouk, Zhao, Zhao,
  \& Zheng}]{alam21}
Alam, S., Aubert, M., Avila, S., {et~al.}
\newblock Completed SDSS-IV extended Baryon Oscillation Spectroscopic Survey:
  Cosmological implications from two decades of spectroscopic surveys at the
  Apache Point Observatory. 2021, Phys. Rev. D, 103, 083533,
  \dodoi{10.1103/PhysRevD.103.083533}

\bibitem[{{Planck Collaboration} {et~al.}(2020{\natexlab{c}}){Planck
  Collaboration}, {Aghanim}, {Akrami}, {Ashdown}, {Aumont}, {Baccigalupi},
  {Ballardini}, {Banday}, {Barreiro}, {Bartolo}, {Basak}, {Benabed}, {Bernard},
  {Bersanelli}, {Bielewicz}, {Bock}, {Bond}, {Borrill}, {Bouchet}, {Boulanger},
  {Bucher}, {Burigana}, {Butler}, {Calabrese}, {Cardoso}, {Carron},
  {Casaponsa}, {Challinor}, {Chiang}, {Colombo}, {Combet}, {Crill}, {Cuttaia},
  {de Bernardis}, {de Rosa}, {de Zotti}, {Delabrouille}, {Delouis}, {Di
  Valentino}, {Diego}, {Dor{\'e}}, {Douspis}, {Ducout}, {Dupac}, {Dusini},
  {Efstathiou}, {Elsner}, {En{\ss}lin}, {Eriksen}, {Fantaye}, {Fernand
  ez-Cobos}, {Finelli}, {Frailis}, {Fraisse}, {Franceschi}, {Frolov},
  {Galeotta}, {Galli}, {Ganga}, {G{\'e}nova-Santos}, {Gerbino}, {Ghosh},
  {Giraud-H{\'e}raud}, {Gonz{\'a}lez-Nuevo}, {G{\'o}rski}, {Gratton},
  {Gruppuso}, {Gudmundsson}, {Hamann}, {Handley}, {Hansen}, {Herranz}, {Hivon},
  {Huang}, {Jaffe}, {Jones}, {Keih{\"a}nen}, {Keskitalo}, {Kiiveri}, {Kim},
  {Kisner}, {Krachmalnicoff}, {Kunz}, {Kurki-Suonio}, {Lagache}, {Lamarre},
  {Lasenby}, {Lattanzi}, {Lawrence}, {Le Jeune}, {Levrier}, {Lewis}, {Liguori},
  {Lilje}, {Lilley}, {Lindholm}, {L{\'o}pez-Caniego}, {Lubin}, {Ma},
  {Mac{\'\i}as-P{\'e}rez}, {Maggio}, {Maino}, {Mandolesi}, {Mangilli},
  {Marcos-Caballero}, {Maris}, {Martin}, {Mart{\'\i}nez-Gonz{\'a}lez},
  {Matarrese}, {Mauri}, {McEwen}, {Meinhold}, {Melchiorri}, {Mennella},
  {Migliaccio}, {Millea}, {Miville-Desch{\^e}nes}, {Molinari}, {Moneti},
  {Montier}, {Morgante}, {Moss}, {Natoli}, {N{\o}rgaard-Nielsen}, {Pagano},
  {Paoletti}, {Partridge}, {Patanchon}, {Peiris}, {Perrotta}, {Pettorino},
  {Piacentini}, {Polenta}, {Puget}, {Rachen}, {Reinecke}, {Remazeilles},
  {Renzi}, {Rocha}, {Rosset}, {Roudier}, {Rubi{\~n}o-Mart{\'\i}n},
  {Ruiz-Granados}, {Salvati}, {Sandri}, {Savelainen}, {Scott}, {Shellard},
  {Sirignano}, {Sirri}, {Spencer}, {Sunyaev}, {Suur-Uski}, {Tauber},
  {Tavagnacco}, {Tenti}, {Toffolatti}, {Tomasi}, {Trombetti}, {Valiviita}, {Van
  Tent}, {Vielva}, {Villa}, {Vittorio}, {Wandelt}, {Wehus}, {Zacchei}, \&
  {Zonca}}]{planck18-5}
{Planck Collaboration}, {Aghanim}, N., {Akrami}, Y., {et~al.}
\newblock {Planck 2018 results. V. CMB power spectra and likelihoods}.
  2020{\natexlab{c}}, \aap, 641, A5, \dodoi{10.1051/0004-6361/201936386}

\bibitem[{{Cooke} {et~al.}(2018){Cooke}, {Pettini}, \& {Steidel}}]{cooke18}
{Cooke}, R.~J., {Pettini}, M., \& {Steidel}, C.~C.
\newblock {One Percent Determination of the Primordial Deuterium Abundance}.
  2018, \apj, 855, 102, \dodoi{10.3847/1538-4357/aaab53}

\bibitem[{{Mossa} {et~al.}(2020){Mossa}, {St{\"o}ckel}, {Cavanna}, {Ferraro},
  {Aliotta}, {Barile}, {Bemmerer}, {Best}, {Boeltzig}, {Broggini}, {Bruno},
  {Caciolli}, {Chillery}, {Ciani}, {Corvisiero}, {Csedreki}, {Davinson},
  {Depalo}, {Di Leva}, {Elekes}, {Fiore}, {Formicola}, {F{\"u}l{\"o}p},
  {Gervino}, {Guglielmetti}, {Gustavino}, {Gy{\"u}rky}, {Imbriani}, {Junker},
  {Kievsky}, {Kochanek}, {Lugaro}, {Marcucci}, {Mangano}, {Marigo}, {Masha},
  {Menegazzo}, {Pantaleo}, {Paticchio}, {Perrino}, {Piatti}, {Pisanti},
  {Prati}, {Schiavulli}, {Straniero}, {Sz{\"u}cs}, {Tak{\'a}cs}, {Trezzi},
  {Viviani}, \& {Zavatarelli}}]{mossa20}
{Mossa}, V., {St{\"o}ckel}, K., {Cavanna}, F., {et~al.}
\newblock {The baryon density of the Universe from an improved rate of
  deuterium burning}. 2020, \nat, 587, 210, \dodoi{10.1038/s41586-020-2878-4}

\bibitem[{{Pan} {et~al.}(2014){Pan}, {Knox}, \& {White}}]{pan14}
{Pan}, Z., {Knox}, L., \& {White}, M.
\newblock {Dependence of the cosmic microwave background lensing power spectrum
  on the matter density}. 2014, \mnras, 445, 2941,
  \dodoi{10.1093/mnras/stu1971}

\bibitem[{Baxter \& Sherwin(2021)}]{baxter21}
Baxter, E.~J., \& Sherwin, B.~D.
\newblock {Determining the Hubble Constant without the Sound Horizon Scale:
  Measurements from CMB Lensing}. 2021, Mon. Not. Roy. Astron. Soc., 501, 1823,
  \dodoi{10.1093/mnras/staa3706}

\bibitem[{{Freedman} {et~al.}(2019){Freedman}, {Madore}, {Hatt}, {Hoyt},
  {Jang}, {Beaton}, {Burns}, {Lee}, {Monson}, {Neeley}, {Phillips}, {Rich}, \&
  {Seibert}}]{freedman19}
{Freedman}, W.~L., {Madore}, B.~F., {Hatt}, D., {Hoyt}, T.~J., {Jang}, I.-S.,
  {Beaton}, R.~L., {Burns}, C.~R., {Lee}, M.~G., {Monson}, A.~J., {Neeley},
  J.~R., {Phillips}, M.~M., {Rich}, J.~A., \& {Seibert}, M.
\newblock {The Carnegie-Chicago Hubble Program. VIII. An Independent
  Determination of the Hubble Constant Based on the Tip of the Red Giant
  Branch}. 2019, arXiv e-prints, arXiv:1907.05922.
\newblock \doarXiv{1907.05922}

\bibitem[{{Wu} {et~al.}(2020){Wu}, {Motloch}, {Hu}, \& {Raveri}}]{wu20}
{Wu}, W.~L.~K., {Motloch}, P., {Hu}, W., \& {Raveri}, M.
\newblock {Hubble constant difference between CMB lensing and BAO
  measurements}. 2020, \prd, 102, 023510, \dodoi{10.1103/PhysRevD.102.023510}

\bibitem[{Abdalla {et~al.}(2022)}]{abdalla22}
Abdalla, E., {et~al.}
\newblock {Cosmology intertwined: A review of the particle physics,
  astrophysics, and cosmology associated with the cosmological tensions and
  anomalies}. 2022, JHEAp, 34, 49, \dodoi{10.1016/j.jheap.2022.04.002}

\bibitem[{{Amon} {et~al.}(2022){Amon}, {Gruen}, {Troxel}, {MacCrann},
  {Dodelson}, {Choi}, {Doux}, {Secco}, {Samuroff}, {Krause}, {Cordero},
  {Myles}, {DeRose}, {Wechsler}, {Gatti}, {Navarro-Alsina}, {Bernstein},
  {Jain}, {Blazek}, {Alarcon}, {Fert{\'e}}, {Lemos}, {Raveri}, {Campos},
  {Prat}, {S{\'a}nchez}, {Jarvis}, {Alves}, {Andrade-Oliveira}, {Baxter},
  {Bechtol}, {Becker}, {Bridle}, {Camacho}, {Carnero Rosell}, {Carrasco Kind},
  {Cawthon}, {Chang}, {Chen}, {Chintalapati}, {Crocce}, {Davis}, {Diehl},
  {Drlica-Wagner}, {Eckert}, {Eifler}, {Elvin-Poole}, {Everett}, {Fang},
  {Fosalba}, {Friedrich}, {Gaztanaga}, {Giannini}, {Gruendl}, {Harrison},
  {Hartley}, {Herner}, {Huang}, {Huff}, {Huterer}, {Kuropatkin}, {Leget},
  {Liddle}, {McCullough}, {Muir}, {Pandey}, {Park}, {Porredon}, {Refregier},
  {Rollins}, {Roodman}, {Rosenfeld}, {Ross}, {Rykoff}, {Sanchez},
  {Sevilla-Noarbe}, {Sheldon}, {Shin}, {Troja}, {Tutusaus}, {Tutusaus},
  {Varga}, {Weaverdyck}, {Yanny}, {Yin}, {Zhang}, {Zuntz}, {Aguena}, {Allam},
  {Annis}, {Bacon}, {Bertin}, {Bhargava}, {Brooks}, {Buckley-Geer}, {Burke},
  {Carretero}, {Costanzi}, {da Costa}, {Pereira}, {De Vicente}, {Desai},
  {Dietrich}, {Doel}, {Ferrero}, {Flaugher}, {Frieman}, {Garc{\'\i}a-Bellido},
  {Gaztanaga}, {Gerdes}, {Giannantonio}, {Gschwend}, {Gutierrez}, {Hinton},
  {Hollowood}, {Honscheid}, {Hoyle}, {James}, {Kron}, {Kuehn}, {Lahav}, {Lima},
  {Lin}, {Maia}, {Marshall}, {Martini}, {Melchior}, {Menanteau}, {Miquel},
  {Mohr}, {Morgan}, {Ogando}, {Palmese}, {Paz-Chinch{\'o}n}, {Petravick},
  {Pieres}, {Romer}, {Sanchez}, {Scarpine}, {Schubnell}, {Serrano}, {Smith},
  {Soares-Santos}, {Tarle}, {Thomas}, {To}, {Weller}, \& {DES
  Collaboration}}]{amon22}
{Amon}, A., {Gruen}, D., {Troxel}, M.~A., {et~al.}
\newblock {Dark Energy Survey Year 3 results: Cosmology from cosmic shear and
  robustness to data calibration}. 2022, \prd, 105, 023514,
  \dodoi{10.1103/PhysRevD.105.023514}

\bibitem[{Secco {et~al.}(2022)Secco, Samuroff, Krause, Jain, Blazek, Raveri,
  Campos, Amon, Chen, Doux, Choi, Gruen, Bernstein, Chang, DeRose, Myles,
  Fert\'e, Lemos, Huterer, Prat, Troxel, MacCrann, Liddle, Kacprzak, Fang,
  S\'anchez, Pandey, Dodelson, Chintalapati, Hoffmann, Alarcon, Alves,
  Andrade-Oliveira, Baxter, Bechtol, Becker, Brandao-Souza, Camacho,
  Carnero~Rosell, Carrasco~Kind, Cawthon, Cordero, Crocce, Davis, Di~Valentino,
  Drlica-Wagner, Eckert, Eifler, Elidaiana, Elsner, Elvin-Poole, Everett,
  Fosalba, Friedrich, Gatti, Giannini, Gruendl, Harrison, Hartley, Herner,
  Huang, Huff, Jarvis, Jeffrey, Kuropatkin, Leget, Muir, Mccullough,
  Navarro~Alsina, Omori, Park, Porredon, Rollins, Roodman, Rosenfeld, Ross,
  Rykoff, Sanchez, Sevilla-Noarbe, Sheldon, Shin, Troja, Tutusaus, Varga,
  Weaverdyck, Wechsler, Yanny, Yin, Zhang, Zuntz, Abbott, Aguena, Allam, Annis,
  Bacon, Bertin, Bhargava, Bridle, Brooks, Buckley-Geer, Burke, Carretero,
  Costanzi, da~Costa, De~Vicente, Diehl, Dietrich, Doel, Ferrero, Flaugher,
  Frieman, Garc\'{\i}a-Bellido, Gaztanaga, Gerdes, Giannantonio, Gschwend,
  Gutierrez, Hinton, Hollowood, Honscheid, Hoyle, James, Jeltema, Kuehn, Lahav,
  Lima, Lin, Maia, Marshall, Martini, Melchior, Menanteau, Miquel, Mohr,
  Morgan, Ogando, Palmese, Paz-Chinch\'on, Petravick, Pieres, Plazas~Malag\'on,
  Rodriguez-Monroy, Romer, Sanchez, Scarpine, Schubnell, Scolnic, Serrano,
  Smith, Soares-Santos, Suchyta, Swanson, Tarle, Thomas, \& To}]{secco22}
Secco, L.~F., Samuroff, S., Krause, E., {et~al.}
\newblock Dark Energy Survey Year 3 results: Cosmology from cosmic shear and
  robustness to modeling uncertainty. 2022, Phys. Rev. D, 105, 023515,
  \dodoi{10.1103/PhysRevD.105.023515}

\bibitem[{{Li} {et~al.}(2023){Li}, {Zhang}, {Sugiyama}, {Dalal}, {Rau},
  {Mandelbaum}, {Takada}, {More}, {Strauss}, {Miyatake}, {Shirasaki}, {Hamana},
  {Oguri}, {Luo}, {Nishizawa}, {Takahashi}, {Nicola}, {Osato}, {Kannawadi},
  {Sunayama}, {Armstrong}, {Komiyama}, {Lupton}, {Lust}, {Miyazaki},
  {Murayama}, {Nishimichi}, {Okura}, {Price}, {Tait}, {Tanaka}, \&
  {Wang}}]{li23}
{Li}, X., {Zhang}, T., {Sugiyama}, S., {et~al.}
\newblock {Hyper Suprime-Cam Year 3 Results: Cosmology from Cosmic Shear
  Two-point Correlation Functions}. 2023, arXiv e-prints, arXiv:2304.00702,
  \dodoi{10.48550/arXiv.2304.00702}

\bibitem[{{Dalal} {et~al.}(2023){Dalal}, {Li}, {Nicola}, {Zuntz}, {Strauss},
  {Sugiyama}, {Zhang}, {Rau}, {Mandelbaum}, {Takada}, {More}, {Miyatake},
  {Kannawadi}, {Shirasaki}, {Taniguchi}, {Takahashi}, {Osato}, {Hamana},
  {Oguri}, {Nishizawa}, {Plazas Malag{\'o}n}, {Sunayama}, {Alonso}, {Slosar},
  {Armstrong}, {Bosch}, {Komiyama}, {Lupton}, {Lust}, {MacArthur}, {Miyazaki},
  {Murayama}, {Nishimichi}, {Okura}, {Price}, {Tait}, {Tanaka}, \&
  {Wang}}]{dalal23}
{Dalal}, R., {Li}, X., {Nicola}, A., {et~al.}
\newblock {Hyper Suprime-Cam Year 3 Results: Cosmology from Cosmic Shear Power
  Spectra}. 2023, arXiv e-prints, arXiv:2304.00701,
  \dodoi{10.48550/arXiv.2304.00701}

\bibitem[{Abbott {et~al.}(2023)}]{abbott23b}
Abbott, T. M.~C., {et~al.}
\newblock {DES Y3 + KiDS-1000: Consistent cosmology combining cosmic shear
  surveys}. 2023.
\newblock \doarXiv{2305.17173}

\bibitem[{Motloch \& Hu(2020)}]{motloch20}
Motloch, P., \& Hu, W.
\newblock Lensinglike tensions in the $Planck$ legacy release. 2020, Phys. Rev.
  D, 101, 083515, \dodoi{10.1103/PhysRevD.101.083515}

\bibitem[{Lemos \& Lewis(2023)}]{lemos23}
Lemos, P., \& Lewis, A.
\newblock {CMB constraints on the early Universe independent of late-time
  cosmology}. 2023, Phys. Rev. D, 107, 103505,
  \dodoi{10.1103/PhysRevD.107.103505}

\bibitem[{Rosenberg {et~al.}(2022)Rosenberg, Gratton, \&
  Efstathiou}]{rosenberg22}
Rosenberg, E., Gratton, S., \& Efstathiou, G.
\newblock {CMB power spectra and cosmological parameters from Planck PR4 with
  CamSpec}. 2022, Monthly Notices of the Royal Astronomical Society, 517, 4620,
  \dodoi{10.1093/mnras/stac2744}

\bibitem[{{Aiola} {et~al.}(2020){Aiola}, {Calabrese}, {Maurin}, {Naess},
  {Schmitt}, {Abitbol}, {Addison}, {Ade}, {Alonso}, {Amiri}, {Amodeo},
  {Angile}, {Austermann}, {Baildon}, {Battaglia}, {Beall}, {Bean}, {Becker},
  {Bond}, {Bruno}, {Calafut}, {Campusano}, {Carrero}, {Chesmore}, {Cho.},
  {Choi}, {Clark}, {Cothard}, {Crichton}, {Crowley}, {Darwish}, {Datta},
  {Denison}, {Devlin}, {Duell}, {Duff}, {Duivenvoorden}, {Dunkley},
  {D{\"u}nner}, {Essinger-Hileman}, {Fankhanel}, {Ferraro}, {Fox}, {Fuzia},
  {Gallardo}, {Gluscevic}, {Golec}, {Grace}, {Gralla}, {Guan}, {Hall},
  {Halpern}, {Han}, {Hargrave}, {Hasselfield}, {Helton}, {Henderson},
  {Hensley}, {Hill}, {Hilton}, {Hilton}, {Hincks}, {Hlo{\v{z}}ek}, {Ho},
  {Hubmayr}, {Huffenberger}, {Hughes}, {Infante}, {Irwin}, {Jackson}, {Klein},
  {Knowles}, {Koopman}, {Kosowsky}, {Lakey}, {Li}, {Li}, {Li}, {Lokken},
  {Louis}, {Lungu}, {MacInnis}, {Madhavacheril}, {Maldonado}, {Mallaby-Kay},
  {Marsden}, {McMahon}, {Menanteau}, {Moodley}, {Morton}, {Namikawa}, {Nati},
  {Newburgh}, {Nibarger}, {Nicola}, {Niemack}, {Nolta}, {Orlowski-Sherer},
  {Page}, {Pappas}, {Partridge}, {Phakathi}, {Prince}, {Puddu}, {Qu}, {Rivera},
  {Robertson}, {Rojas}, {Salatino}, {Schaan}, {Schillaci}, {Sehgal}, {Sherwin},
  {Sierra}, {Sievers}, {Sifon}, {Sikhosana}, {Simon}, {Spergel}, {Staggs},
  {Stevens}, {Storer}, {Sunder}, {Switzer}, {Thorne}, {Thornton}, {Trac},
  {Treu}, {Tucker}, {Vale}, {Van Engelen}, {Van Lanen}, {Vavagiakis},
  {Wagoner}, {Wang}, {Ward}, {Wollack}, {Xu}, {Zago}, \& {Zhu}}]{aiola20}
{Aiola}, S., {Calabrese}, E., {Maurin}, L., {et~al.}
\newblock {The Atacama Cosmology Telescope: DR4 Maps and Cosmological
  Parameters}. 2020, arXiv e-prints, arXiv:2007.07288.
\newblock \doarXiv{2007.07288}

\bibitem[{{Calabrese} {et~al.}(2008){Calabrese}, {Slosar}, {Melchiorri},
  {Smoot}, \& {Zahn}}]{calabrese08}
{Calabrese}, E., {Slosar}, A., {Melchiorri}, A., {Smoot}, G.~F., \& {Zahn}, O.
\newblock {Cosmic microwave weak lensing data as a test for the dark universe}.
  2008, \prd, 77, 123531, \dodoi{10.1103/PhysRevD.77.123531}

\bibitem[{{Smith} {et~al.}(2009){Smith}, {Cooray}, {Das}, {Dor{\'e}}, {Hanson},
  {Hirata}, {Kaplinghat}, {Keating}, {Loverde}, {Miller}, {Rocha}, {Shimon}, \&
  {Zahn}}]{smith09}
{Smith}, K.~M., {Cooray}, A., {Das}, S., {Dor{\'e}}, O., {Hanson}, D.,
  {Hirata}, C., {Kaplinghat}, M., {Keating}, B., {Loverde}, M., {Miller}, N.,
  {Rocha}, G., {Shimon}, M., \& {Zahn}, O.
\newblock {Gravitational Lensing}. 2009, in American Institute of Physics
  Conference Series, Vol. 1141, American Institute of Physics Conference
  Series, 121--178, \dodoi{10.1063/1.3160886}

\bibitem[{{Abazajian} {et~al.}(2015){Abazajian}, {Arnold}, {Austermann},
  {Benson}, {Bischoff}, {Bock}, {Bond}, {Borrill}, {Calabrese}, {Carlstrom},
  {Carvalho}, {Chang}, {Chiang}, {Church}, {Cooray}, {Crawford}, {Dawson},
  {Das}, {Devlin}, {Dobbs}, {Dodelson}, {Dor{\'e}}, {Dunkley}, {Errard},
  {Fraisse}, {Gallicchio}, {Halverson}, {Hanany}, {Hildebrandt}, {Hincks},
  {Hlozek}, {Holder}, {Holzapfel}, {Honscheid}, {Hu}, {Hubmayr}, {Irwin},
  {Jones}, {Kamionkowski}, {Keating}, {Keisler}, {Knox}, {Komatsu}, {Kovac},
  {Kuo}, {Lawrence}, {Lee}, {Leitch}, {Linder}, {Lubin}, {McMahon}, {Miller},
  {Newburgh}, {Niemack}, {Nguyen}, {Nguyen}, {Page}, {Pryke}, {Reichardt},
  {Ruhl}, {Sehgal}, {Seljak}, {Sievers}, {Silverstein}, {Slosar}, {Smith},
  {Spergel}, {Staggs}, {Stark}, {Stompor}, {Vieregg}, {Wang}, {Watson},
  {Wollack}, {Wu}, {Yoon}, \& {Zahn}}]{abazajian15b}
{Abazajian}, K.~N., {Arnold}, K., {Austermann}, J., {et~al.}
\newblock {Neutrino physics from the cosmic microwave background and large
  scale structure}. 2015, Astroparticle Physics, 63, 66,
  \dodoi{10.1016/j.astropartphys.2014.05.014}

\bibitem[{Fukuda {et~al.}(1998)Fukuda, Hayakawa, Ichihara, Inoue, Ishihara,
  Ishino, Itow, Kajita, Kameda, Kasuga, Kobayashi, Kobayashi, Koshio, Miura,
  Nakahata, Nakayama, Okada, Okumura, Sakurai, Shiozawa, Suzuki, Takeuchi,
  Totsuka, Yamada, Earl, Habig, Kearns, Messier, Scholberg, Stone, Sulak,
  Walter, Goldhaber, Barszczxak, Casper, Gajewski, Halverson, Hsu, Kropp,
  Price, Reines, Smy, Sobel, Vagins, Ganezer, Keig, Ellsworth, Tasaka,
  Flanagan, Kibayashi, Learned, Matsuno, Stenger, Takemori, Ishii, Kanzaki,
  Kobayashi, Mine, Nakamura, Nishikawa, Oyama, Sakai, Sakuda, Sasaki, Echigo,
  Kohama, Suzuki, Haines, Blaufuss, Kim, Sanford, Svoboda, Chen, Conner,
  Goodman, Sullivan, Hill, Jung, Martens, Mauger, McGrew, Sharkey, Viren,
  Yanagisawa, Doki, Miyano, Okazawa, Saji, Takahata, Nagashima, Takita,
  Yamaguchi, Yoshida, Kim, Etoh, Fujita, Hasegawa, Hasegawa, Hatakeyama,
  Iwamoto, Koga, Maruyama, Ogawa, Shirai, Suzuki, Tsushima, Koshiba, Nemoto,
  Nishijima, Futagami, Hayato, Kanaya, Kaneyuki, Watanabe, Kielczewska, Doyle,
  George, Stachyra, Wai, Wilkes, \& Young}]{fukuda98}
Fukuda, Y., Hayakawa, T., Ichihara, E., {et~al.}
\newblock Evidence for Oscillation of Atmospheric Neutrinos. 1998, Phys. Rev.
  Lett., 81, 1562, \dodoi{10.1103/PhysRevLett.81.1562}

\bibitem[{Ahmad {et~al.}(2002)Ahmad, Allen, Andersen, D.Anglin, Barton, Beier,
  Bercovitch, Bigu, Biller, Black, Blevis, Boardman, Boger, Bonvin, Boulay,
  Bowler, Bowles, Brice, Browne, Bullard, B\"uhler, Cameron, Chan, Chen, Chen,
  Chen, Cleveland, Clifford, Cowan, Cowen, Cox, Dai, Dalnoki-Veress, Davidson,
  Doe, Doucas, Dragowsky, Duba, Duncan, Dunford, Dunmore, Earle, Elliott,
  Evans, Ewan, Farine, Fergani, Ferraris, Ford, Formaggio, Fowler, Frame,
  Frank, Frati, Gagnon, Germani, Gil, Graham, Grant, Hahn, Hallin, Hallman,
  Hamer, Hamian, Handler, Haq, Hargrove, Harvey, Hazama, Heeger, Heintzelman,
  Heise, Helmer, Hepburn, Heron, Hewett, Hime, Howe, Hykawy, Isaac, Jagam,
  Jelley, Jillings, Jonkmans, Kazkaz, Keener, Klein, Knox, Komar, Kouzes,
  Kutter, Kyba, Law, Lawson, Lay, Lee, Lesko, Leslie, Levine, Locke, Luoma,
  Lyon, Majerus, Mak, Maneira, Manor, Marino, McCauley, McDonald, McDonald,
  McFarlane, McGregor, Meijer~Drees, Mifflin, Miller, Milton, Moffat, Moorhead,
  Nally, Neubauer, Newcomer, Ng, Noble, Norman, Novikov, O'Neill, Okada,
  Ollerhead, Omori, Orrell, Oser, Poon, Radcliffe, Roberge, Robertson,
  Robertson, Rosendahl, Rowley, Rusu, Saettler, Schaffer, Schwendener,
  Sch\"ulke, Seifert, Shatkay, Simpson, Sims, Sinclair, Skensved, Smith, Smith,
  Spreitzer, Starinsky, Steiger, Stokstad, Stonehill, Storey, Sur, Tafirout,
  Tagg, Tanner, Taplin, Thorman, Thornewell, Trent, Tserkovnyak, Van~Berg,
  Van~de Water, Virtue, Waltham, Wang, Wark, West, Wilhelmy, Wilkerson, Wilson,
  Wittich, Wouters, \& Yeh}]{ahmad02}
Ahmad, Q.~R., Allen, R.~C., Andersen, T.~C., {et~al.}
\newblock Direct Evidence for Neutrino Flavor Transformation from
  Neutral-Current Interactions in the Sudbury Neutrino Observatory. 2002, Phys.
  Rev. Lett., 89, 011301, \dodoi{10.1103/PhysRevLett.89.011301}

\bibitem[{Aker {et~al.}(2022)}]{aker22}
Aker, M., {et~al.}
\newblock {Direct neutrino-mass measurement with sub-electronvolt sensitivity}.
  2022, Nature Phys., 18, 160, \dodoi{10.1038/s41567-021-01463-1}

\bibitem[{{CMB-S4 Collaboration} {et~al.}(2019){CMB-S4 Collaboration},
  {Abazajian}, {Addison}, {Adshead}, {Ahmed}, {Allen}, {Alonso}, {Alvarez},
  {Anderson}, {Arnold}, {Baccigalupi}, \& et~al.}]{abazajian19}
{CMB-S4 Collaboration}, {Abazajian}, K., {Addison}, G., {Adshead}, P., {Ahmed},
  Z., {Allen}, S.~W., {Alonso}, D., {Alvarez}, M., {Anderson}, A., {Arnold},
  K.~S., {Baccigalupi}, C., \& et~al.
\newblock {CMB-S4 Science Case, Reference Design, and Project Plan}. 2019,
  arXiv e-prints, arXiv:1907.04473.
\newblock \doarXiv{1907.04473}

\bibitem[{{Gerbino} {et~al.}(2022){Gerbino}, {Grohs}, {Lattanzi}, {Abazajian},
  {Blinov}, {Brinckmann}, {Chen}, {Djurcic}, {Du}, {Escudero}, {Hagstotz},
  {Kelly}, {Lorenz}, {Loverde}, {Mart{\'\i}nez-Mirav{\'e}}, {Mena}, {Meyers},
  {Pettus}, {Saviano}, {Suliga}, {Takhistov}, {T{\'o}rtola}, {Valle}, \&
  {Wallisch}}]{gerbino22}
{Gerbino}, M., {Grohs}, E., {Lattanzi}, M., {et~al.}
\newblock {Synergy between cosmological and laboratory searches in neutrino
  physics}. 2022, arXiv e-prints, arXiv:2203.07377,
  \dodoi{10.48550/arXiv.2203.07377}

\bibitem[{{Stompor} \& {Efstathiou}(1999)}]{stompor99}
{Stompor}, R., \& {Efstathiou}, G.
\newblock {Gravitational lensing of cosmic microwave background anisotropies
  and cosmological parameter estimation}. 1999, \mnras, 302, 735,
  \dodoi{10.1046/j.1365-8711.1999.02174.x}

\bibitem[{Handley(2021)}]{handley21}
Handley, W.
\newblock {Curvature tension: evidence for a closed universe}. 2021, Phys. Rev.
  D, 103, L041301, \dodoi{10.1103/PhysRevD.103.L041301}

\bibitem[{Di~Valentino {et~al.}(2019)Di~Valentino, Melchiorri, \&
  Silk}]{divalentino19}
Di~Valentino, E., Melchiorri, A., \& Silk, J.
\newblock {Planck evidence for a closed Universe and a possible crisis for
  cosmology}. 2019, Nature Astron., 4, 196, \dodoi{10.1038/s41550-019-0906-9}

\bibitem[{Millea \& Seljak(2022)}]{millea22}
Millea, M., \& Seljak, U. c.~v.
\newblock Marginal unbiased score expansion and application to CMB lensing.
  2022, Phys. Rev. D, 105, 103531, \dodoi{10.1103/PhysRevD.105.103531}

\bibitem[{Legrand \& Carron(2022)}]{legrand22}
Legrand, L., \& Carron, J.
\newblock Lensing power spectrum of the cosmic microwave background with deep
  polarization experiments. 2022, Phys. Rev. D, 105, 123519,
  \dodoi{10.1103/PhysRevD.105.123519}

\bibitem[{{Bianchini} \& {Millea}(2023)}]{bianchini23}
{Bianchini}, F., \& {Millea}, M.
\newblock {Inference of gravitational lensing and patchy reionization with
  future CMB data}. 2023, \prd, 107, 043521,
  \dodoi{10.1103/PhysRevD.107.043521}

\bibitem[{{Hui} {et~al.}(2018){Hui}, {Ade}, {Ahmed}, {Aikin}, {Alexander},
  {Barkats}, {Benton}, {Bischoff}, {Bock}, {Bowens-Rubin}, {Brevik}, {Buder},
  {Bullock}, {Buza}, {Connors}, {Cornelison}, {Crill}, {Crumrine}, {Dierickx},
  {Duband}, {Dvorkin}, {Filippini}, {Fliescher}, {Grayson}, {Hall}, {Halpern},
  {Harrison}, {Hildebrandt}, {Hilton}, {Irwin}, {Kang}, {Karkare}, {Karpel},
  {Kaufman}, {Keating}, {Kefeli}, {Kernasovskiy}, {Kovac}, {Kuo}, {Lau},
  {Larsen}, {Leitch}, {Lueker}, {Megerian}, {Moncelsi}, {Namikawa},
  {Netterfield}, {Nguyen}, {O'Brient}, {Ogburn}, {Palladino}, {Pryke},
  {Racine}, {Richter}, {Schwarz}, {Schillaci}, {Sheehy}, {Soliman},
  {St.~Germaine}, {Staniszewski}, {Steinbach}, {Sudiwala}, {Teply}, {Thompson},
  {Tolan}, {Tucker}, {Turner}, {Umilta}, {Vieregg}, {Wandui}, {Weber}, {Wiebe},
  {Willmert}, {Wong}, {Wu}, {Yang}, {Yoon}, \& {Zhang}}]{hui18}
{Hui}, H., {Ade}, P.~A.~R., {Ahmed}, Z., {et~al.}
\newblock {BICEP Array: a multi-frequency degree-scale CMB polarimeter}. 2018,
  in \procspie, Vol. 10708, Society of Photo-Optical Instrumentation Engineers
  (SPIE) Conference Series, 1070807, \dodoi{10.1117/12.2311725}

\bibitem[{{Ade} {et~al.}(2021){Ade}, {Ahmed}, {Amiri}, {Barkats}, {Thakur},
  {Bischoff}, {Beck}, {Bock}, {Boenish}, {Bullock}, {Buza}, {Cheshire},
  {Connors}, {Cornelison}, {Crumrine}, {Cukierman}, {Denison}, {Dierickx},
  {Duband}, {Eiben}, {Fatigoni}, {Filippini}, {Fliescher}, {Goeckner-Wald},
  {Goldfinger}, {Grayson}, {Grimes}, {Hall}, {Halal}, {Halpern}, {Hand},
  {Harrison}, {Henderson}, {Hildebrandt}, {Hilton}, {Hubmayr}, {Hui}, {Irwin},
  {Kang}, {Karkare}, {Karpel}, {Kefeli}, {Kernasovskiy}, {Kovac}, {Kuo}, {Lau},
  {Leitch}, {Lennox}, {Megerian}, {Minutolo}, {Moncelsi}, {Nakato}, {Namikawa},
  {Nguyen}, {O'Brient}, {Ogburn}, {Palladino}, {Prouve}, {Pryke}, {Racine},
  {Reintsema}, {Richter}, {Schillaci}, {Schwarz}, {Schmitt}, {Sheehy},
  {Soliman}, {Germaine}, {Steinbach}, {Sudiwala}, {Teply}, {Thompson}, {Tolan},
  {Tucker}, {Turner}, {Umilt{\`a}}, {Verg{\`e}s}, {Vieregg}, {Wandui}, {Weber},
  {Wiebe}, {Willmert}, {Wong}, {Wu}, {Yang}, {Yoon}, {Young}, {Yu}, {Zeng},
  {Zhang}, {Zhang}, \& {Bicep/Keck Collaboration}}]{bicep2keck21}
{Ade}, P.~A.~R., {Ahmed}, Z., {Amiri}, M., {et~al.}
\newblock {Improved Constraints on Primordial Gravitational Waves using Planck,
  WMAP, and BICEP/Keck Observations through the 2018 Observing Season}. 2021,
  \prl, 127, 151301, \dodoi{10.1103/PhysRevLett.127.151301}

\bibitem[{Pordes {et~al.}(2007)}]{pordes07}
Pordes, R., {et~al.}
\newblock {The Open Science Grid}. 2007, J. Phys. Conf. Ser., 78, 012057,
  \dodoi{10.1088/1742-6596/78/1/012057}

\bibitem[{{Sfiligoi} {et~al.}(2009){Sfiligoi}, {Bradley}, {Holzman},
  {Mhashilkar}, {Padhi}, \& {Wurthwein}}]{sfiligoi09}
{Sfiligoi}, I., {Bradley}, D.~C., {Holzman}, B., {Mhashilkar}, P., {Padhi}, S.,
  \& {Wurthwein}, F.
\newblock The Pilot Way to Grid Resources Using glideinWMS. 2009, in 2, Vol.~2,
  2009 WRI World Congress on Computer Science and Information Engineering,
  428--432, \dodoi{10.1109/CSIE.2009.950}

\end{thebibliography}

\end{document}